%% file: main.tex
\documentclass[acmsmall,screen]{acmart}

\AtBeginDocument{%
  \providecommand\BibTeX{{%
    \normalfont B\kern-0.5em{\scshape i\kern-0.25em b}\kern-0.8em\TeX}}}

\setcopyright{acmcopyright}

\acmJournal{CSUR}

\usepackage{multirow}
\usepackage{stfloats}
\usepackage[T1]{fontenc}
\usepackage{dfadobe} 
\usepackage{xcolor}
\usepackage{array}
\usepackage{pifont}%
\usepackage{soul}
\usepackage{changepage}
\usepackage[absolute,overlay]{textpos}
\definecolor{ticgreen}{rgb}{0.55, 0.71, 0.0}
\renewcommand{\checkmark}{\ding{51}}%
\newcommand{\xmark}{\ding{55}}%
\usepackage{lscape,lipsum,graphicx} %
\usepackage{pdflscape}
\usepackage[absolute]{textpos}
\usepackage{fancyhdr}
\usepackage{afterpage}
\definecolor{pink}{rgb}{1.0,0.55,0.75}
\definecolor{purple}{rgb}{0.52,0.2,0.925}

\newcommand{\new}[1]{\textcolor{black}{#1}}
\newcommand{\csur}[1]{\textcolor{black}{#1}}

\begin{document}

\title{Multimodality in VR: A survey}

\author{Daniel Martin}
\affiliation{%
  \institution{Universidad de Zaragoza, I3A}
  \city{Zaragoza}
  \country{Spain}}
\authornote{Both authors contributed equally.}
\email{danims@unizar.es}

\author{Sandra Malpica}
\authornotemark[1]
\affiliation{%
  \institution{Universidad de Zaragoza, I3A}
  \city{Zaragoza}
  \country{Spain}}
\email{smalpica@unizar.es}

\author{Diego Gutierrez}
\affiliation{%
  \institution{Universidad de Zaragoza, I3A}
  \city{Zaragoza}
  \country{Spain}}
\email{diegog@unizar.es}

\author{Belen Masia}
\affiliation{%
  \institution{Universidad de Zaragoza, I3A}
  \city{Zaragoza}
  \country{Spain}}
\email{bmasia@unizar.es}

\author{Ana Serrano}
\affiliation{%
  \institution{Universidad de Zaragoza, I3A}
  \city{Zaragoza}
  \country{Spain}}
\email{anase@unizar.es}

\renewcommand{\shortauthors}{D. Martin, and S. Malpica, and D. Gutierrez, and B. Masia, and A. Serrano}
\begin{abstract}
  Virtual reality (VR) is rapidly growing, with the potential to change the way we create and consume content. In VR, users integrate multimodal sensory information they receive, to create a unified perception of the virtual world. In this survey, we \new{review} the body of work addressing multimodality in VR, and its role and benefits in user experience, together with different applications that leverage multimodality in many disciplines. These works thus encompass several fields of research, and demonstrate that multimodality plays a fundamental role \new{in VR}, enhancing the experience, improving overall performance, and yielding unprecedented abilities in skill and knowledge transfer.
\end{abstract}



\maketitle

\input{n1_Introduction}

\input{n2_Scope_Structure}

\input{n3_Fidelity_Environment}

\input{n4_Self_Perception}

\input{n5_Saliency_Guidance}

\input{n6_Interaction}

\input{n7_Tricking}

\input{n8_Applications}

\input{n9_Conclusions}

\begin{acks}
This work has received funding from the European Research Council (ERC) under the European Union’s Horizon 2020 research and innovation programme (project CHAMELEON, Grant no. 682080). This work has received funding from the European Union’s Horizon 2020 research and innovation programme under the Marie Skłodowska-Curie grant agreement No 765121 and No 956585. Additionally, Daniel Martin and Sandra Malpica were supported by a Gobierno de Aragon (2020-2024 and 2018-2022) predoctoral grant.
\end{acks}

\bibliographystyle{ACM-Reference-Format}
\bibliography{references}

\end{document}

%% file: n1_Introduction.tex
\section{Introduction}
\label{sec:n1_intro}

Virtual Reality (VR) is inherently different from traditional media since it introduces additional degrees of freedom, a wider field of view, more sophisticated sound spatialization, or even gives users control of the camera. \new{VR immersive setups (such as head-mounted displays (HMDs) or CAVE-like systems)} thus have the potential to change the way in which content is consumed, increasing realism, immersion, and engagement. This has \new{impacted} 
many application areas such as education and training~\cite{checa2019review}, rehabilitation and neuroscience~\cite{wilson2015use,sano2015reliability}, or virtual cinematography \cite{Serrano_VR-cine_SIGGRAPH2017}. \new{One of the key aspects of these systems lies in their ability to reproduce sensory information from different modalities (mainly visual and auditory, but also haptic, olfactory, gustatory, or proprioceptive), giving them an unprecedented potential.}

Although visual stimuli tend to be the predominant source of information for humans~\cite{spence2017,burns2005}, additional sensory information helps to increase our understanding of the world. Our brain integrates different sources of %
\new{sensory feedback}
including both external stimuli (visual, auditory, or haptic information) and internal stimuli (vestibular or proprioceptive cues), thus creating a coherent, stable perception of objects,  events, and oneself. The unified experience of the world as we perceive it therefore emerges from these multimodal cues~\cite{Prinz2006-PRIITM, shams2010crossmodal}. %
These different sources of information must be correctly synchronized to be perceived as belonging together~\cite{nidiffer2018,powers2016}, and synchronization sensitivity varies depending on the context, task and individual~\cite{eg2015}. In general, different modalities will be perceived as coming from a single event or object as long as their temporal incongruency is shorter than their corresponding window of integration~\cite{malpica2020crossmodal}. 

\begin{figure}[tbp]
    \centering
    \includegraphics[width=0.80\columnwidth]{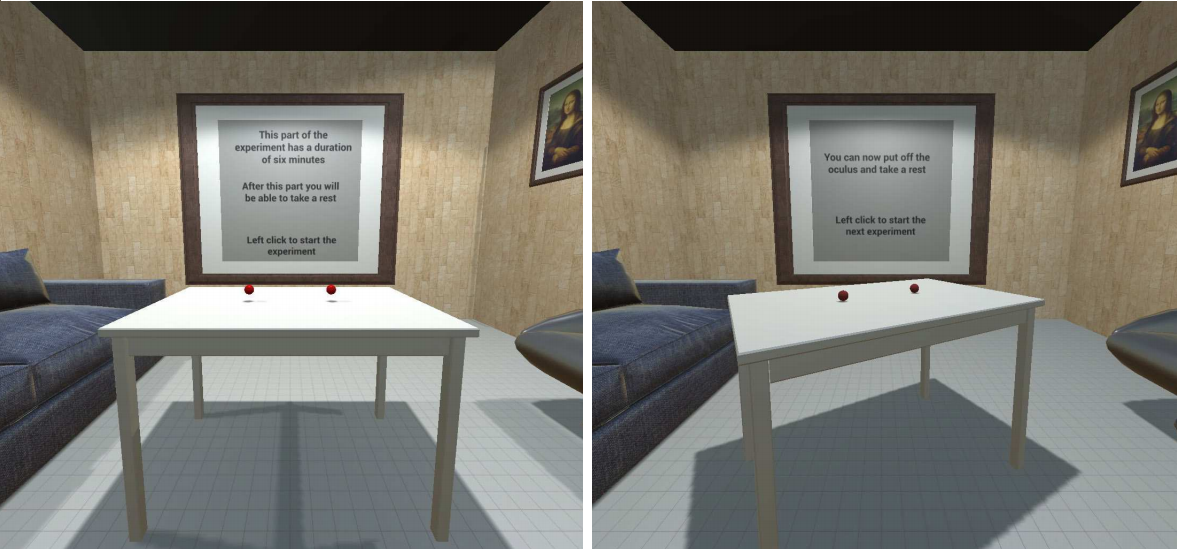}%
    \caption{VR can be used to systematically analyze the interactions of multimodal information. In this example, the influence of auditory signals in the perception of visual motion is studied~\cite{malpica2020crossmodal}. \new{The authors found that different temporal synchronization profiles affected how the stimuli were perceived: When the visual (red balls moving) and auditory (an impact sound) stimuli were correctly synchronized, users perceived a unified event, in particular a collision between both balls.}} 
    \label{fig:malp_synch}
    \vspace{-1.15em}
\end{figure}

When exploring virtual environments, the presence of stimuli from multiple sources and senses \new{(i.e., multimodality) and their potential overlaps (i.e., crossmodality),} may also enhance the \new{final} experience~\cite{melo2020multisensory}. 
Many works have described techniques to integrate some of these stimuli to produce more engaging VR experiences, or to analyze the rich interplay of the different senses. 
For instance, leveraging the window of integration mentioned above may alleviate hardware limitations and lag time,  producing the illusion of real-time performance; this is particularly useful when different modalities are reproduced at different refresh rates~\cite{cheng2019haptic}. 
Moreover, VR is also inherently well suited to systematically study the integration process of multimodal stimuli~\cite{diggs2018}, and analyze the complex interactions that occur when combining different stimuli~\cite{malpica2020crossmodal} (see Figure~\ref{fig:malp_synch}).

In this survey we provide an in-depth review of multimodality in VR. \new{Sensory modalities include information from the five senses: visual for sight, auditory for hearing, olfactory for smell, gustatory for taste, and haptic and thermal for touch.}
Apart from the five senses, we also consider proprioception, which can be defined as the sense of self-movement and body position, and has been defined as the \textit{sixth sense}~\cite{cole2007affective,tuthill2018proprioception}. We synthesize the existing body of knowledge with a particular focus on the \textit{interaction} between sensory modalities \new{focusing on visual, auditory, haptic and proprioceptive feedback}; in addition, we offer an extensive overview of existing VR applications that directly take multimodality into account.

\subsection{\new{The five senses}}
\label{subsec:1_fivesenses}
\new{The way we perceive the world is defined by the five senses: sight, hearing, smell, taste, and touch. Vision is the dominant sense when retrieving information of our surroundings~\cite{rock64}. We are capable of understanding complex scenes with varying visual patterns, we can detect moving objects in our peripheral view, and we are highly sensitive to light~\cite{tinsley2016direct}. However, we tend to focus our visual attention in a narrow region of frontal space~\cite{spence2017}. In that sense, we rely on hearing to retrieve information from unseen locations. Auditory stimuli can grab our attention irrespective of our orientation, and we are good at filtering out particular sounds in a noisy environment (e.g., the cocktail party phenomenon~\cite{arons1992review}). The sense of touch includes different aspects: haptic, kinaesthetic (related with proprioception), temperature, pressure, and pain sensing. Touch occurs across the whole body, although our hands are our primary interface for this sense. Finally, the senses of smell and taste are closely related. They are often linked to emotions or memories, and can even trigger aversive reactions~\cite{miranda2012taste}. Most importantly, besides the particularities of each different sense, and as we will see through this review, our multiple senses influence each other.}

\begin{table*}[]
\centering
\resizebox{0.97\textwidth}{!}{
\def\arraystretch{1.2}%
\begin{tabular}{|c|c|c|c|c|c|c|c|}
\hline
\textbf{Manufacturer  and model} & \textbf{Resolution per eye} & \textbf{\begin{tabular}[c]{@{}c@{}}Positional \\ tracking\end{tabular}} & \textbf{\begin{tabular}[c]{@{}c@{}}Max. refresh \\ rate (Hz)\end{tabular}} & \textbf{\begin{tabular}[c]{@{}c@{}}Field of view \\ (degrees)\end{tabular}} & \textbf{Display type} & \textbf{Integrated audio} & \textbf{Price} \\ \hline
FOVE                             & 1280 x 1440                 & Yes                                                                     & 70                                                                         & 100                                                                         & OLED                  & No - Jack 3.5mm           & \$600          \\ \hline
HP Reverb - Pro                  & 2160 x 2160                 & Inside-out                                                              & 90                                                                         & 114                                                                         & LCD                   & Built-in headphones       & \$649          \\ \hline
HTC VIVE Pro                     & 1440 x 1600                 & Yes                                                                     & 90                                                                         & 110                                                                         & AMOLED                & Built-in headphones       & \$799          \\ \hline
HTC VIVE Pro 2                   & 2448 x 2448                 & Yes                                                                     & 120                                                                        & 120                                                                         & LCD                   & Built-in headphones       & \$800          \\ \hline
Oculus Rift S                    & 1280 x 1440                 & Inside-out                                                              & 80                                                                         & 110                                                                         & LCD                   & In-line speakers          & \$399          \\ \hline
Oculus Quest 2                   & 1832 x 1920                 & Inside-out                                                              & 120                                                                        & 104                                                                         & LCD                   & Stereo speakers           & \$399          \\ \hline
Samsung Odyssey                  & 1440 x 1600                 & Inside-out                                                              & 90                                                                         & 110                                                                         & AMOLED                & Built-in headphones       & \$500          \\ \hline
PlayStation VR                   & 960 x 1080                  & Outside-in                                                              & 120                                                                        & 100                                                                         & OLED                  & No - Jack 3.5mm           & \$299          \\ \hline
Valve Index                      & 1440 x 1600                 & Yes                                                                     & 144                                                                        & 130                                                                         & LCD                   & No - Jack 3.5mm           & \$999          \\ \hline
Varjo VR-3                       & 2880 x 2720                 & Yes                                                                     & 90                                                                         & 115                                                                         & uOLED                 & Off-ear speakers          & \$3195         \\ \hline
\end{tabular}
}
\vspace{0.2cm}
\caption{\csur{Overview of predominant current HMD devices. For each of them, we include the resolution per eye, whether they provide positional tracking, their maximum refresh rate (in Hz), their field of view (FoV, in degrees), the type of display, and a current estimate of the final consumer price. The better specs (in terms of refresh rate and FoV) offered by Valve Index come at a higher cost, while other manufacturers opt for cheaper HMDs, potentially more affordable to consumers. 
} }
\label{tab:hardware_dispAudio}
\vspace{-2.15em}
\end{table*}

\subsection{Proprioception}
\label{subsec:1_proprioceptive}
Proprioception arises from static (position) and dynamic (motion) information~\cite{burns2005}. It plays a key role in the concept of self, and has been more traditionally defined as "awareness of the spatial and mechanical status of the musculoskeletal framework"~\cite{valori2020proprioceptive}. Proprioceptive information comes mainly from mechanosensory neurons next to muscles, tendons and joints, although other senses can induce proprioceptive sensations as well. A well-known example are visual cues inducing the phantom limb illusion~\cite{ramachandran1996}.

Proprioception plays an important role in VR as well. On the one hand, it helps provide the subjective sensation of \textit{being there} ~\cite{slater1993presence, sadowski2002presence}.
On the other hand, proprioception is tied to cybersickness, since simulator sickness is strongly related to the consistency between visual, vestibular, and proprioceptive information; significant conflicts between them could potentially lead to discomfort~\cite{laviola2000discussion, mcgill2017passenger}.

\subsection{Reproducing sensory modalities in VR}

\label{subsec:1_hardware}

 Important efforts have been made in VR so that all the modalities previously mentioned can be integrated. Visual and auditory feedback are the most commonly used, and almost all consumer-level devices integrate these modalities. \csur{There is currently a wide variety of manufacturers providing different HMD systems to enjoy VR at consumer level. Each of them offers devices with different capabilities and specifications, at different costs. 
 Table~\ref{tab:hardware_dispAudio} compiles an overview of the most relevant devices currently in the market. An open issue in VR is latency~\cite{elbamby2018toward}: newer HMD models feature higher refresh rates, as well as significantly increased spatial resolution. 
 }
\csur{Usually, those} displays feature a field-of-view (FoV) slightly smaller than that allowed by human peripheral vision\csur{. However, }new stereoscopic rendering techniques allow to present content in 3D, and therefore perception of materials, depth, or many other cues can be achieved through visual cues~\cite{armbruster2008depth}. %
 \emph{Auditory} feedback\csur{, which is often integrated in the HMD as a built-in feature~(Table \ref{tab:hardware_dispAudio}),} is generally enabled by speakers or headphones, and spatial audio rendering techniques also support our perception of space in virtual environments~\cite{andreasen2019auditory}, even enhancing perceived visual properties~\cite{malpica2020crossmodal}. %
 \emph{Haptic} feedback is still in an exploratory phase, and can be achieved through a variety of sensors, including wearables~\cite{pezent2020explorations,salazar2020altering}, physical accessories~\cite{strandholt2020knock}, ultrasounds~\cite{marchal2020can}, controller devices~\cite{strasnick2018haptic,whitmire2018haptic}, rotary components~\cite{je2019aero}, or electric muscle stimulation~\cite{lopes2018adding}. Other resources like fans, hear lamps, or even spray bottles have been used to provide additional tactile stimuli in VR~\cite{wilberz2020facehaptics}. \csur{Recently, advances in ultrasound sensor technology have resulted in the creation of a novel haptic device that will allow for mid-air force around virtual objects and interactions~\cite{long2014rendering}; this device is about to reach the consumer market~\cite{Emergeio}.} 
 \emph{Olfactory} stimuli can be provided through smell cartridges or specific hardware~\cite{herrera2014development}, and electric stimulation of taste buds has been tested to generate flavors~\cite{ranasinghe2013simulating}. How all these different feedback modalities can be integrated depends on the particular context, with some algorithms and techniques already proposed for the most common sensory combinations, including audiovisual or audiohaptic stimuli~\cite{magalhaes2020physics}.

\begin{figure*}[t]
	\centering
	\mbox{} \hfill
	\includegraphics[width=0.85\linewidth]{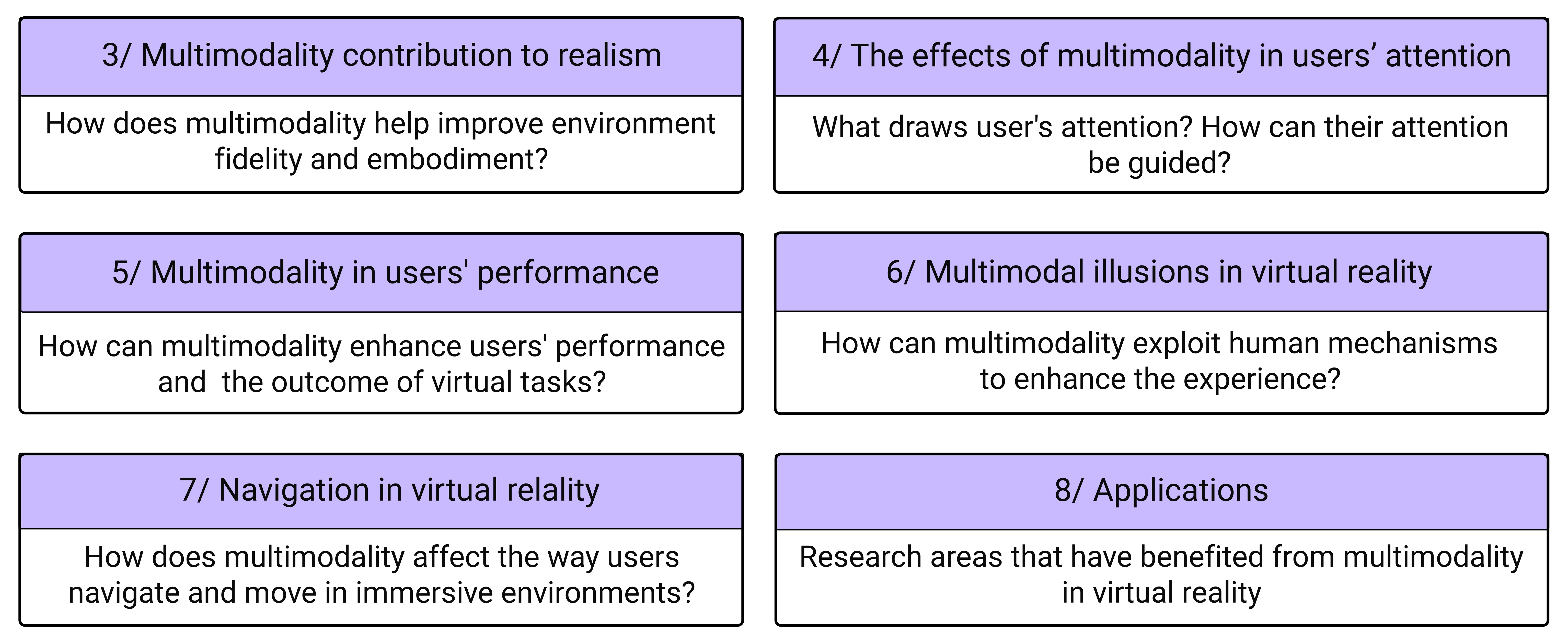}
    \hfill \mbox{}
	\caption{
	Structure of this state-of-the-art report. We divide it into \new{different} areas of the VR experience in which multimodality can play a key role: user immersion, presence, and realism of the experience (Section~\ref{sec:n3_fidelity}); user attention when exploring the virtual environment (Section~\ref{sec:n5_guidance}); user performance when completing tasks (Section~\ref{sec:n6_performance}); multimodal perceptual illusions that can be leveraged in VR (Section~\ref{sec:n7_manipulation}); \new{and navigational effects of multimodality in virtual environments~ (Section~\ref{sec:new_7_nav})}. Finally, we review different applications where multimodality has been shown to improve the end goal (Section~\ref{sec:n8_applications}), and finalize with a discussion on the need for multimodality, and open avenues of research (Section~\ref{sec:n9_conclusions}).
	}
	\vspace{-1.15em}
	\label{fig:schema}
\end{figure*}

\subsection{Related surveys}
\label{subsec:1_reviews}

\csur{Our perception of the world depends on the integration of information from multiple senses. Several works have reviewed the influence of individual senses \textit{in isolation}, including sight~\cite{richardt2020capture}, sound~\cite{serafin2018sonic}, or touch~\cite{laycock2007survey}, which are the three key modalities in VR (see Table~\ref{tab:supertable}).}
\csur{Several surveys exist focusing on particular aspects of VR. Rubio et al.~\cite{rubio2017immersive} systematically reviewed advances in communication, interaction, and simulation in VR, pointing out that the key factors to generate appealing virtual experiences include a good interactivity, representation, gameplay, and narrative. %
The latter, narrative, has been explored in depth from a cinematic perspective~\cite{rothe2019guidance} \csur{since content creators no longer have the same level of control over how the viewer attention is directed}: \csur{To find new ways of guiding viewer's attention, the} authors reviewed current attention techniques in virtual environments, either unimodal or multimodal, emphasizing how auditory cues can be critical, and the still unexploited potential of haptic devices (see Section~\ref{subsec:6_4_entertainment}) } %

\csur{Other surveys target specific applications of VR, such as education or medicine~\cite{rosen1996evolution}. For the specific case of clinical medicine, Li et al.~\cite{li2017application} found that most of the works in the literature leverage the capabilities of haptic devices to simulate real, clinical tasks, in line with our insights (see Section~\ref{subsec:6_1_medicine}). %
Freina et al.~\cite{freina2015literature} reviewed works focused on using VR in education, concluding that it increases the learner's involvement and motivation, which are enhanced with multimodality (Section ~\ref{subsec:6_2_learning}).}

\csur{Some works are concerned particularly with cognitive aspects in multimodal environments. Hecht et al.~\cite{hecht2006multimodal} briefly studied how integrating multiple sources can increase presence, enhance attention and improve response time. %
Koelewijn et al.~\cite{koelewijn2010attention} focused on surveying works related to low-level, audiovisual interactions, concluding that both multisensory integration and attentional processes take place and can interact at multiple stages in the brain. However, multisensory overload can sometimes lead to a preference on simpler environments if not handled correctly; we delve deeper into this in Section~\ref{sec:n3_fidelity},.}  %
Other surveys have studied multimodality
in \textit{traditional media}, including cognition ~\cite{spence2010prior}, interaction~\cite{jaimes2007multimodal}, human-computer interfaces~\cite{dumas2009multimodal, gurkok2012brain}, or fusion and integration techniques~\cite{atrey2010multimodal}. 
\csur{Closer to the present work, Melo et al.~\cite{melo2020multisensory} systematically studied the impact of multisensory stimuli in virtual experiences. Their study suggests that 85\% of works tackling multimodality in VR report positive impacts, with only 1\% of them reporting negative impacts. They also reported how multisensory experiences in VR are mainly applied in the health domain, science and engineering, teaching, or machinery; which is in line with our reports in Section~\ref{sec:n8_applications}. While they report that these are the more common applications among the 105 studies they surveyed, we provide a discussion on how multimodality is impacting each of these fields.} %

However, and different from all these works, in this survey we focus on the integration of multimodal information \new{and the benefits and experiences that can be achieved that way}, and compile a large body of works studying \new{not only those positive effects, but also their applicability into different disciplines.}

\subsection{\new{The challenges of multimodality}}
\label{subsec:1_challenge}
\new{One of the main challenges when considering a fully multimodal immersive experience is the gaps of empirical knowledge that exist in this field. As stated before, in this survey our main focus lies in the visual, auditory, haptic and proprioceptive modalities. 
This is partly related to the fact that many modalities and their interactions remain unexplored, and there is still much to learn about them. Moreover, the available data on multimodality in VR (both referring to multimodal stimuli and to user data while experiencing multimodal VR) is scarce at best. 
It is also important to consider the window of integration. The necessity of synchronizing different modalities implies the need for real-time, high fidelity computation. Hardware processing limitations might also imply a constrain in what multimodal techniques should be used in different scenarios. Moreover, not all VR headsets are equally prepared to support multimodality: Although most of them can give audiovisual feedback, proprioception and haptic feedback are sometimes limited, while olfactory and gustatory feedback are usually not found at all in consumer-level headsets. For example, most smartphone-based VR headsets do not include controllers, hindering the possibility of including haptic feedback. Many other basic VR systems are not able to track translations either (i.e., only have three degrees of rotatory freedom), which limits proprioceptive feedback.
However, one of the most critical risks with multimodality is its definition itself. Although multimodality has the potential to improve user experience, increase immersion, or even improve performance in certain tasks, multimodal applications have to be very carefully designed, making sure that each modality has its function or purpose. This is not a trivial task, since the dimensionality of the problem grows with each added modality. Otherwise, additional modalities might distract and overload the final user, hampering the experience, and diminishing user satisfaction.} %

%% file: n2_Scope_Structure.tex
\section{Scope and organization of this survey}
\label{sec:n2_scope}

In this \new{survey}
we provide an in-depth review of the most significant works devoted to explore the role and effects of multimodality in \new{virtual reality}. We gather knowledge about how multiple sensory modalities interact and affect the perception, the creation, and the interaction with the virtual experience. 

The structure of this survey can be seen in Figure~\ref{fig:schema}. Since our focus is not on any specific part of the VR pipeline, but rather on the \emph{VR experience} for the user, we have identified several areas of the VR experience in which multimodality plays a key role. First, Section~\ref{sec:n3_fidelity} is devoted to the realism of the VR experience, which is tied to immersion and the sense of presence that the user experiences. Second, Section~\ref{sec:n5_guidance} looks into how multimodality can affect the attentive process of the user in the virtual environment, determining how they explore the environment and what drives their attention within it. Third, Section~\ref{sec:n6_performance} delves into works that demonstrate how multimodality can help the user in completing certain tasks, essentially improving user performance in the virtual environment. 

Additionally, there are a number of works devoted to analyzing multimodal perceptual illusions and their perception in VR environments. These, which we compile in Section~\ref{sec:n7_manipulation}, can be leveraged by future techniques to improve any of the aforementioned areas of the VR experience.
\new{Some other works have tackled the problem of navigation in VR, which is another integral part of the virtual experience, and Section~\ref{sec:new_7_nav} encompasses them. A complete perspective of all these sections is also included in Table~\ref{tab:supertable}.} Finally, we devote Section~\ref{sec:n8_applications} to reviewing application areas that have benefited from the use of multimodal virtual experiences, and conclude (Section~\ref{sec:n9_conclusions}) with a discussion of the potential of multimodality in VR, and interesting avenues of future research.

%% file: n3_Fidelity_Environment.tex
\section{The effects of multimodality in perceived realism}
\label{sec:n3_fidelity}

Perceived realism elicits
realistic responses in immersive virtual environments~\cite{slater2009visual} 
, and is tied to the overall perception of the experience. 
There are two key 
factors that can lead to users responding in a realistic manner
: the place illusion and the plausibility illusion~\cite{slater2009place}. The former, also called ``presence'' 
, defines the sensation of ``being there'', and is dependent on sensorimotor information, whilst the latter refers to the illusion that the scenario that is apparently happening is actually taking place, and is determined by the ability of the system to produce events that relate to the user, i.e., the overall credibility of the scenario being depicted in comparison with the user's expectations. Slater~\cite{slater2009place} %
argued that participants respond realistically to an immersive VR environment when these two factors 
are present. 
\new{Similar observations were made in telepresence systems~\cite{steuer1992defining}, where sensorially-rich mediated environments were proved to actually elicit more realistic responses. }

\new{Increasing the feeling of presence can therefore enhance the experience by eliciting more realistic responses from the users, and as aforementioned, increasing the perceived realism has a positive impact in the feeling of presence. This actually depends on both the virtual environment where the user is placed, and its own representation in there. As happens in the real world~\cite{gonccalves2019impact}, all human modalities play a fundamental role, and must be correctly integrated, to construct a coherent notion of the both the virtual environment, and the self. In this section, we will therefore focus on how multimodal cues can affect perceived realism, by affecting both the perception of the environment, and the perception of the self.}

\subsection{Perception of the environment}

\begin{figure}[tbp]
    \centering
    \includegraphics[width=0.85\columnwidth]{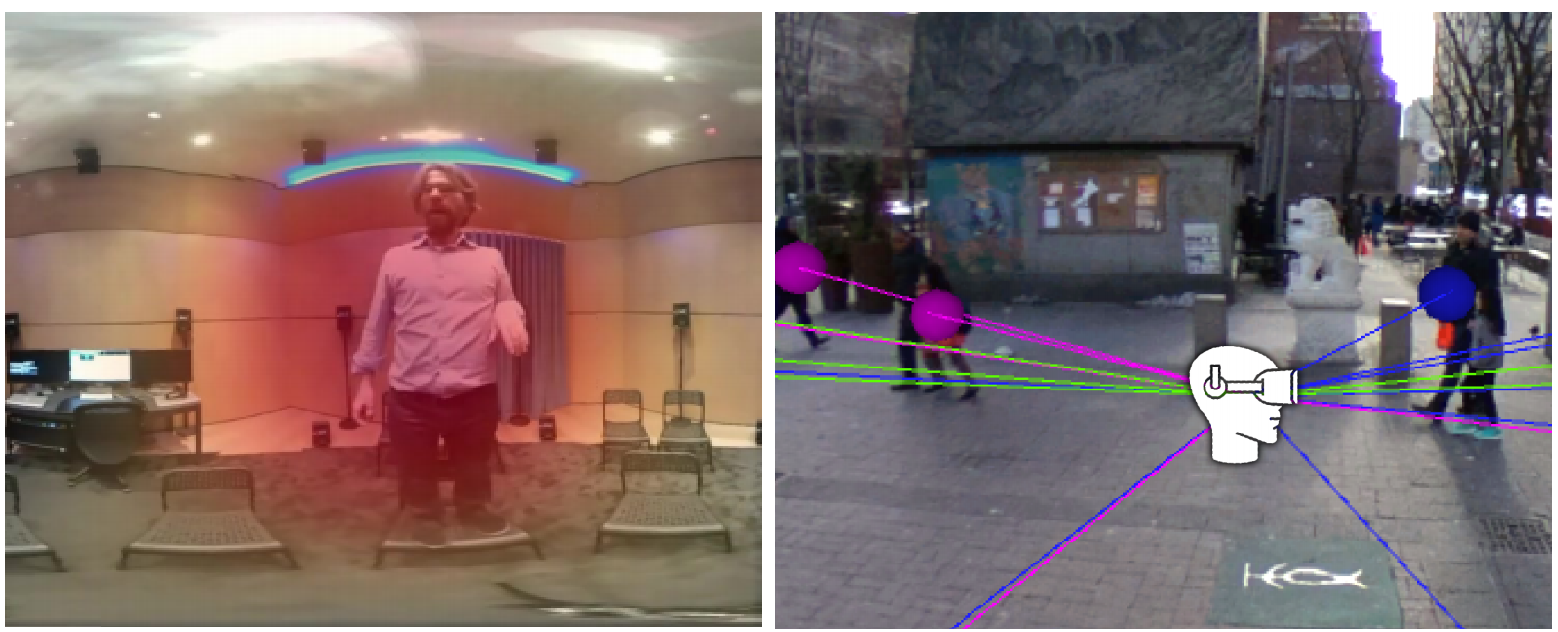}%
    \caption{Including correct and coherent auditory information in the virtual environment has been proved to increase realism and immersion. \textit{Left}: A system that automatically generates ambisonic information that creates a smoother acoustic \new{experience} for the scene~\cite{morgado2018self}. \textit{Right}: A framework to include auditory information into 360º panoramas depending on the elements in the scene~\cite{huang2019audible}. In both cases, their validation experiments yield users' preference when auditory information is included, and an overall increase in the perceived realism and immersion.}
    \label{fig:realism}
    \vspace{-0.95em}
\end{figure}

The perceived realism of virtual environments is a key concern when designing virtual experiences, therefore many works have been devoted to investigate how multimodality and crossmodality 
can indeed help achieve sensorially-rich %
experiences. \new{While multimodality refers to the binding of different inputs from multiple sensory modalities, crossmodality involves interactions between different sensory modalities that influence the perception of one another~\cite{lalanne2004crossmodal, spence2009crossmodal}}.  
Chalmers et al.~\cite{chalmers2009towards} discussed how crossmodal effects in human multisensory perception can be exploited to selectively deliver high-fidelity virtual environments, for instance, rendering with higher visual quality those items related to the current auditory information of the scene, allowing to reduce computational costs in unattended regions of the virtual environment. \csur{This work also reports that humans perceive sensory information with more or less attention depending on the task they are executing (i.e., some task require more attention to particular types of stimuli), or on if they have already been preconditioned to that kind of virtual environments (e.g., they are used to it).} 
Traditionally, sound has proven to facilitate visual perception, including enabling a better understanding of the environment, yielding a more comfortable experience, or even increasing performance of visual-related tasks~\cite{iachini2012multisensory,shams2008benefits}. \csur{Seitz et al.~\cite{seitz2006sound} conducted a ten-day experiment where two groups of people were trained for auditory-visual motion-detection tasks, one with only visual, and the other with audiovisual stimuli. 
Although all of them improved their performance over time, those trained with multimodal stimuli showed significantly better performances.}%
Various works have been thus devoted to this audiovisual integration: %
Morgado et al.~\cite{morgado2018self} presented a system that generates ambisonic audio for 360º panoramas, so that auditory information is represented in a spherical, smoother way (see Figure~\ref{fig:realism}, left). %
Similarly, Huang et al.~\cite{huang2019audible} proposed a system that automatically adds spatialized sounds to create more realistic environments (see Figure~\ref{fig:realism}, right), validating by means of user studies the overall preference of this solution in terms of realism. Indeed, different soundscapes \new{(a sound or combination of sounds created from an immersive environment)} %
are able to increase the sense of presence in VR~\cite{serafin2004sound}, and as Liao et al.~\cite{liao2020data} studied, combining visual and auditory zeitgebers \new{(periodically occurring natural phenomena which act as cues in the regulation of biological rhythms),}
which act like synchronizers, could actually enhance presence, even influencing time perception. \csur{All these previous works suggest that using auditory information, either spatialized or not, enhances the realism of the experience, although some of them warn about the potential backfire of increasing the cognitive load, which can negatively impact users' confidence~\cite{jung2020impact}.}

However, 
\csur{multimodal integration can also present some drawbacks:}
Akhtar and Falk~\cite{akhtar2017} surveyed current audiovisual quality assessment and found that auditory information may cause discomfort and decrease the quality of the virtual experience~\cite{ruotolo2013immersive}. To avoid negative effects during multimodal integration, different sensory cues should be not only realistic, but also coherent to the environment and between them.

\new{Proprioception also plays an important role in eliciting realism, as it constributes to the feeling of the user being there.}
\new{Although some works have demonstrated that some manipulations in virtual movement directions and distances can be unnoticeably performed \csur{(either by manipulating the environment itself through the game engine or by modifying the real-to-virtual mapping of users' movement)}~\cite{Serrano2020_VR-LateralMotion, langbehn2018}, users tend to expect their virtual movements to match their real ones, to maintain a coherent experience.}
\new{In this line, Mast and Oman~\cite{mast2004top} studied the so-called visual reorientation illusions: When the environment is rotated \new{above a given noticeable threshold} in any axis, users can perceive that the expected vertical axis does not match the virtual one, and conflicts between visual and vestibular cues may arise, \new{potentially causing motion sickness}. Although the effect of this illusion is stronger for elder users~\cite{howard2000visually}, an incoherent spatial estimation in VR can potentially diminish the perceived realism}.

\new{Including additional modalities can also enhance environment realism. In particular, giving realistic feedback with respect to what users expect to happen actually increases plausibility. Normand et al.~\cite{normand2011multisensory} showed that it is possible to induce a body distortion illusion by synchronous visual-tactile and visual-motor correlations (see Figure~\ref{fig:correlations}). In a similar fashion, Hoffman compared the realism of virtually touching an object with that of touching it physically at the same time~\cite{hoffman1998physically_a}, yielding a significant increase in perceived realism when the object was physically touched too. Similar results were obtained with taste and olfactory cues~\cite{hoffman1998physically_b}: They found a preference on smelling and physically biting a chocolate bar in contrast to only virtually biting it.  The level of presence achieved depends on the different combinations of sensory feedback, and multi-sensory systems have been proved to be superior to traditional audio-visual virtual systems in terms of the sense of presence and user preference~\cite{jung2020impact}. Similar conclusions were obtained by Hecht et al. ~\cite{hecht2006multimodal}, who reported that multimodality led to a faster start of the cognitive process, which ultimately contributed to an enhanced sense of presence. \csur{However, and even if the benefits of multimodal integration are widely known and shared between researchers and practitioners, there is still much to learn about the limits and drawbacks of multisensory integration, and studying up to what point multimodal interaction can be safely applied to increase perceived realism in different scenarios remains an interesting future avenue. }}

%% file: n4_Self_Perception.tex
\subsection{Perception of the self}

Virtual experiences are designed for humans, and in many occasions, users are provided with a virtual representation of themselves. This is a very effective way of establishing their presence in the virtual environment, hence contributing to the place illusion~\cite{slater2009place}. This representation does not need to be visually realistic, but it has to be coherent enough with the users' actions \new{and expectations} to maintain the consistency of the experience. 
\new{In the following, }we review different works that have leveraged multimodality in \new{virtual reality} to achieve \new{consciousness of the self}
and embodiment, and therefore to create realistic representations of the users.

Having the feeling of being in control of oneself is possibly one of the main characteristics that VR offers~\cite{slater2009place}. \new{The feeling of presence is possible without being in control; however, being able to control a virtual body highly increases this illusion~\cite{schuemie2001research}.} %
The sense of embodiment gathers the feeling of owning, controlling, and being inside a body. As Kilteni et al.~\cite{kilteni2012sense} reported, this depends on various subcomponents, namely 
\new{the} sense of self-location (a determined volume in space where one feels to be located), \new{the} sense of agency (having the subjective experience of action, control, intention, motor selection and the conscious experience of will), and \new{the} sense of body ownership (having one’s self-attribution of a body, implying that the body is the source of the experienced sensations). \new{Other factors like the proximity of virtual objects to the body also have an effect on the sense of embodiment~\cite{seinfeld2020impact}.} All these concepts (such as presence or embodiment) are intrinsic characteristics that VR can achieve, and they yield the self-consciousness feeling that makes VR so different from other media. 

Multimodality has been largely studied as a means to enhance those sensations. \new{Particularly, presence} is tied to the integration of multiple modalities, and many works have demonstrated how it is increased when multiple sensory information is \new{combined~\cite{sanchez2005presence}}, as opposed to unimodal (i.e., only visual) systems~\cite{jung2020impact}. \csur{For instance, Gonçalves et al.~\cite{gonccalves2019impact} designed an experiment where three groups of people were exposed to virtual environments including different amount of modalities in the presented stimuli; %
and reported how users experiencing more modalities reported a higher involvement. Moreover, they remark the positive impact of including haptic feedback in an experience.} Blanke et al.~\cite{blanke2015behavioral} discussed the relevance of a series of principles to achieve a correct sensation of bodily self-consciousness, requiring body-centered perception (hand, face, and trunk), and integrating proprioceptive, vestibular, and visual bodily inputs, along with spatio-temporal multisensory information. Sakhardande et al.~\cite{sakhardande2020exploring} presented a systematic study to compare the effect of tactile, visual, visual-motor, and olfactory stimuli on body association in VR, with the latter having the strongest effect on body association. \new{Similar insights were proposed by Pozeg et al.~\cite{pozeg2015those}, who demonstrated the importance of first-person visual-spatial viewpoints for the integration of visual-tactile stimuli, in this case for the sense of leg ownership.}
The main factors to build embodiment and body-ownership in VR have been widely studied~\cite{maselli2013building}. \csur{Spanlang et al.~\cite{spanlang2014build} presented technical guidelines to create a core virtual embodiment system, defining three key aspects: (i) a VR module to handle creation, management, and rendering of all virtual entities, (ii) a head-tracking module to map real movements to the virtual environment, and (iii) a display module to present all that environment. However, designing experiences that are too realistic can have negative aspects and be a drawback in certain specific cases:}
\new{For example, group pressure of alien virtual avatars can result in users performing potentially harmful actions towards others that they would not normally carry out~\cite{neyret2020embodied}}. %

\begin{figure}[tbp]
    \centering
    \includegraphics[width=0.85\columnwidth]{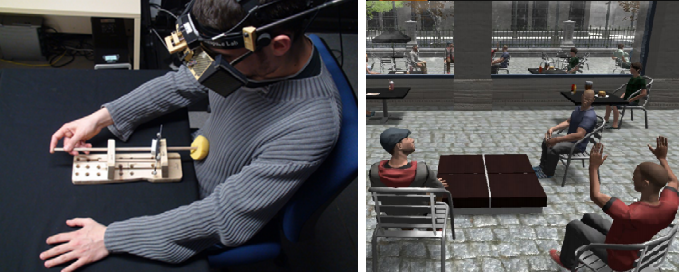}%
    \caption{\textit{Left}: \new{Synchronizing different modalities increases the feeling of presence and the perception of the self. Moreover, multimodality can even create a distortion of that perception:}
    Normand et al.~\cite{normand2011multisensory} presented a study where a body distortion illusion is achieved by synchronous visual-tactile and visual-motor correlations. \textit{Right}: Some works have studied \new{how} different physical and behavioral factors can directly affect\new{, and even manipulate,} embodiment~\cite{neyret2020embodied}\new{, and thereofre, the perception of the self}.
    }
    \label{fig:correlations}
    \vspace{-1.2em}
\end{figure}

 The sense of moving (which depends on agency and body ownership, as previously mentioned) is also 
 \new{key to achieve} self-consciousness. Kruijff et al.~\cite{kruijff2016your} presented a work showing that adding walking-related auditory, visual, and vibrotactile cues could all enhance the participants' sensation of self-motion and presence. Various works have been presented in this line, e.g., investigating the integration of tendon vibrations to give standing users the impression of physically walking~\cite{kouakoua2020rhythmic}.
 Sometimes \new{locomotion} is not possible, and it has to be externally generated, e.g., by means of a virtual walking system for sitting observers using only passive sensations such as optic flow and foot vibrations~\cite{matsuda2020perception}. \new{However, these techniques are likely to} creating the well-known self-motion illusion: \new{Although users are not actually moving, their brain unconsciously assumes they are moving, and their body sometimes generates postural responses~\cite{dichgans1978visual} to control their stability}. %
 \new{Meyer et al.~\cite{meyer2013modulation} studied the impact of having multimodal (visual, auditory, and haptic) anchor points in the virtual environment in users' postural sway. \csur{They report how incongruent cues diminish perceived realism. However, they also remark on the complexity of providing dynamic tactile signals in VR, which leaves an interesting research line in how to exploit tactile cues to increase presence.}}
 \new{Some other works have also explored alternatives for cases when locomotion is not feasible, for instance} proposing and evaluating a virtual walking system for sitting observers using only passive sensations such as optic flow and foot vibrations~\cite{matsuda2020perception}.

Other modalities 
\new{may also play} an important role in users' self-consciousness:
\new{Several works has shown that multimodality can dramatically increase the sense of presence~\cite{gallace2012multisensory}, although confidence levels for certain tasks are higher in traditional (i.e., audio-visual) virtual environments, due to a higher cognitive load~\cite{jung2020impact}.} \new{Besides additional modalities, other factors such as immersion and emotion have been analyzed and argued to have a clear impact on the sense of presence~\cite{banos2004immersion}.} \new{In particular, audiovisual content eliciting emotional responses (like sadness) can increase engagement and presence, somehow bypassing the immersive effects of specific displays.} 

\csur{As reported in some of the aforementioned works, }multimodality 
presents some challenges and limitations: Gallace et al.~\cite{gallace2012multisensory} focused on 
the ones associated with the simultaneous stimulation of multiple senses, including the senses of touch, smell, taste, and even nioceptive (i.e., painful) sense, given the cognitive limitations in the human sensory perception bandwidth when users have to divide their attention between multiple sensory modalities. \new{Moreover, situations where some modalities violate interpersonal space may also lead to diminishing presence and comfort~\cite{wilcox2006personal}}. Ultimately, achieving user's self-consciousness depends on finding the right balance between different multimodal cues, and the users' comfort, confidence, and capacity to integrate them. \csur{Establishing guidelines towards this balance remains one of the most interesting avenues in multimodal interaction.}

%% file: n5_Saliency_Guidance.tex
\section{The effects of multimodality in users' attention}
\label{sec:n5_guidance}

When users are exploring or interacting with a virtual environment, different elements or events can draw their attention. Visual attention influences the processing of visual information, since it induces gaze to be directed to the regions that are considered more interesting or relevant (salient regions). The saliency of different regions results from a combination of top-down attentional modulation mechanisms \new{(task-based)} %
and the \new{bottom-up} multisensory information these regions provide \new{(feature-based)}, creating an integrated saliency map of the environment~\cite{treue2003visual}. As discussed in the previous section, VR setups may produce  realistic responses and interactions, which can be different %
from traditional media due to the differences in perceived realism and interaction methods. Therefore, some works have been devoted to understanding saliency and users' attention in VR, offering some key insights about head-gaze coordination and users' exploratory behavior in VR. For example, Sitzmann et al.~\cite{Sitzmann_TVCG_VR-saliency} detected the \emph{equator bias} when users are freely exploring omnistereo panoramas: They observed a bias towards gazing at the central latitude of the scene (equator bias), which often corresponds to the horizon plane.

\begin{figure}[tbp]
    \centering
    \includegraphics[width=0.85\columnwidth]{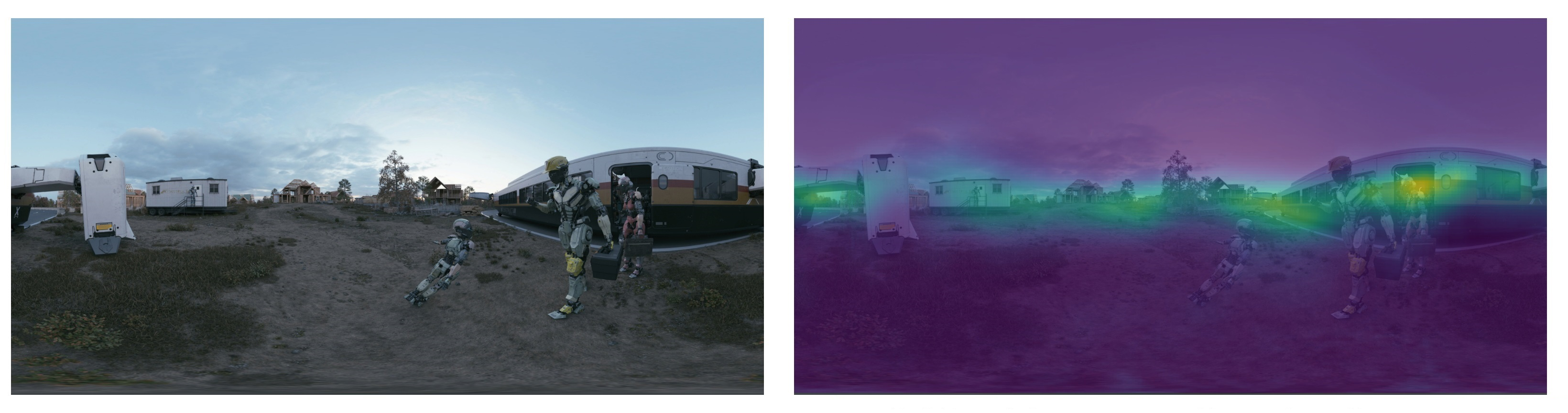}%
    \caption{Saliency %
    maps show the likelihood of users directing their attention to each part of the scene. Most of the current literature has been devoted to estimating saliency in unimodal, visual stimuli. This image shows the recent visual saliency estimation method proposed by Martin et al.~\cite{martin20saliency} (\textit{Left}: Input panorama. \textit{Right}: Estimated saliency). It has been shown that each sensory modality has the potential of influencing users' attentional behavior, therefore, there is a need for further exploration of multimodal saliency in VR.}
    \label{fig:saliency_figure}
    \vspace{-1.25em}
\end{figure}

\new{Saliency has been widely studied, }both inside and outside the VR field~\cite{xu2020state}. Users are more likely to turn their attention and interact with those regions of the scene that provide more sensory information. Therefore, knowing a priori which parts of the scene are more salient may help anticipating how users will behave. So far, most of the works on saliency in VR have been carried out following a unimodal perspective ~\cite{zhu2018prediction, martin20saliency}: \new{Although several senses can be considered to create a saliency map, }they all leverage users' head position and gaze orientation %
to create probabilistic maps indicating the chances of a user looking at each part of the virtual scene (see Figure~\ref{fig:saliency_figure}). %
\new{Based on the study of visual cues}, various works have presented systems able to predict users' gaze, depending on the environment %
and also on user's previous behavior ~\cite{hu2020dgaze,martin2021scangan360}.

Multimodality in saliency estimation has been only tackled in traditional \new{media}: The integration of visual and auditory information \new{for saliency prediction in videos} has been widely explored~\cite{de2010model, min2020multimodal}. All these approaches work under the assumption of audiovisual correlation: Moving elements were the source of the auditory cues. \new{In a different approach, }Evangelopoulos et al.~\cite{evangelopoulos2013multimodal} proposed the addition of text information in form of subtitles \new{when speech was present in the auditory stream.} %
\new{In their work, saliency was considered as} a top-down process, since the interpretation of the subtitles, a complex cognitive task, can distract viewer's attention from other parts of the scene. 

Multimodality in saliency prediction for VR still remains in early phases, and only very few works have been devoted to it. Chao et al.~\cite{chao2020audio} proposed the first work that studies user behavior (including \new{saliency }%
corresponding to sound source locations, viewing navigation congruence between observers, and %
\new{the distribution of gaze behavior}) in virtual environments containing both visual and auditory cues (including both monaural and ambisonic sounds). However, there are still many open avenues for future research: Visual saliency and gaze prediction in VR is still in an early phase, and the effects of auditory cues in saliency in virtual scenarios remain to be further explored. Auditory cues in VR may produce more complex effects and interactions than in traditional scenarios, since sound sources are not always in the user's field of view, and there might be several competing audiovisual cues. Additionally, investigating how other senses interact and predominate in saliency and attention can be useful for many applications, specially for content creation. \csur{Furthermore, with the proliferation of data-driven methods, it is also crucial to elaborate datasets that encompass enough variety of multimodal stimuli to support the formulation of new multimodal attentional models.}

Although there is still much to learn about how multimodal cues compete and alter users' behavior, it is well known that multimodality itself has consequences on how users behave \new{in immersive environments~\cite{tsakiris2017multisensory, chao2020audio}} 
. One of the main difficulties when designing and creating content for VR lies on the fact that users' typically have control over the camera, and therefore each user may pay attention to different regions of the scene and create a different experience~\cite{Serrano_VR-cine_SIGGRAPH2017, maranes2020exploring}.
Therefore, it is usually hard to make assumptions about users' behavior and attention. To support the creation of engaging experiences that convey the creators' intentions, multimodality can be exploited, so that cues from different modalities can induce specific behaviors and even guide users' attention.

\begin{figure}[tbp]
    \centering
    \includegraphics[width=0.80\columnwidth]{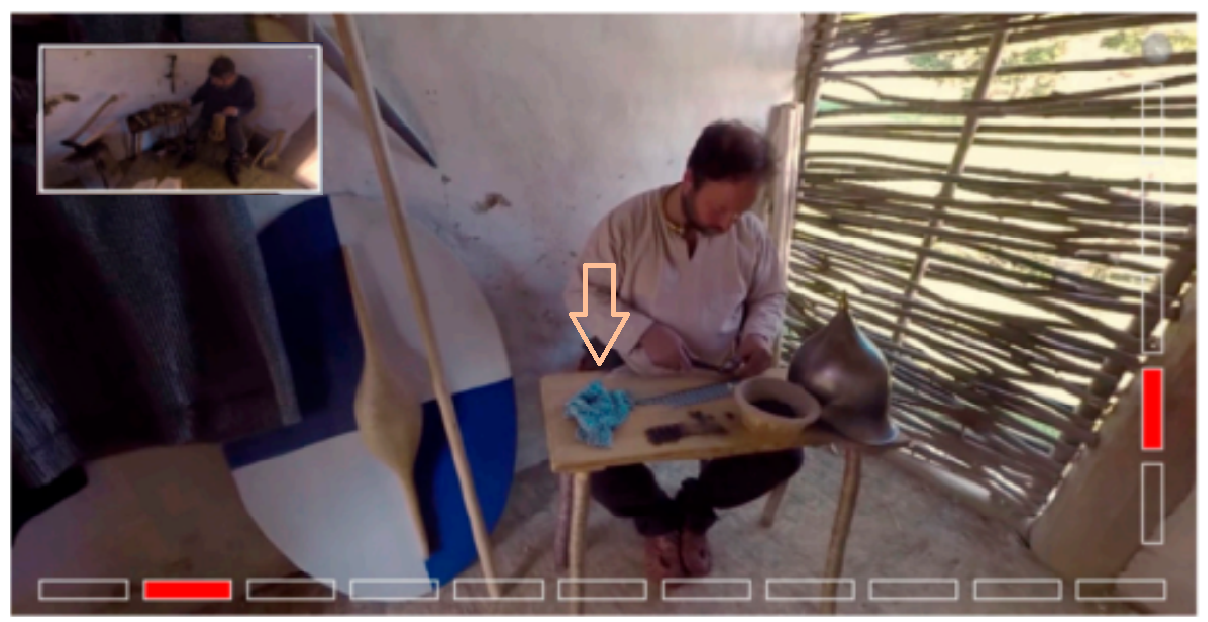}%
    \caption{Examples of visual guidance methods in VR, adapted from Rothe et al.'s review on users' guidance for cinematic content~\cite{rothe2019guidance}. Three visual guidance techniques are presented in this image: Arrows pointing regions of interest, picture-in-picture techniques that show information of rear regions, and the position of a point of interest marked with red bars. Most of these techniques are intrusive, hence they may break immersion and realism. With the addition of multimodal cues, \new{guidance can be facilitated while maintaining a positive user experience.}} %
    \label{fig:rothe_guid}
    \vspace{-1.15em}
\end{figure}

For the case of attention, understanding and guiding users attention in VR has been a hot topic during the last years. Various works have explored the use of visual guiding mechanisms, such as %
central arrows and peripheral flicker to guide attention in panoramic videos~\cite{schmitz2020directing}. The recent work of Wallgrun et al.~\cite{wallgrun2020comparison} compares different visual guiding mechanisms to guide attention in 360º environments~\cite{wallgrun2020comparison}. Lin et al.~\cite{lin2017outside} proposed a picture-in-picture method that includes insets of regions of interest that are not in the current \new{field of view, }%
so users are aware of all the elements that surround them. %
Inducing the users to direct their attention to a specific part of the scene has also been explored, for example, using focus assistance techniques~\cite{lin2017tell}, such as indicating the direction of the relevant part, or automatically \new{orienting} the world so that users' do not miss that part of the experience. Following this line, Gugenheimer et al.~\cite{gugenheimer2016swivrchair} presented a motorized swivel chair to rotate users until they were focusing on the relevant part of the scene, while Nielsen et al.~\cite{nielsen2016missing} forced virtual body orientation to guide users attention to the most relevant region. Other techniques directly let the viewer press a button to immediately reorient the scene to the part containing the relevant information~\cite{pavel2017shot}. It is worth mentioning that this kind of techniques have to be taken into consideration with caution, since they can cause dizziness or discomfort due to visual-vestibular conflicts. We refer the reader to Rothe et al.'s work~\cite{rothe2019guidance} for a complete survey about guidance in VR (see Figure~\ref{fig:rothe_guid}).

However, guidance techniques are not necessarily constrained to visual manipulations. Multimodality can be also exploited to guide, focus and redirect attention in VR, in many cases achieving more subtle, less intrusive methods. This is important to maintain the users experience, as intrusive methods can alter the sense of presence, immersion, or suspension of disbelief (the temporary acceptance as believable of events or places that would ordinarily be seen as incredible).
As shown in previous sections, sound can help enhance the virtual experience. Besides, it can also be used to manipulate or guide users attention. Rothe et al.~\cite{rothe2017diegetic} demonstrated that the attention of the viewer could be effectively directed by sound and movements, and later~\cite{rothe2018guiding} investigated and compared three methods for implicitly guiding attention: Lights, movements, and sounds, showing that sounds elicit users' exploratory behavior, while moving lights can also easily draw attention. Other works have explored various unobtrusive techniques combining auditory and visual information, showing that auditory cues 
indeed reinforce users' attention being drawn towards specific parts of the environment~\cite{brown2016directing}. \new{Similar insights were obtained by Masia et al.~\cite{masiaSound2021}, who investigated the impact of directional sound during cinematic cuts in VR, finding that in the presence of directional sound cues, users converge much faster to the main action after a cut, even if the sound is missaligned with the region of interest. Given the importance of including sound in VR experiences, }Bala et al.~\cite{bala2018cue} presented a software for adding sound to panoramic videos, and studied how sound helped people direct their attention. Later, they examined the use of sound spatialization for orientation purposes~\cite{bala2019elephant}. \new{In particular, they found that full spatial manipulation of sound (e.g., locating music in a visual region of interest) helped guiding attention}. In a similar fashion,
some works have studied how to design sound to influence attention in VR~\cite{salselas2020}, and how decision making processes are affected by auditory and visual cues of diegetic (i.e., sounds emanating for the virtual environment itself) and non-diegetic (i.e., sounds that do not originate from the virtual environment itself) origins~\cite{ccamci2019exploring}. %
However, non-diegetic cues need to be analyzed and presented carefully: The work by Peck et al.~\cite{peck2009evaluation} showed that a distractor audio can be successful at fostering users' head rotations (and thus redirection); however, users considered this method as unnatural. It has been also suggested that too many sound sources in a VR cinematic video can produce clutter, and therefore hinder the identification of relevant sound sources in the movie~\cite{rothe2019guidance}. \csur{How to use multimodality for guiding users' attention has many open possibilities for further investigation. In this context, establishing guidelines regarding which senses to use, how to combine them, and up to what extent each of them can surpass the others remains for now a complex, unresolved task.}%

%% file: n6_Interaction.tex
\section{Multimodality in users' performance}
\label{sec:n6_performance}

Understanding how users perform different tasks in VR is key for developing better interfaces and experiences. Although in many cases task performance highly depends on the users' skills and experience, there are many scenarios where multimodality can play an important role in this aspect: By integrating multiple sensory information we can mimic better the real world, and this can 
\new{lead to higher performance in different scenarios, comparable to real life.} Additionally, \new{multimodal} VR technologies are becoming a very powerful tool for training \new{and education}, specially in scenarios that can be expensive, or \new{even} dangerous, in the real life. In those cases, multimodality can help completing some tasks in a shorter period, or with a higher accuracy~\cite{hecht2006multimodal}.

The effects of multimodality in task performance have been largely studied in traditional media. Lovelace et al.~\cite{lovelace2003irrelevant} demonstrated how the presence of a task-irrelevant light enhances the detectability of a brief, low-intensity sound. This behavior also holds in the inverse direction: Concurrent auditory stimuli could enhance the ability to detect brief visual events~\cite{noesselt2008sound}. Therefore, integrating audiovisual cues may diminish the risk of users losing some relevant information. In a similar line, Van der Burg et al.~\cite{van2008pip} reported that a simple auditory \emph{pip} drastically decreased detection time for a synchronized visual stimuli. These effects are not only present in audiovisual stimuli: Tactile-visual interactions also affect search times for visual stimuli~\cite{van2009poke}. \csur{In most of these work, the experiments were carried out in laboratory conditions with simple stimuli, and therefore studying their applicability and limitations in more complex scenarios remains an interesting avenue.}
Furthermore, Maggioni et al.~\cite{maggioni2018smell} studied the potential of smell for conveying and recalling information. They compared the effectiveness of visual and olfactory cues, 
and their combination in this task, and demonstrated that olfactory cues indeed improved users' confidence and performance. Therefore, the integration of multiple cues has been widely proved to be effective in terms of detectability and efficiency. 
As Hecht et al.~\cite{hecht2006multimodal} studied, \new{this improvement in terms of performance also holds for multimodal VR:} When there are multiple senses involved, users start their cognitive process faster, thus they can pay attention to more cues and details, resulting in a richer, more complete and coherent experience. 

Performance in spatial tasks \new{can be} greatly benefited from \new{multimodality~\cite{gao2020visualechoes,ammi2014intermodal}.} %
Auditory cues are extremely useful in spatial tasks in VR,
and therefore have been widely explored:
The effect of sound beacons in navigation performance when no visual cues are available has been explored~\cite{walker2003effect}, with some works proving that navigation when no visual information is available is possible using only auditory cues~\cite{groehn2001some}. Other works have exploited this, proposing a \new{novel technique to visualize sounds, similar to how echolocation would work in animals,} 
which improved the space perception in VR thanks to the integration of auditory and visual information~\cite{rosenkvist2019hearing}, or combining the spatial information contained in echoes to benefit visual tasks requiring spatial reasoning~\cite{gao2020visualechoes}. Other senses have also been explored with the goal of enhancing spatial search tasks: Ammi and Katz~\cite{ammi2014intermodal} proposed a method coupling auditory and haptic information to \new{enhance spatial reasoning, and thus} improving performance in search tasks. 
Direct interaction tasks can be also enhanced by multimodality: Auditory stimuli has been proved to facilitate touching a virtual object outside user’s field of view, hence creating a more natural interaction~\cite{kimura2020auditory}. Egocentric interaction is also likely to happen, and proprioception plays an important role on those cases. Poupyrev et al.~\cite{poupyrev1998egocentric} presented a formal study comparing virtual hand and virtual pointer as interaction metaphors, in object selection and positioning experiments, yielding that indeed both techniques were suitable for different interaction scenarios. 

\new{As aforementioned, when developing VR experiences requiring users' to complete some tasks, the integration of multiple modalities can increase their performance and spatial reasoning, leading to better, more consistent results. 
Furthermore, adding certain modalities (e.g., olfactory or haptic information) is not always easy, specially at consumer level. \csur{Enabling these modalities within current consumer-level devices (Table~\ref{tab:hardware_dispAudio}) remains a future avenue that would not only greatly benefit multimodality in terms of performance, but it would also improve the whole experience. In spite of that,} in some cases, combining several modalities can lead to the opposite effect, suppressing or diminishing some abilities~\cite{malpica2020}, hence special care must be paid when designing multimodal experiences (Section~\ref{subsec:1_challenge}).}

%% file: n7_Tricking.tex
\section{Multimodal illusions in VR}
\label{sec:n7_manipulation}

\begin{figure}[tbp]
    \centering
    \includegraphics[width=0.80\columnwidth]{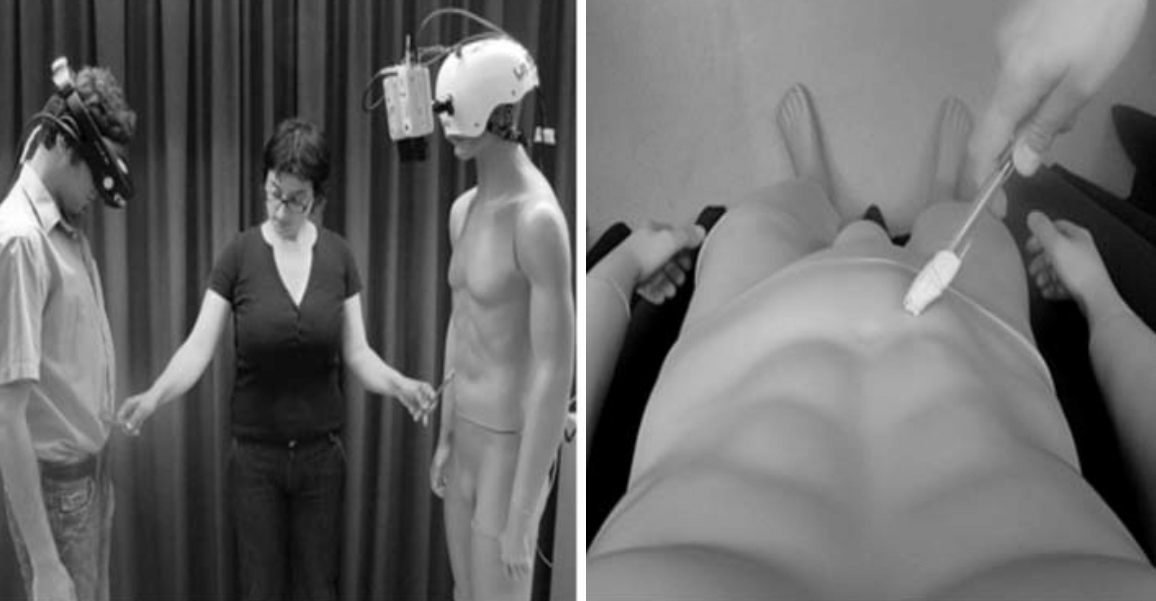}%
    \caption{Multimodality illusions can change the way users perceive both themselves and the environment. For instance, Petkova et al.~\cite{petkova2008} studied how proprioceptive and haptic cues could lead to body ownership illusions.}
    \label{fig:illusion}
    \vspace{-1.85em}
\end{figure}

Multimodality can be leveraged to trick the self perception of the users, or to alter how they perceive the world around them, by means of facilitatory or inhibitory (suppressive) effects, which can have direct implications on how users behave in the virtual environment.
Being able to manipulate the experience can be very useful in certain contexts and applications: For instance, sometimes it can be useful to guide the user towards a particular aspect of the virtual environment without disrupting the experience (e.g., in cinematography or videogames). A forced guidance could lead to reduced a immersion feeling, or even rupture of the suspension of disbelief. In other cases, physical space is constrained, and manipulating users' movement may allow to reduce the necessary physical space to complete a task~\cite{Serrano2020_VR-LateralMotion}. Manipulating the experience can be also useful to reduce simulator sickness, \csur{for instance, by means of manipulating camera control depending on some characteristics such as velocity, acceleration, or scene depth}~\cite{hu2019reducing}. Although this can be done using a single modality, the use of multimodal cues can improve the effectiveness of these techniques.

Illusion refers to an incorrect perception or interpretation of a real, external stimulus. It can lead to interpreting reality in several ways. Any healthy person can experiment illusions without experiencing any pathological condition. However, not every person is affected in the same way by an illusion. 
Illusions can have physiological (e.g., an after-image caused by a strong light~\cite{ito2012}) or cognitive (e.g., the Rubin vase~\cite{peatfield2015}) components. They have been widely studied, as understanding illusions yields valuable information about the limitations of human senses, and helps understanding the underlying neural mechanisms that help create the perception of the outside world. Moreover, illusions can allow to alter users responses to certain tasks, even increasing performance. \csur{For instance, Chauvel et al.~\cite{chauvel2015} conducted some experiments where non-golfers practiced putting golf balls, some of them with manipulated holes to enhance their visual acuity. Those who trained under these conditions showed a more effective learning outcome, and a better performance when trying in real-life scenarios}. In this subsection we will focus on multimodal illusions or effects, or how illusions in other senses can affect visual perception. For visual only illusions, we refer the reader to The Oxford Compendium of Visual Illusions~\cite{shapiro2016}.

Multimodal illusions can be useful for boosting accuracy in certain tasks. For example, multisensory cues can improve depth perception when using handheld devices \new{by simulating tactile responses when holding, or interacting with a virtual object with a force feedback system}~\cite{bouguila2000,swapp2006}. Using a small number of worn haptic devices, Glyn et al.~\cite{glyn2016} improved  spatial awareness in virtual environments without the need of creating physical prototypes. Instead of applying contact (haptic feedback) at the exact physical point of the users body that was touching a virtual object, they used a small, fixed set of haptic devices to convey the same information. Their work was based on the funnelling illusion~\cite{barghout2009}, in which the perceived point of contact can be manipulated by adjusting relative intensities of adjacent tactile devices. Visuo-haptic illusions allow not only to better perceive the virtual space, but also to feel certain virtual object properties, like weight, that are not easy to \csur{simulate. Even further, these properties can be unnoticeably altered when combining multiple sensory information. Carlon~\cite{carlon2018} showed that users' perception of heaviness can be unnoticeably altered when manipulating their movements in a virtual environment}. 

The rubber hand illusion is an illusion where users are induced to feel like a rubber hand is part of their body. In VR, proprioceptive and haptic cues can lead to a similar feeling induced either for an arm~\cite{yuan2010} or for the whole body~\cite{petkova2008} (see Figure~\ref{fig:illusion}). %
\new{Similarly, proprioception can also be altered by modifying the virtual avatar (i.e., distorting the position or length of the virtual arms and hands) while retaining body ownership, allowing users to explore a bigger area of the virtual environment with their body~\cite{feuchtner2017extending}.}
Regarding audiovisual illusions, the well known McGurk effect has been replicated in VR. The McGurt effect happens when the audio of a syllable is paired with visual stimuli of a second syllable, raising the perception of a third, different syllable. This illusion has been used to study how audio spatialization affects speech perception, suggesting that sounds can be located at different positions and still create a correct speech experience~\cite{siddig2019,abubakr2019}. It was also found that the spatial mismatch does not affect immersion levels, suggesting that computational resources devoted to audio localization could be decreased without affecting the overall user experience. Another interesting audiovisual illusion that appears both in conventional media and VR is the ventriloquist effect, where auditory stimuli coming from a distant source seem to emerge from an actors' lips. The best located or dominant modality (usually vision)
\new{overrides the spatial information of the weak modality } %
giving raise to the apparent translation of sound \new{to the location of the visual stimulus}~\cite{alais2004ventriloquist}. In this sense, auditory stimuli are affected by visual cues~\cite{sarlat2006ventriloquism}, with visual stimuli influencing the processing of binaural directional cues of sound localization. In a complementary way, auditory perception can also act as a support for visual perception, orienting users to regions of interest outside the field of view~\cite{kyto2015ventriloquist}. Not every audiovisual illusion has to do with speech. In the sound-induced flash illusion~\cite{shams2005sound}, a single flash paired with two brief sounds was perceived as two separate flashes. The reverse illusion also happened when two flashes were concurrent with a single beep, raising the perception of a single flash. 

\begin{figure}[tbp]
    \centering
    \includegraphics[width=0.85\columnwidth]{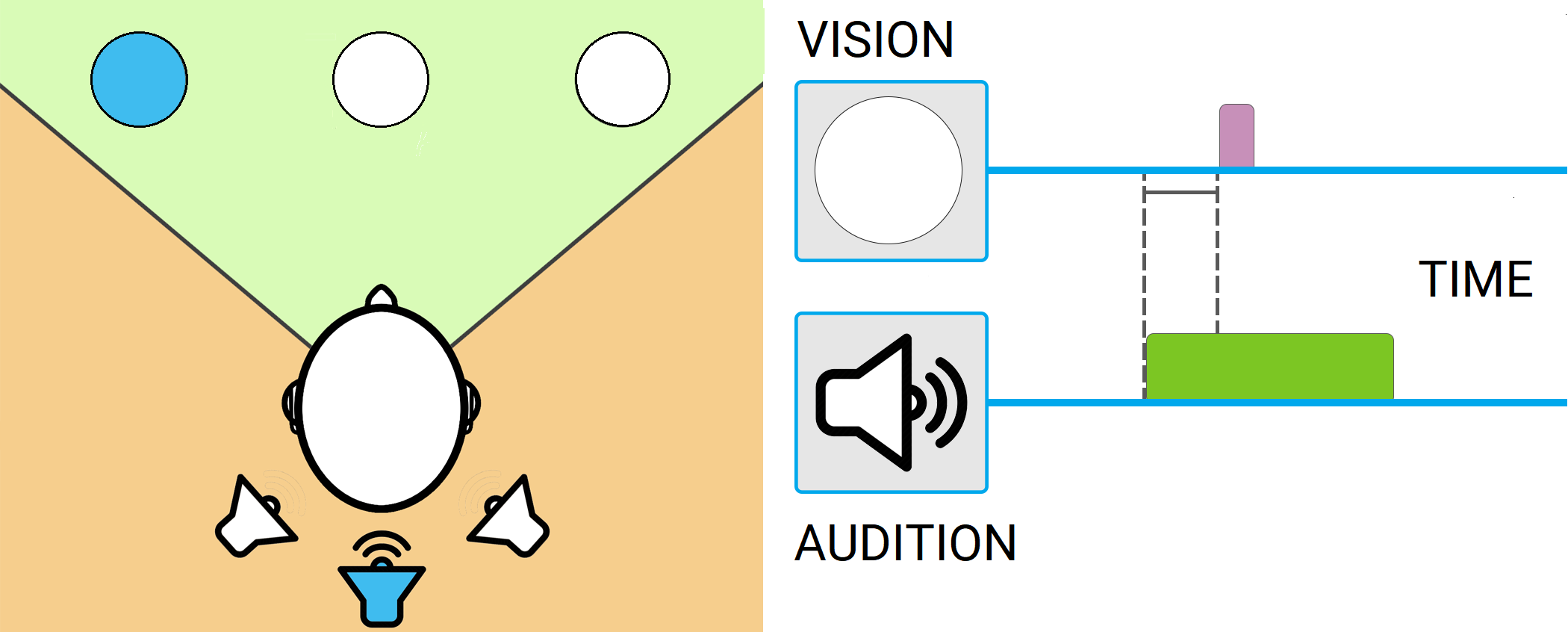}%
    \caption{Spatiotemporal layout of an auditory-triggered effect that degrades visual perception. Sound cues located outside the field of view are concurrent with the appearance of a visual target (inside the field of view), causing the visual target to be missed by participants even when they are directly fixating on it. Figure adapted from Malpica et al.~\cite{malpica2020}.}
    \label{fig:malpica_sound}
    \vspace{-1.0em}
\end{figure}

In addition to illusions, in which new stimuli are sometimes created, there is also the phenomenon of perceptual suppression, in which one stimulus is no longer (completely or partially) perceived due to an external circumstance. For example, visual suppression is often present in the human visual system. The human brain has evolved to discard visual information when needed to maintain a coherent and stable image of the surrounding environment. Two good examples of visual suppression are blinks and saccades~\cite{bolte2015subliminal,sun2018}, which avoid the processing of blurry information without causing perceptual breaks. Perceptual suppression has been demonstrated and used both in conventional media and in VR~\cite{sun2018}, usually allowing for environmental changes without the users awareness, which is useful in many applications, such as navigation in VR. It has also been studied how stimuli of a given modality can alter or suppress information of a different modality, usually visual. In particular, for traditional media, both auditory~\cite{hidaka2015sound} or haptic~\cite{ide2013} stimuli can suppress visual stimuli. Functional imaging studies~\cite{laurienti2002} suggest that crossmodal suppression occurs at neural levels, involving sensory cortices of different modalities. Crossmodal suppression has not been widely studied in VR. However, a recent study~\cite{malpica2020} shows that auditory stimuli can degrade visual performance in VR using a specific spatiotemporal layout (see Figure ~\ref{fig:malpica_sound}). \csur{Nonetheless, there is still much to investigate about traditional illusions and whether they still hold in virtual environments. The interaction between senses, and in particular the predominance of some of them against the rest, may also diminish or enhance these phenomena, and therefore remains an interesting avenue for future work.} We \csur{thus} believe that a deeper study of crossmodal interactions, both facilitatory and inhibitory, could greatly benefit VR applications, as well as increase our knowledge on sensory perceptual processing in humans.

%% file: n8_Applications.tex
\section{Multimodality in navigation}
\label{sec:new_7_nav}

As discussed in previous sections, agency has an effect in the feeling of realism in a virtual experience, \new{and it} is achieved when users feel that their avatar responses are coherent with their real actions, \new{which has a direct implication in body ownership (which also depends on other factors ~\cite{waltemate2018impact}).} 
\new{One characteristic that makes VR intrinsically different from any other traditional media, and that contributes to the users' feeling of control of themselves, is its ability to reproduce each movement of the user into the virtual world. Virtual environments naturally elicit exploration, which usually requires the user to move across the virtual environment. In many cases, movement is heavily constrained by the physical space available~\cite{Serrano2020_VR-LateralMotion}, and therefore a complete 1:1 reproduction of the movement is not feasible.}

\new{Enabling full locomotion in a VR application (i.e., allowing the user to freely move in the virtual space) would increase the possibilities of the virtual experience.
However, designers and practitioners are aware of the limited size of physical spaces in which users can consume VR.} 
Redirected walking techniques (RDW) emerged in the pursuit of alleviating this limitation: These techniques propose different ways to subtly or overtly manipulate either the user or the environment during locomotion, in order to allow the exploration of virtual worlds larger than the available physical space. Nilsson et al.~\cite{nilsson201815} presented an overview of research works in this field since redirected walking was first practically demonstrated. Nevertheless, most of these works rely on visual manipulations: Some of them exploit only visual cues or mechanisms, such as saccades~\cite{sun2018} or blinks~\cite{langbehn2018} to perform inadvertent discrete manipulations, whereas others exploit continuous manipulations that remain unnoticed by users~\cite{Serrano2020_VR-LateralMotion,matsumoto2020detection}. However, these previous works do not exploit cues from other sensory modalities. As we have presented along this work, integrating multiple senses can take these kind of techniques a step further.

\new{Rhythmic auditory stimuli affects how we move~\cite{maculewicz2016investigation}, and auditory stimuli can be therefore used to actively manipulate our motion perception. }Serafin et al.~\cite{serafin2013estimation} described two psychophysical experiments showing that humans can unknowingly be virtually turned about 20\% more or 12\% less than their physical rotation by using auditory stimuli: With no visual information available, and with an alarm sound as the only informative cue, users' could not reliably discriminate whether their physical rotations had been smaller or larger than the virtual ones.
Nogalski and Fohl~\cite{nogalski2016acoustic} presented a similar experiment, aiming for detection thresholds for acoustic redirected walking, in this case by means of wave field synthesis: By designing a scenario surrounded by speakers, and with no visual information available, they demonstrated that some curvature gains can be applied when users walk towards, or turn away from some sound source.
Their work yielded similar rotation detection thresholds of $\pm20\%$, which is additionally in line with other works, proving the ability of acoustic signals to manipulate users' movements~\cite{feigl2017acoustical} and the potential benefits of using auditory stimuli in complex navigational tasks. \csur{Later, Rewkowski et al.~\cite{rewkowski2019evaluating} confirmed that RDW with auditory distractors can be safely used in complex navigational tasks such as crossing streets and avoiding obstacles}. Nilsson et al.~\cite{nilsson2016estimation} revealed similar detection thresholds for conditions involving moving or static correlated audio-visual stimuli. Additionally, Nogalski and Fohl~\cite{nogalski2017curvature} summarized how users behavior significantly varies between audio-visual and auditory only stimuli, with the latter yielding more pronounced and less constant curvatures than with audio-visual information. 

Many other sensory modalities can be used both to manipulate user's virtual movement, \new{improving agency and therefore leading to a more natural navigation.}
Hayashi et al.~\cite{hayashi2019redirected} presented a technique that allows to manipulate the mapping of the user's physical jumping distance and direction. Jumping is an action strongly correlated to proprioception, but it is usually unfeasible due to the available physical space. Manipulating the virtual distance when jumping can allow users to physically jump even when space is constrained, hence proprioceptive cues and realism can be maintained in the experience. Campos et al.~\cite{campos2012multisensory} also introduced an integration of visual and proprioceptive cues for travelled distance perception, demonstrating that body-based cues contributed to walked distance estimation, attributable to vestibular inputs. Matsumoto et al.~\cite{matsumoto2019unlimited} presented a combination of redirected walking techniques with visuo-haptic interaction and a path planning algorithm. \new{Haptic feedback directly applied to feet can also influence audiovisual self-motion illusions~\cite{nilsson2012haptically}.} Exogenous%
cues \new{(i.e., any external information coming from the environment)} can also play a role in these kind of manipulations. Feng et al.~\cite{feng2016effect} examined the effects, influence and interactions of multi-sensory cues during non-fatiguing walking, including movement directional wind, footstep vibrations, and footstep sounds, yielding results that evidenced the improvement on user experience and realism when these cues were available.

In some cases, motion is not possible at all, hence it is necessary to generate an external, visual motion. This self-motion illusion is commonly known as vection, and sometimes leads to some postural responses (pursuing a correct vestibular and proprioceptive integration of information). It has been demonstrated that auditory cues increase vection strength in comparison with purely visual cues~\cite{keshavarz2014combined}, and that moving sounds enhance circular vection~\cite{riecke2009moving}. Moreover, vection may also depend on the environment itself: Meyer et al.~\cite{meyer2013modulation} explored which factors actually modulate those postural responses, and showed that real and virtual foreground objects serve as static visual, auditory and haptic reference points. \csur{Some of the experiments in these works were carried out under rigidly controlled setups and in laboratory conditions, %
and therefore may not apply to free viewing or other complex conditions. Exploring the effectiveness (or degrading effects) of these insights in more complex scenarios can be an interesting future avenue for research. Finally, the effects of other senses besides auditory and haptics in navigation remain unexplored.}

\section{Applications}
\label{sec:n8_applications}

We have reviewed different aspects of multimodality in VR, as well as crossmodal interactions between the different sensory modalities, \new{together with different achievable effects.}
\new{A summary of all those benefits that multimodality can lead to in VR can be seen in Table~\ref{tab:supertable}}.
Different disciplines have leveraged these benefits to enhance different VR applications, showing that multimodality can indeed deliver more realistic and immersive VR experiences. 
\new{While the application scenarios range across many disciplines, here we focus on applications of multimodality to three areas where VR has made a critical impact: medicine, training and education, and entertainment.}

\subsection{Medicine}
\label{subsec:6_1_medicine}
The potential uses of VR for medical applications have been studied for decades, and research on this field has evolved as the virtual technologies have done so. Satava et al.~\cite{satava1999virtual} presented a review about how VR \csur{has become an integral technology of medicine both for profesionals and patients alike: from medical image visualization and preoperative planning to teaching and simulation, including teleinterventions and rehabilitation.}
Other works focused more deeply on the use of VR in the areas of surgical planning, interoperative navigation, and surgical simulations~\cite{rosen1996evolution, satava1998current}. \csur{The possibility of virtualizing a real human body previously scanned and watching it from a far more realistic, immersive perspective than through a conventional display is of great use for health professionals.} This has been possible, to a large extent, due to the increasingly photorealistic representation of the anatomy (both in terms of physical tissue properties and of physiological parameters) that virtual environments are achieving. 

\begin{figure}[tbp]
    \centering
    \includegraphics[width=0.85\columnwidth]{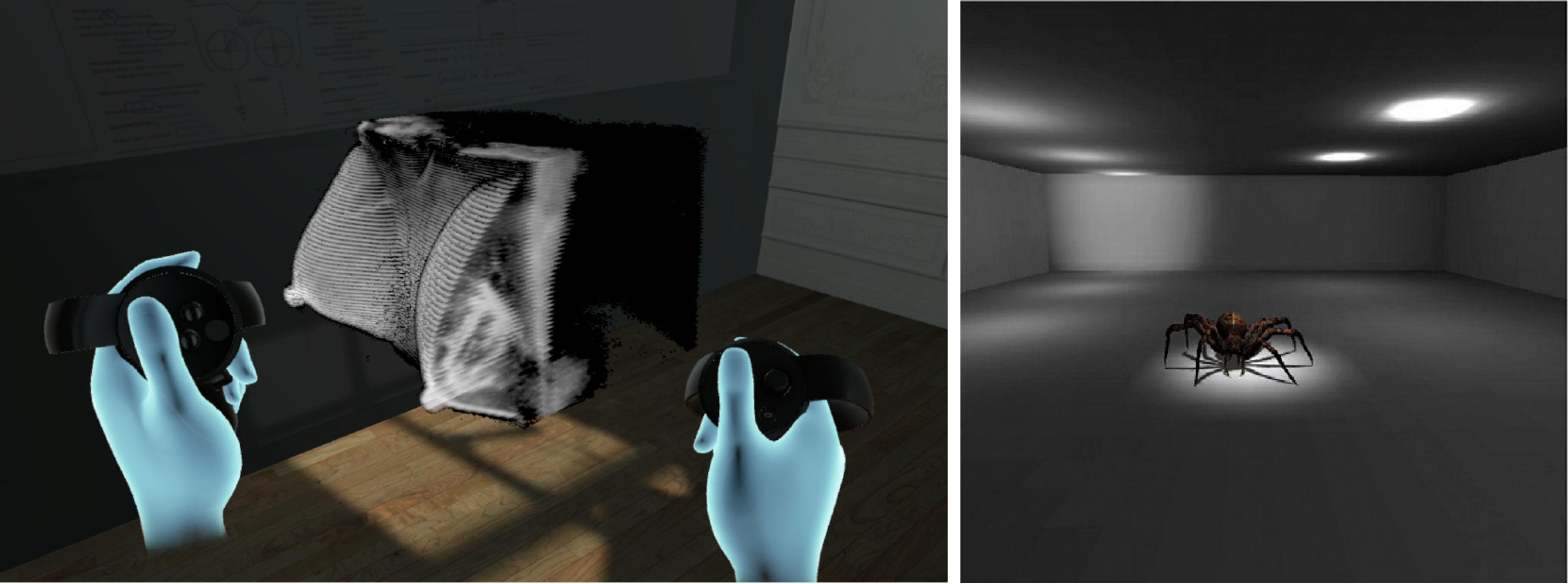}%
    \caption{Two representative examples of different applications of multimodality in medicine. \textit{Left:} Data visualization and manipulation frameworks~\cite{prange2018medical} are important in medical and surgical education and training, and multimodality may enhance the realism and immersion, thus achieving better learning transfer. \textit{Right:} Multimodal VR is a key tool for phobia treatments, since it is able to create realistic environments that face users against their fears, without actually exposing them~\cite{shiban2013effect}.}
    \label{fig:examples_medicine}
    \vspace{-0.6em}
\end{figure}

One of the most pervasive applications of VR in medicine is training, since it can provide a realistic environment for training without the risks of its real counterpart. The enhanced realism and immersion that multimodality provides can lead to improved training and education. 
Lu et al.~\cite{lu2010multimodal} presented an audio-visual platform for medical education purposes. One step further, multimodal setups including haptics have been proposed for medical surgery training, where the realism of the feedback significantly improved the learning effect, for both virtual~\cite{hutchins2005design} and augmented~\cite{harders2007multimodal} reality interfaces. In the area of medical visualization, Prange et al.~\cite{prange2018medical} also exploited virtual environments and presented a multimodal medical 3D image system where users could walk freely inside a room and interact with the system by means of speech, and manipulate patients' information with gestures (see Figure~\ref{fig:examples_medicine}). 
\new{Multimodal VR applications} are however not constrained to medical training and visualization areas: Psychological research relying on VR has also experienced an unprecedented growth, as Wilson and Soranzo reviewed~\cite{wilson2015use}, emphasizing both the advantages \csur{(e.g., greater control over stimulus presentation, safe exposure to adverse conditions, etc.)} and challenges \csur{(e.g., VR-induced side effects)} of VR in this area. Similarly, Bohil et al. studied the latest advances in VR technology and their applications to neuroscience research~\cite{bohil2011virtual}, \csur{highlighting its high compatibility with medical imaging technologies (such as functional magnetic resonance imaging - fMRI), which allow for a high degree of ecological validity and control over the tHerapeutic experience}.

\new{Other areas that have leveraged the benefits of multimodality are}
rehabilitative medicine and psychiatry, where significant progress has been made.
Psychiatric therapies can benefit from multimodality, since different aspects of behavioral syndromes can be extensively analyzed in virtual environments: Given the suitability of VR to manipulate \new{the virtual world and control certain tasks},
it has proven to be a fitting paradigm to treat diseases like OCD~\cite{cipresso2013break} or Parkinson's disease~\cite{cipresso2014virtual}.
\new{Phobia treatment is one of the main areas leveraging the benefits of multimodal environments:} The realism that multimodality offers over visual-only VR experiences enhances these experiences, and increases the effectiveness of the treatment. In addition, VR allows exposing patients to their fears in a safe and highly controlled way, minimizing any potential risks of exposure therapy. Shiban et al. ~\cite{shiban2013effect} studied the effect of multiple context exposure on renewal 
in spider phobia (see Figure~\ref{fig:examples_medicine}), \new{suggesting that exposure in multiple contexts improves the generalizability of exposure to a new context, therefore helping patients to reduce the chances of future relapses}. The work of Hoffman et al.~\cite{hoffman2003interfaces} went a step further: They explored not only whether VR exposure therapy reduces fear of spiders, but also concluded that giving patients the illusion of physically touching the virtual spider increases treatment effectiveness. Muhlberger et al.~\cite{muhlberger2003efficacy} studied the effect of VR in the treatment of fear of flying, exploiting not only visual and acoustic cues, but also proprioceptive information, since motion simulation may increase realism and help induce fear. Later, they studied the long-term effect of the exposure treatment~\cite{muhlberger2006one}, proving its efficacy in treating the fear of flying. The effect of auditory feedback has been studied in other domains, such as the particular case of agoraphobic patients~\cite{viaud2006high}, where multimodality increases patients' immersion feeling, hence facilitating emotional responses. However, those techniques should be applied with caution, since large exposures to VR scenarios may hinder patients' ability to distinguish between the real and the virtual world~\cite{ichimura2001investigation}, leading to the disorder known as Chronic Alternate-World Disorder (CAWD).

\begin{table*}[t!]
\centering
\resizebox{\textwidth}{!}{
\def\arraystretch{1.85}%
\begin{tabular}{|c|c|c|c|c|c|c|}
\hline
\multirow{2}{*}{\textbf{Application}}                                       & \multirow{2}{*}{\textbf{Example work}} & \multicolumn{4}{c|}{\textbf{Additional involved senses (other than vision)}}                        & \multirow{2}{*}{\textbf{Brief description}}                    \\ \cline{3-6}
                                                                            &                                & \textbf{Audition} & \textbf{Proprioception} & \textbf{Haptics} & \textbf{Other} &                                                                \\ \hline
\multirow{2}{*}{Rehabilitation}                                             & Fordell et al.~\cite{fordell2016rehatt}         & \textbf{\color{red}{\xmark}}          & \textbf{\color{ticgreen}{\checkmark}}       & \textbf{\color{ticgreen}{\checkmark}}& \textbf{\color{red}{\xmark}}       & Chronic neglect treatment, with a force feedback interface.    \\ \cline{2-7} 
                                                                            & Sano et al.~\cite{sano2015reliability}            & \textbf{\color{ticgreen}{\checkmark}} & \textbf{\color{ticgreen}{\checkmark}}       & \textbf{\color{ticgreen}{\checkmark}}& \textbf{\color{red}{\xmark}}       & Multimodal sensory feedback to reduce phantom limb.            \\ \hline
\multirow{3}{*}{\begin{tabular}[c]{@{}c@{}}Phobia treatment\end{tabular}} & Viaud et al.~\cite{viaud2006high}           & \textbf{\color{ticgreen}{\checkmark}} & \textbf{\color{red}{\xmark}}                & \textbf{\color{red}{\xmark}}         & \textbf{\color{red}{\xmark}}       & Effects of auditory feedback in agoraphobic patients.          \\ \cline{2-7} 
                                                                            & Mülberger et al.~\cite{muhlberger2003efficacy, muhlberger2006one}       & \textbf{\color{ticgreen}{\checkmark}} & \textbf{\color{ticgreen}{\checkmark}}       & \textbf{\color{red}{\xmark}}         & \textbf{\color{red}{\xmark}}       & Multimodality  short- and long-term effects on fear of flight. \\ \cline{2-7} 
                                                                            & Hoffman et al.~\cite{hoffman2003interfaces}         & \textbf{\color{ticgreen}{\checkmark}} & \textbf{\color{red}{\xmark}}                & \textbf{\color{ticgreen}{\checkmark}}& \textbf{\color{red}{\xmark}}       & Illusions of touching to reduce fear of spiders.               \\ \hline
OCD therapy                                                                 & Cipresso et al.~\cite{cipresso2013break}        & \textbf{\color{red}{\xmark}}          & \textbf{\color{ticgreen}{\checkmark}}       & \textbf{\color{red}{\xmark}}         & \textbf{\color{red}{\xmark}}       & Different instructions to analyze behavioral syndromes.        \\ \hline
Medical data visualization                                                  & Prange et al.~\cite{prange2018medical}          & \textbf{\color{ticgreen}{\checkmark}} & \textbf{\color{ticgreen}{\checkmark}}       & \textbf{\color{ticgreen}{\checkmark}}& \textbf{\color{red}{\xmark}}       & Visualize and manipulate patients' medical data in 3D.         \\ \hline
\multirow{2}{*}{Surgery training}                                           & Hutchins et al.~\cite{hutchins2005design}        & \textbf{\color{ticgreen}{\checkmark}} & \textbf{\color{red}{\xmark}}                & \textbf{\color{ticgreen}{\checkmark}}& \textbf{\color{red}{\xmark}}       & Medical training simulator with haptic feedback.               \\ \cline{2-7} 
                                                                            & Harders et al.~\cite{harders2007multimodal}         & \textbf{\color{red}{\xmark}}          & \textbf{\color{red}{\xmark}}                & \textbf{\color{ticgreen}{\checkmark}}& \textbf{\color{red}{\xmark}}       & Medical training simulator with AR features.                   \\ \hline 
Medical education                                                           & Lu et al.~\cite{lu2010multimodal}              & \textbf{\color{ticgreen}{\checkmark}} & \textbf{\color{red}{\xmark}}                & \textbf{\color{red}{\xmark}}         & \textbf{\color{red}{\xmark}}       & Virtual platform to educate on medicine.                       \\ \hline
\end{tabular}
}
\vspace{0.2cm}
\caption {Example works of different medical applications where multimodality plays an important role.}
\label{tab:app_medicina}
\vspace{-2em}
\end{table*}
 
Rehabilitation has also leveraged 
advances in VR, yielding impressive results. Sano et al.~\cite{sano2015reliability} demonstrated that phantom limb pain (which is the sensation of an amputated limb still attached) was reliably reduced when multimodal sensory feedback was included in the VR therapy of patients with brachial plexus avulsion or arm amputation. Fordell et al.~\cite{fordell2016rehatt} presented a treatment method for chronic neglect, where a VR forced feedback interface provided sensorimotor activation in the contra-lesional arm, \new{which combined with visual scanning training, yielded improvements in activities of daily life that recquiring spatial attention, and an improvement in transfer to real life.}
\new{Moreover, spatialized sound was also beneficial to improve rehabilitation of postural control dysfunction~\cite{wang2019virtual}.}

Table~\ref{tab:app_medicina} compiles examples leveraging multimodal VR for medical applications. \csur{As for the future, VR has the potential to serve medicine even in extreme situations. Virtual care has become an option to foster personalized connections between doctors and patients when in-person appointments are not possible, continuously adapting to the realities of the COVID-19 pandemic~\cite{FutureMedicine}.}

\subsection{Education and training}
\label{subsec:6_2_learning}

Training and education are areas in which VR holds great promise, and in which it has already begun to show its capabilities:  
Jensen and Konradsen presented a review on the use of VR headsets for training and education, and showed that in many cases, better learning transfer can be achieved in this medium compared to traditional media~\cite{jensen2018review}.

In education, VR has been widely studied as a new paradigm for teaching: Designing ad-hoc environments helps create adequate scenarios for each learning purpose, \new{hence facilitating the transfer of knowledge to real life scenarios.}
Stojvsic et al.~\cite{stojvsic2019virtual} reviewed the literature on VR applications in education, and conducted a small study showing that teachers perceived benefits in introducing immersive technologies, since students were more motivated and immersed in the topic of interest. At the same time, childhood education processes have been shown to be improved by leveraging multimodality \new{in virtual environments}, by means of human-computer interaction methods \new{including feedback and interaction from multiple modalities}~\cite{christopoulos2009multimodal}, or somatic interaction (hand gestures and body movements)~\cite{fernandes2015bringing, alves2016exploring}.

Many frameworks \new{regarding VR in teaching and education} have been studied \new{and evaluated},
demonstrating that using virtual manipulatives (i.e., virtual interaction paradigms) \new{which} provide multimodal interactions \new{actually} yields richer perceptual experiences than classical methodologies in the cases of \new{e.g.,} mathematics learning~\cite{paek2012impact} or chemistry education~\cite{ali2014effect}. \new{In the case of the latter,} a virtual multimodal laboratory was designed, where the user could perform chemistry experiments like in the real world, through a 3D interaction interface with also audio-visual feedback, which indeed improved the learning capabilities of students.
Similarly, Tang et al.~\cite{tang2010multimodal} introduced an immersive multimodal virtual environment supporting interactions with 3D deformable models through haptic devices, where not only gestures were replicated but also touching forces were correctly simulated, hence generating realistic scenarios. \new{One step further,} Richard et al.~\cite{richard2006bmulti} surveyed existing works including haptic or olfactory feedback in the field of education, and described a simulation VR platform that provides haptic, olfactory, and auditory feedback, which they tested in various teaching scenarios, demonstrating they affected student engagement and learning positively, \new{and obtaining similar insights as other reviews in educational scenarios, such as in STEAM (science, technology, engineering, arts and mathematics) classrooms~\cite{taljaard2016review}}.

It is worth mentioning that multimodality can also help alleviate sensory impairments, since environments can be designed to maximize the use of the non-affected senses. Following this idea, Yu and Brewster~\cite{yu2002multimodal} studied the strengths of a multimodal interface (i.e., with speech interactions) against traditional tactile diagrams in conveying information to visually impaired and blind people, showing benefits of this approach in terms of the accuracy obtained by users.

\begin{table*}[t!]
\centering
\renewcommand{\arraystretch}{1.2}
\resizebox{\textwidth}{!}{
\def\arraystretch{1.85}%
\begin{tabular}{|c|c|c|c|c|c|c|}
\hline
\multirow{2}{*}{\textbf{Application}} & \multirow{2}{*}{\textbf{Example work}}      & \multicolumn{4}{c|}{\textbf{Additional involved senses (other than vision)}}                        & \multirow{2}{*}{\textbf{Brief description}} \\ \cline{3-6}
                                      &                                     & \textbf{Audition} & \textbf{Proprioception} & \textbf{Haptics} & \textbf{Other} &                                             \\ \hline
\multirow{6}{*}{Education}            & Christopoulos and Gaitatzes~\cite{christopoulos2009multimodal} & \textbf{\color{ticgreen}{\checkmark}} & \textbf{\color{ticgreen}{\checkmark}}       & \textbf{\color{red}{\xmark}}   & \textbf{\color{red}{\xmark}} & Children education on history               \\ \cline{2-7} 
                                      & Alves et al.~\cite{alves2016exploring}                & \textbf{\color{red}{\xmark}}    & \textbf{\color{ticgreen}{\checkmark}}       & \textbf{\color{red}{\xmark}}   & \textbf{\color{red}{\xmark}} & Serious games for children education on history               \\ \cline{2-7} 
                                      & Ali et al.~\cite{ali2014effect}                  & \textbf{\color{ticgreen}{\checkmark}} & \textbf{\color{red}{\xmark}}          & \textbf{\color{red}{\xmark}}   & \textbf{\color{red}{\xmark}} & Children education on chemistry             \\ \cline{2-7} 
                                      & Tang et al.~\cite{tang2010multimodal}                 & \textbf{\color{red}{\xmark}}    & \textbf{\color{red}{\xmark}}          & \textbf{\color{ticgreen}{\checkmark}}& \textbf{\color{red}{\xmark}} & Education on deformable materials           \\ \cline{2-7} 
                                      & Lu et al.~\cite{lu2010multimodal}                   & \textbf{\color{ticgreen}{\checkmark}} & \textbf{\color{red}{\xmark}}          & \textbf{\color{red}{\xmark}}   & \textbf{\color{red}{\xmark}} & Education on medicine                       \\ \cline{2-7} 
                                      & Richard et al.~\cite{richard2006bmulti}              & \textbf{\color{ticgreen}{\checkmark}} & \textbf{\color{red}{\xmark}}          & \textbf{\color{ticgreen}{\checkmark}}& \color{ticgreen}{Olfactory}     & Education on physics                        \\ \hline
Accessibility in education            & Yu and Brewster~\cite{yu2002multimodal}             & \textbf{\color{ticgreen}{\checkmark}} & \textbf{\color{red}{\xmark}}          & \textbf{\color{ticgreen}{\checkmark}}& \textbf{\color{red}{\xmark}} & Accessibility for blind people               \\ \hline
Serious games                         & Deng et al.~\cite{deng2014multimodality}                 & \textbf{\color{red}{\xmark}}    & \textbf{\color{red}{\xmark}}          & \textbf{\color{ticgreen}{\checkmark}}& \textbf{\color{red}{\xmark}} & Review on multimodality for serious games   \\ \hline
\multirow{5}{*}{Skill training}       & Gopher~\cite{gopher2012skill}                      & \textbf{\color{ticgreen}{\checkmark}} & \textbf{\color{red}{\xmark}}          & \textbf{\color{ticgreen}{\checkmark}}& \textbf{\color{red}{\xmark}} & Review on multimodality for skill training  \\ \cline{2-7} 
                                      & Boud et al.~\cite{boud2000virtual}                 & \textbf{\color{red}{\xmark}}    & \textbf{\color{red}{\xmark}}          & \textbf{\color{ticgreen}{\checkmark}}& \textbf{\color{red}{\xmark}} & Skill training for industrial processes     \\ \cline{2-7} 
                                      & Crison et al.~\cite{crison2005virtual}               & \textbf{\color{ticgreen}{\checkmark}} & \textbf{\color{red}{\xmark}}          & \textbf{\color{ticgreen}{\checkmark}}& \textbf{\color{red}{\xmark}} & Skill training for industrial processes     \\ \cline{2-7} 
                                      & Ha et al.~\cite{ha2009virtual}               & \textbf{\color{red}{\xmark}} & \textbf{\color{red}{\xmark}}          & \textbf{\color{ticgreen}{\checkmark}}& \textbf{\color{red}{\xmark}} & Skill training for virtual prototyping     \\ \cline{2-7} 
                                      & MacDonald et al.~\cite{macdonald2002intelligibility}            & \textbf{\color{ticgreen}{\checkmark}} & \textbf{\color{red}{\xmark}}          & \textbf{\color{red}{\xmark}}   & \textbf{\color{red}{\xmark}} & Skill training for air traffic control      \\ \hline
\end{tabular}
}
\vspace{0.2cm}
\caption {Example works of different education and training applications where multimodality plays an important role.}
\label{tab:app_education}
\vspace{-2em}
\end{table*}

\new{One widespread technique to enhance learning leverages the so-called \textit{serious games}, which enable learning by means of interactive, yet enriching video-games.}. Checa and Bustillo~\cite{checa2019review} reviewed the use of immersive VR serious games in this context, and their possitive effect on learning processes and transfer. %
Multimodal VR can \new{actually} benefit the learning process of these learning-based \new{serious} games~\cite{deng2014multimodality}, since multisensory feedback can enhance many of the cognitive processes involved. Covaci et al.~\cite{covaci2018multisensory} presented a multisensory educational game to investigate how olfactory stimuli could contribute to users' learning experience: It made the experience more enjoyable, but also led to an improvement in users' performance and overall learning. 

\new{As aforementioned, multimodal VR has potential in the transfer of knowledge. Given this, it is well suited for simulating and training complex and usually expensive real-life skills requiring high cognitive loads}. Gopher~\cite{gopher2012skill} highlighted how virtual multimodal training conditions give better results when compared with traditional training conditions in many domains, including sports, rehabilitation, industry, or surgery; with the latter being the core of Van der Meijden et al's work~\cite{van2009value}, which reviewed the use of haptic feedback for surgery training, concluding how the addition of this information yields positive assessments in the majority of the cases and even reduce surgical errors. Transferring learning from training simulators to real life situations is one of the most relevant parts of the learning process, and multimodality has been proved to enhance it~\cite{lathan2002using}.
 
In the manufacturing industry, many processes require learning specific skills,
and multimodal virtual environments can offer new ways of training. Some works have studied 
the usability of VR for a manufacturing application such as the assembly of components into a final product, where proprioception and haptic manipulation was required~\cite{boud2000virtual}. \new{Other works} have proposed a virtual system dedicated to train workers in the use and programming of milling machines, offering visual, audio and haptic (force) feedback~\cite{crison2005virtual}, also replacing the use of conventional mechanical milling machines.
Since fine motor skills can be transferred to the performance of manual tasks, other studies have analyzed the effectiveness of virtual training in the specific case of industry in contrast to real-life training~\cite{poyade2013motor}. At the end, \new{the aforementioned works on virtual skill training}
agree that virtual training could replace real training, since learning is correctly transferred, and the \new{virtual counterparts are usually less expensive and time-consuming.}

\new{VR is also extremely helpful for assembly and maintenance processes (e.g., virtual prototyping~\cite{de1999virtual}), since it provides a cheap method to directly inspect, interact with, and modify 3D prototypes without the need of a physical industrial manufacturing process~\cite{seth2011virtual}. In this context, haptic feedback might be crucial to provide feedback in assembly simulations~\cite{ha2009virtual}.}

\new{Other complex tasks can also benefit from multimodal virtual training. MacDonald et al.~\cite{macdonald2002intelligibility} focused on the air traffic control problem, and evaluated the relevant aspects of the auditory modality to improve the detection of sonic warnings, including the best design patterns to maximize performance, signal positioning, and optimal distances on the interaural axis depending on the sound amplitudes. Real-time acoustic spatialized simulation can be also used in architecture, when designing acoustic isolation, or studying how sound will be propagated through an indoor environment~\cite{vorlander2015virtual}.}

All \new{the works mentioned in this section} concluded that multimodality offered higher user engagement than unimodal or traditional environments, leading to a better experience and learning transfer. \new{Training in virtual environments has proven to be useful, specially in contexts that are hard or expensive to replicate in real life. On the other hand, and while VR training is effective, the lack of a particular modality (e.g., haptic feedback when learning to manipulate pumps~\cite{winther2020design}) could diminish the effectiveness of VR with regard to traditional \textit{hands-on} experiences.} \csur{Hence, it is important to include all the useful sensory information that is needed for each particular experience, and make it as realistic and reliable as possible.}
A list of some representative applications of multimodal VR in training and education can be found in Table~\ref{tab:app_education}.

\subsection{Entertainment}
\label{subsec:6_4_entertainment}

Entertainment is undergoing an important revolution with the re-emergence of VR as a new medium: As VR devices become more affordable, their use at consumer level is rapidly increasing. Leisure by means of VR videogames, cinematography, or narrative experiences is becoming increasingly common, and creating realistic, engaging experiences is the main goal of content creators. Multimodality can be instrumental in improving both realism and engagement. %

\begin{figure}[tbp]
    \centering
    \includegraphics[width=0.80\columnwidth]{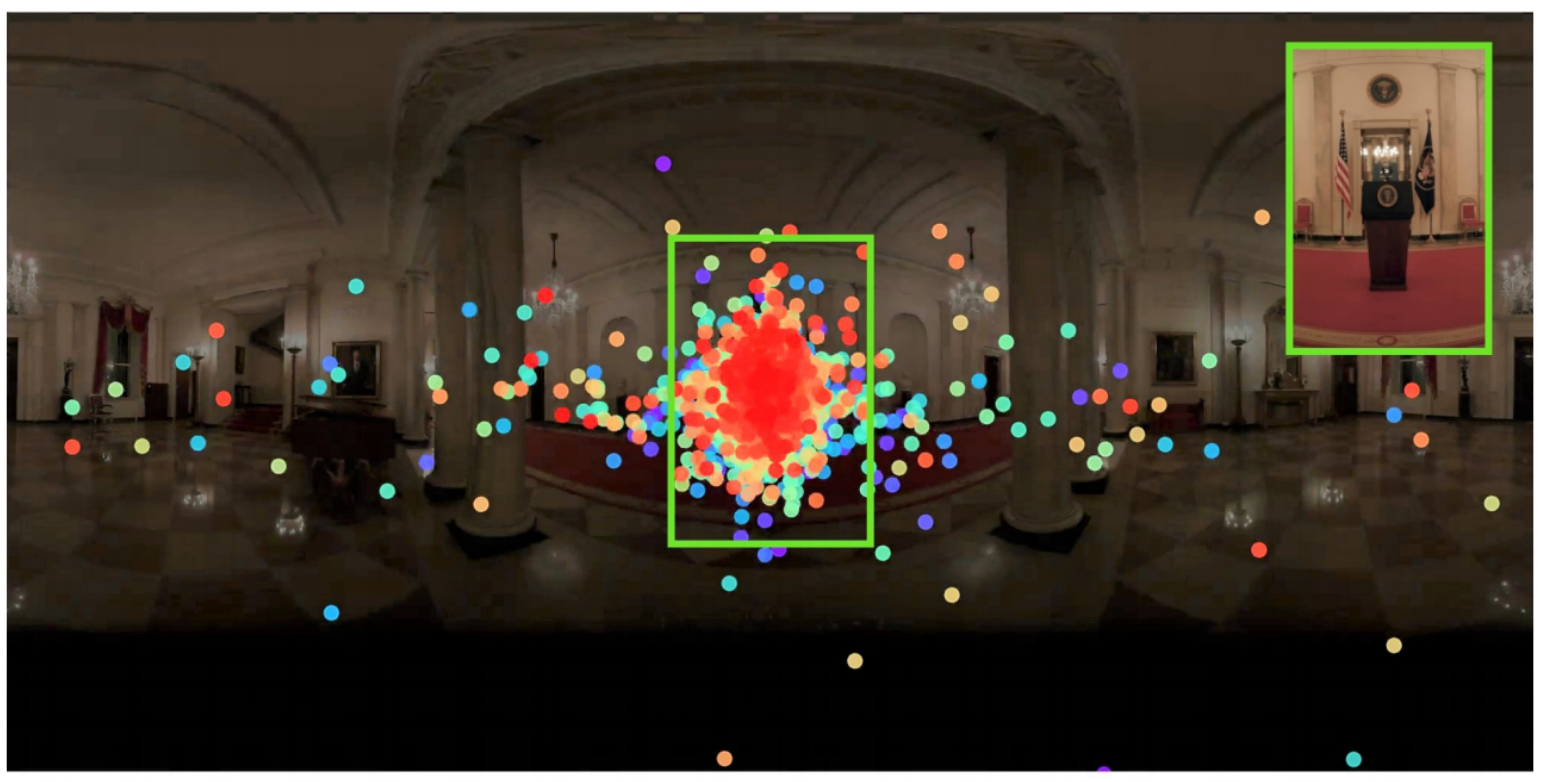}%
    \caption{Representative image of the work of  Mara\~nes et al.~\cite{maranes2020exploring}, where they analyze users' gaze behavior during visualization of VR cinematic content. One of the key open problems in VR is the generation of engaging virtual experiences that meet users' expectations. To that end, it is necessary to understand users' behavior in such virtual experiences.}
    \label{fig:examples_entertainment}
    \vspace{-0.6cm}
\end{figure}

Videogames allow users to interact with a virtual environment, controlling characters or avatars that respond based on their actions. Traditional videogames have leveraged narrative characteristics to connect with the player, to immerse them in the virtual world, so that the experience feels more engaging. With the appearance of VR, immersive games are evolving: Higher realism, and stronger feelings of presence and agency can potentially be achieved now with this technology. 

Nesbitt and Hoskens~\cite{nesbitt2008multi} hypothesized that integrating information from different senses could assist players in their performance. They evaluated visual, auditory and haptic information combinations, and although no significant performance improvement was achieved, players reported improved immersion, confidence and satisfaction in the multisensory cases. Since haptic devices may enhance the experience, some works have devoted in developing different toolkits to offer these interactions in VR (e.g., vibrotactile interactions~\cite{martinez2014vitaki}), whilst other works have exploited somatic interactions, including not only haptic but whole proprioceptive cues. Alves et al.~\cite{alves2016exploring} studied user experience in games which included hand gestures and body movements, identifying problems and potential uses of \new{gestural interaction devices} in an integrated manner 
Many narrative experiences may require the user to have the feeling of walking, and it may be one of the hardest scenarios to get a realistic response, since multiple sensory information is combined. In this scope, some works investigated the addition of multisensory walking-related cues in locomotion~\cite{kruijff2016your}, showing that adding auditory cues (i.e., footstep sounds), visual cues (i.e., head motions during walking), and vibrotactile cues (under participants' feet) could all enhance participants' sensation of self-motion (vection) and presence. Sometimes, full locomotion is not permitted, however realism can still be achieved: Colley et al.~\cite{colley2015skiing} went a step further in exploiting body proprioception, presenting a work that proposed using a HMD in skiing and snowboarding training while the user was on a real slope, so that proprioceptive cues were completely realistic. %

Although many of the current VR videogames exploit audiovisual and somatic cues (which are the easiest to provide with current technology), some have tried to work with additional cues. As in previously mentioned learning processes, some works have explored the use of olfactory cues to investigate how enabling olfaction can contribute to users' learning performance, engagement, and quality of experience~\cite{covaci2018multisensory}, \new{although this modality still remains in an early exploratory phase.} 

In a similar manner, gustatory cues have been studied in several works. Arnold et al.~\cite{arnold2017you} presented a game involving eating real food to survive, which combined with the capture and reproduction of chewing sounds increased the realism of the experience. Following this line, Mueller et al.~\cite{mueller2018towards} highlighted the potential technologies and designs to support eating as a form of play.

\new{Multisensory feedback can enhance many of the high and complex cognitive processes involved in VR~\cite{deng2014multimodality}. Particularly, multimodality can trigger different emotional responses in immersive games: Kruijff et al.~\cite{kruijff2015enhancing} investigated those emotional effects and proposed guidelines that can be applied to reproduce diverse emotional responses in multimodal games.}
\begin{table*}[tbp]
\centering
\resizebox{\textwidth}{!}{
\def\arraystretch{1.85}%
\begin{tabular}{|c|c|c|c|c|c|c|}
\hline
\multirow{2}{*}{\textbf{Application}}                                                       & \multirow{2}{*}{\textbf{Example work}} & \multicolumn{4}{c|}{\textbf{Additional involved senses (other than vision)}}                        & \multirow{2}{*}{\textbf{Brief description}}         \\ \cline{3-6}
                                                                                            &                                & \textbf{Audition} & \textbf{Proprioception} & \textbf{Haptics} & \textbf{Other}           &                                                     \\ \hline
\multirow{3}{*}{Videogames}                                                                 & Nesbitt et al.~\cite{nesbitt2008multi}         & \textbf{\color{ticgreen}{\checkmark}}       & \textbf{\color{red}{\xmark}}                & \textbf{\color{ticgreen}{\checkmark}}     & \textbf{\color{red}{\xmark}}  & Multimodality to assist players' performance        \\ \cline{2-7} 
                                                                                            & Martinez et al.~\cite{martinez2014vitaki}        & \textbf{\color{red}{\xmark}}                & \textbf{\color{red}{\xmark}}                & \textbf{\color{ticgreen}{\checkmark}}     & \textbf{\color{red}{\xmark}}  & Vibrotactile toolkit for immersive videogames       \\ \cline{2-7} 
                                                                                            & Alves et al.~\cite{alves2016exploring}           & \textbf{\color{red}{\xmark}}                & \textbf{\color{ticgreen}{\checkmark}}       & \textbf{\color{red}{\xmark}}              & \textbf{\color{red}{\xmark}}   & Serious games for children education on history     \\ \hline
\multirow{2}{*}{\begin{tabular}[c]{@{}c@{}}Physical activity simulation\end{tabular}}     & Kruijff et al.~\cite{kruijff2016your}         & \textbf{\color{ticgreen}{\checkmark}}       & \textbf{\color{ticgreen}{\checkmark}}       & \textbf{\color{ticgreen}{\checkmark}}     & \textbf{\color{red}{\xmark}}  & Walking simulation for leisure applications         \\ \cline{2-7} 
                                                                                            & Colley et al.~\cite{colley2015skiing}          & \textbf{\color{red}{\xmark}}                & \textbf{\color{ticgreen}{\checkmark}}       & \textbf{\color{red}{\xmark}}              & \textbf{\color{red}{\xmark}}  & Proprioceptive cues to simulate skiing              \\ \hline
\multirow{2}{*}{\begin{tabular}[c]{@{}c@{}}Cognitive and emotional effects\end{tabular}} & Kruijff et al.~\cite{kruijff2015enhancing}         & \textbf{\color{ticgreen}{\checkmark}}       & \textbf{\color{red}{\xmark}}                & \textbf{\color{ticgreen}{\checkmark}}     & \color{ticgreen}{Olfactory}       & Study of emotional responses in virtual experiences \\ \cline{2-7} 
                                                                                            & Deng et al.~\cite{deng2014multimodality}            & \textbf{\color{ticgreen}{\checkmark}}       & \textbf{\color{red}{\xmark}}                & \textbf{\color{ticgreen}{\checkmark}}     & \textbf{\color{red}{\xmark}}   & Cognitive load and processes in serious games       \\ \hline
\multirow{2}{*}{Narrative experiences}                                                      & Rothe et al..~\cite{rothe2017diegetic, rothe2018guiding}          & \textbf{\color{ticgreen}{\checkmark}}       & \textbf{\color{red}{\xmark}}                & \textbf{\color{red}{\xmark}}              & \textbf{\color{red}{\xmark}}  & Attention guidance in narrative experiences         \\ \cline{2-7} 
                                                                                            & Ranasinghe et al.~\cite{ranasinghe2018season}      & \textbf{\color{ticgreen}{\checkmark}}       & \textbf{\color{red}{\xmark}}                & \textbf{\color{ticgreen}{\checkmark}}     & \color{ticgreen}{Olfactory}       & Enhancing engagement in narrative experiences       \\ \hline
\end{tabular}
}
\vspace{0.1cm}
\caption {Example works of different applications in entertainment where multimodality plays an important role.}
\label{tab:app_entertainment}
\vspace{-0.8cm}
\end{table*}
Within the wide area of entertainment, cinematographic and narrative experiences in VR have been emerging during the last years. 
\new{As explored in Section~\ref{sec:n5_guidance}, guiding users' attention is specially challenging in virtual environments, where users cannot see the whole scenario at once. Although some traditional continuity editing rules may still apply~\cite{Serrano_VR-cine_SIGGRAPH2017}, and visual cuts may impact users' behavior~\cite{maranes2020exploring}, the presence of directional sounds can also influence how users explore immersive environments~\cite{masiaSound2021}, thus special attention must be paid to sound design when considering narrative experiences in VR~\cite{gospodarek2019sound}. 
To explore how different cues \new{may actually} define how users drive their attention in cinematic VR, Rothe et al.~\cite{rothe2018guiding} investigated implicitly guiding the attention of the viewer by means of lights, movements, and sounds, integrating auditory and visual modalities, while Ranasinghe et al. ~\cite{ranasinghe2018season} proposed adding olfactory and haptic (thermal and wind) stimuli to virtual narrative experiences, in order to achieve enhanced sensory engagement}. 
A compilation of representative applications of multimodal VR in entertainment can be found in Table~\ref{tab:app_entertainment}.

%% file: n9_Conclusions.tex
\section{Conclusions}
\label{sec:n9_conclusions}

Virtual reality can dramatically change the way we create and consume content in many aspects of
our everyday life, including entertainment, training, design and manufacturing, communication, or advertising. In the last years, it has been rapidly growing and evolving as a field, with the thrust of impressive technical innovations in both acquisition and visualization hardware and software.
However, if this new medium is going to succeed, it will be based on its ability to create \emph{compelling user experiences}. The interaction between different sensory modalities \new{(such as the five senses, or proprioception)} has always been of interest to content creators, but in a VR setting, in which the user

\clearpage

\thispagestyle{empty}

\begin{landscape}

\begin{textblock*}{10cm}(0.75cm,7.5cm) %
\rotatebox{90}{%
   D. Martin, and S. Malpica, and D. Gutierrez, and B. Masia, and A. Serrano
}
\end{textblock*}

\begin{table}[]
\centering
\begin{adjustwidth}{-0.77cm}{}
\resizebox{1.63\textwidth}{!}{
\def\arraystretch{1.5}%
\begin{tabular}{|c|c|c|c|c|c|c|} 
\hline
\multirow{2}{*}{\textbf{Effect}}                                          & \multirow{2}{*}{\textbf{How}}                                                                                                                                 & \multicolumn{4}{c|}{\textbf{Sensory modalities (other than visual)}}       & \multirow{2}{*}{\textbf{Related works}}            \\ 
\cline{3-6}
                                                                          &                                                                                                                                                               & \textbf{Auditory} & \textbf{Haptic} & \textbf{Proprioc.} & \textbf{Other} &                                                    \\ 
\hline
\multirow{4}{*}{Better perception and understanding of the environment (Section~\ref{sec:n3_fidelity})}    & 
\def\arraystretch{1}\begin{tabular}{@{}c@{}}Integrating the visual appearance of an object or scene \\ in the VE with a realistic and coherent sound of the same object \end{tabular}                                      & \textbf{\color{ticgreen}{\checkmark}}                 &     \textbf{\color{red}{\xmark}}            &       \textbf{\color{red}{\xmark}}             &      \textbf{\color{red}{\xmark}}          & \cite{malpica2020crossmodal,iachini2012multisensory}                                \\ 
\cline{2-7}
                                                                          & Correctly spatializing audio in a 360 environment                                                                                                             & \textbf{\color{ticgreen}{\checkmark}}                 &  \textbf{\color{red}{\xmark}}               &        \textbf{\color{red}{\xmark}}            &     \textbf{\color{red}{\xmark}}           & \cite{morgado2018self,huang2019audible,walker2003effect,groehn2001some,gao20192}\\ 
\cline{2-7}
                                                                          & \def\arraystretch{1}\begin{tabular}{@{}c@{}} Enhancing realism by touching a physical \\ object and a virtual object at the same time \end{tabular}                                                                             &    \textbf{\color{red}{\xmark}}               & \textbf{\color{ticgreen}{\checkmark}}               & \textbf{\color{ticgreen}{\checkmark}}                  & \textbf{\color{ticgreen}{Olfactory, gustatory}}      & \cite{hoffman1998physically_a,hoffman1998physically_b}\\ 
\cline{2-7}
                                                                          & \def\arraystretch{1}\begin{tabular}{@{}c@{}} Proprioceptive cues (e.g. a hand instead of a pointer) \\ are suitable for interaction scenarios \end{tabular}                                                                    &     \textbf{\color{red}{\xmark}}              &      \textbf{\color{red}{\xmark}}           & \textbf{\color{ticgreen}{\checkmark}}                  &    \textbf{\color{red}{\xmark}}            & \cite{poupyrev1998egocentric}\\ 
\hline
\multirow{2}{*}{Increase in spatial awareness (Section~\ref{sec:n3_fidelity}, \ref{sec:new_7_nav})}                            & Training users to navigate a VE using auditory reflections and reverberation                                                                                  & \textbf{\color{ticgreen}{\checkmark}}                 &      \textbf{\color{red}{\xmark}}           &     \textbf{\color{red}{\xmark}}               &       \textbf{\color{red}{\xmark}}         & \cite{andreasen2019auditory}                                         \\ 
\cline{2-7}
                                                                          & \def\arraystretch{1}\begin{tabular}{@{}c@{}}  Creating sensory illusions with worn haptic devices \\ that map different points in the VE to the body \end{tabular}                                                            &        \textbf{\color{red}{\xmark}}           & \textbf{\color{ticgreen}{\checkmark}}               & \textbf{\color{ticgreen}{\checkmark}}                  &     \textbf{\color{red}{\xmark}}           & \cite{glyn2016}\\ 
\hline
\multirow{5}{*}{Increase in the feeling of presence, realism or immersion (Section~\ref{sec:n3_fidelity})} & Being able to control a virtual body                                                                                                                           &      \textbf{\color{red}{\xmark}}             &        \textbf{\color{red}{\xmark}}         & \textbf{\color{ticgreen}{\checkmark}}                  &        \textbf{\color{red}{\xmark}}        & \cite{schuemie2001research}\\ 
\cline{2-7}
                                                                          & 
\def\arraystretch{1}\begin{tabular}{@{}c@{}} Body-centered perception (hand, face and trunk) achieved via spatio-temporal\\  multisensory information integration within peripersonal space \end{tabular}                                                                                              & \textbf{\color{ticgreen}{\checkmark}}                 & \textbf{\color{ticgreen}{\checkmark}}               & \textbf{\color{ticgreen}{\checkmark}}                  & \textbf{\color{ticgreen}{Olfactory}}            & \cite{blanke2015behavioral,petkova2008,sakhardande2020exploring}\\ 
\cline{2-7}
                                                                          & \def\arraystretch{1}\begin{tabular}{@{}c@{}} Achieving partial body ownership considering first \\  person visuo-spatial viewpoint and anatomical similarity \end{tabular}                                                     &         \textbf{\color{red}{\xmark}}          & \textbf{\color{ticgreen}{\checkmark}}               & \textbf{\color{ticgreen}{\checkmark}}                  &    \textbf{\color{red}{\xmark}}            & \cite{pozeg2015those,yuan2010}\\ 
\cline{2-7}
                                                                          & 
\def\arraystretch{1}\begin{tabular}{@{}c@{}} Integrating as much multisensory information as possible (spatialized sound, \\  vibrations, wind, real objects, physical movement, scent) coherently with the VE\end{tabular}                                                                          & \textbf{\color{ticgreen}{\checkmark}}                 & \textbf{\color{ticgreen}{\checkmark}}               & \textbf{\color{ticgreen}{\checkmark}}                  & \textbf{\color{ticgreen}{Olfactory}}            & \cite{gonccalves2019impact,jung2020impact,hecht2006multimodal}\\ 
\cline{2-7}
                                                                          & Use of soundscapes or zeitgebers in the virtual scenes                                                                                                        & \textbf{\color{ticgreen}{\checkmark}}                 &   \textbf{\color{red}{\xmark}}              &       \textbf{\color{red}{\xmark}}             &      \textbf{\color{red}{\xmark}}          & \cite{serafin2004sound,liao2020data}\\ 
\hline
\new{Self-motion} (also related to presence) (Section~\ref{sec:new_7_nav})                             & \def\arraystretch{1}\begin{tabular}{@{}c@{}} Adding walking-related multimodal cues, integrating \\ vibrations in standing or even sitting users \end{tabular}                                                                & \textbf{\color{ticgreen}{\checkmark}}                 & \textbf{\color{ticgreen}{\checkmark}}               & \textbf{\color{ticgreen}{\checkmark}}                  &       \textbf{\color{red}{\xmark}}         & \cite{kruijff2016your,kouakoua2020rhythmic,matsuda2020perception,meyer2013modulation}\\ 
\hline
\multirow{2}{*}{Modification of the saliency of the environment (Section~\ref{sec:n5_guidance})}                                      & \def\arraystretch{1}\begin{tabular}{@{}c@{}} Users tend to pay attention to regions with more \\  sensory information (mostly explored in traditional media). \end{tabular}                                                   & \textbf{\color{ticgreen}{\checkmark}}                 &      \textbf{\color{red}{\xmark}}           &     \textbf{\color{red}{\xmark}}               &       \textbf{\color{red}{\xmark}}         & \cite{de2010model,kayser2005mechanisms,de2009computational,nakajima2013incorporating,duangudom2007using,min2020multimodal}\\ 
\cline{2-7}
                                                                          & Auditory and visual stimuli are spatio-temporally correlated to increase saliency                                                                             & \textbf{\color{ticgreen}{\checkmark}}                 &       \textbf{\color{red}{\xmark}}          &      \textbf{\color{red}{\xmark}}              &     \textbf{\color{red}{\xmark}}           & \cite{chao2020audio}\\ 
\hline
\multirow{2}{*}{Guidance or direction of attention (Section~~\ref{sec:n5_guidance})}                           & Using diegetic sound and moving cues to trigger exploratory behavior                                                                                          & \textbf{\color{ticgreen}{\checkmark}}                 &      \textbf{\color{red}{\xmark}}           &         \textbf{\color{red}{\xmark}}           &        \textbf{\color{red}{\xmark}}        & \cite{rothe2017diegetic,rothe2018guiding,salselas2020,ccamci2019exploring,peck2009evaluation}\\ 
\cline{2-7}
                                                                          & 
                                        \def\arraystretch{1}\begin{tabular}{@{}c@{}} Spatialized auditory stimuli can also direct attention to specific parts \\ of the environment, including those outside the field of view \end{tabular}                                                            & \textbf{\color{ticgreen}{\checkmark}}                 &        \textbf{\color{red}{\xmark}}         &     \textbf{\color{red}{\xmark}}               &          \textbf{\color{red}{\xmark}}      & \cite{brown2016directing,bala2018cue,bala2019elephant,kyto2015ventriloquist}          \\ 
\hline
\multirow{7}{*}{Improvement of \new{user} performance (Section~\ref{sec:n6_performance})}                                   & 
\def\arraystretch{1}\begin{tabular}{@{}c@{}}Simultaneous stimuli of irrelevant sensory modalities increase detection of \\  target stimulus and can also decrease search times or increase memory retention\end{tabular}
   & \textbf{\color{ticgreen}{\checkmark}}                 & \textbf{\color{ticgreen}{\checkmark}}               &            \textbf{\color{red}{\xmark}}        & \textbf{\color{ticgreen}{Olfactory}}            & \cite{lovelace2003irrelevant,noesselt2008sound,van2009poke,maggioni2018smell}\\ 
\cline{2-7}
                                                                          & \def\arraystretch{1}\begin{tabular}{@{}c@{}} Auditory stimuli improve spatial awareness and thus \\ reducing search times and improving spatial reasoning \end{tabular}                                                           & \textbf{\color{ticgreen}{\checkmark}}                 &         \textbf{\color{red}{\xmark}}        &            \textbf{\color{red}{\xmark}}        &       \textbf{\color{red}{\xmark}}         & \cite{walker2003effect,groehn2001some,rosenkvist2019hearing,gao2020visualechoes}\\ 
\cline{2-7}
                                                                          & \def\arraystretch{1}\begin{tabular}{@{}c@{}} Cognitive process starts faster in the presence of multimodality, \\ allowing users to pay attention to more cues and details \end{tabular}                                      & \textbf{\color{ticgreen}{\checkmark}}                 & \textbf{\color{ticgreen}{\checkmark}}               &        \textbf{\color{red}{\xmark}}            &         \textbf{\color{red}{\xmark}}       & \cite{hecht2006multimodal}\\ 
\cline{2-7}
                                                                          & \def\arraystretch{1}\begin{tabular}{@{}c@{}} Combination of audio-haptic tempos to convey spatial information \\ and relevant stimuli increase performance in search tasks \end{tabular}                                        & \textbf{\color{ticgreen}{\checkmark}}                 & \textbf{\color{ticgreen}{\checkmark}}               &       \textbf{\color{red}{\xmark}}             &      \textbf{\color{red}{\xmark}}          & \cite{ammi2014intermodal}\\ 
\cline{2-7}
                                                                          & Auditory stimuli facilitate touch of virtual objects outside the field of view                                                                                & \textbf{\color{ticgreen}{\checkmark}}                 & \textbf{\color{ticgreen}{\checkmark}}               &         \textbf{\color{red}{\xmark}}           &       \textbf{\color{red}{\xmark}}         & \cite{kimura2020auditory}\\ 
\cline{2-7}
                                                                          & \def\arraystretch{1}\begin{tabular}{@{}c@{}}  Improvement of depth perception via simulated \\ touch with a force-feedback system \end{tabular}                                                                               &         \textbf{\color{red}{\xmark}}          & \textbf{\color{ticgreen}{\checkmark}}               &         \textbf{\color{red}{\xmark}}           &       \textbf{\color{red}{\xmark}}         & \cite{bouguila2000,swapp2006}\\ 
\cline{2-7}
                                                                          & \def\arraystretch{1}\begin{tabular}{@{}c@{}} Increasing the space that the user can explore \\ with their body modifying the virtual avatar \end{tabular}                                                                    &         \textbf{\color{red}{\xmark}}          &            \textbf{\color{red}{\xmark}}     & \textbf{\color{ticgreen}{\checkmark}}                  &      \textbf{\color{red}{\xmark}}          & \cite{feuchtner2017extending,feuchtner2018ownershift}\\ 
\hline
Simulation of physical properties (Section~\ref{sec:n3_fidelity}, ~\ref{sec:n6_performance})                                        & Weight or touch simulation by haptic feedback                                                                                                                 &       \textbf{\color{red}{\xmark}}            & \textbf{\color{ticgreen}{\checkmark}}               & \textbf{\color{ticgreen}{\checkmark}}                  &        \textbf{\color{red}{\xmark}}        & \cite{carlon2018,samad2019pseudo}\\ 
\hline
\end{tabular}
}
\end{adjustwidth}
\vspace{0.2cm}
\caption {Illustrative examples of different effects that can be achieved through multimodality in VR.}
\label{tab:supertable}
\end{table}

\end{landscape}

\clearpage

\noindent is immersed in an alternative reality, the importance of multimodal sensory input plays a more relevant role, \new{since the feedback either from any modality, or from the combination of multiple of them, affects the final experience}. In fact, it becomes both a possible liability, if not handled properly, and a potential strength, that if adequately leveraged can boost realism, help direct user attention, or improve user performance. Throughout this survey, we have summarized not only the main lines of research in these areas, but also outlined 
relevant insights for future directions in each of them.

While making use of multimodal setups can provide benefits to the experience, it also increases costs and complexity. From the point of view of the hardware, however, audiovisual integration is almost always present in current systems \csur{(see Table~\ref{tab:hardware_dispAudio})}, and this is also the case for proprioception (except for smartphone-based and related headsets). Most controllers also include some kind of haptic feedback, although in this case it is quite simple and rudimentary, with ample room for improvement and sophistication in consumer-level systems. \csur{While research-level technology in haptics is quite advanced, transforming it into consumer-level solutions has been and still is a challenge, due to systems complexity, durability, or cost. We are currently witnessing the first attempts at providing more sophisticated haptic interactions with simpler, consumer-level hardware by using ultrasound; and certainly more advances are to be expected in this area, given the importance of haptics to the multimodal experience.} \csur{Taste and smell are almost untapped in terms of hardware. Unlike the case of touch, where haptics is abundantly explored at research level, these senses are in their infancy from a research standpoint as well. Thus, special effort should be made towards developing hardware that is able to simulate compelling stimuli for these underexplored senses.}  From the point of view of the software, inclusion of multimodal input increases the bandwidth and computational resources needed, both current stumbling blocks of VR experiences, particularly collaborative ones. Thus, compression techniques and computational optimizations (both hardware and software-based) are two of the most active areas of research in VR that would also help an increased use of multimodal input. At the same time, works have shown that multimodal input can help maintain realism and immersion with lower quality visual input, so it can also be an advantage in these areas.
Additionally, even if it implies an increase in cost and complexity, and depending on the final application scenario, these increased costs may still be more than advantageous if the alternative is setting up a similar, real scenario, in, e.g., emergency or medical training.

The inherent increased complexity \csur{resulting from the interaction between sources} also poses a challenge for researchers in this area. %
\csur{We have reviewed a number of studies analyzing the interaction of \emph{two} sensory modalities. Most of them were based on constrained experiments under laboratory conditions. However, the final goal of VR is to be present at consumer level, where more complex phenomena and interactions are likely to happen. Thus, lifting constraints on the experimental conditions, and exploring to what extent the insights found generalize and hold in free-viewing scenarios with more confounding factors, remains a critical avenue of future work, which undoubtedly needs to be built on the findings from controlled, constrained experiments.} Works exploring three or more modalities are more rare. The integration of input from multiple senses has been an open area of research for over a century, partly because of the curse of dimensionality into which one runs when tackling this problem: The size of the parameter space grows exponentially and soon becomes intractable. Even when the data was available, deriving models to explain it has been a challenge, and analytical models often failed short to explain phenomena outside the particular scenario and parameter space explored, partly because of their lack of generality, partly because the type of data gathered can be very sensitive to the particular experimental setup. Current data-driven approaches certainly provide a new tool to address the problem, and some works have already started to rely on them, as is the case with audiovisual attention modeling. For this to be a solid path forward, however, we need public, carefully-crafted datasets that can be used by the community and in benchmarks, and we need reproducible experimental setups. Incidentally, VR is in itself a great experimental scenario for reproducibility, as opposed to physical, real-world setups.

Being aware of how the different sensory inputs interact thus helps researchers and practitioners in the field in two ways: In a first level, it aids them to create believable, successful experiences with the limited hardware and software resources available to them. 
At the next level, they can leverage the way the different sensory inputs will interact to overcome some of the limitations imposed by the hardware and software available, and even to improve the \emph{design} of such hardware and software.
As multimodal interactions become known and well understood, they can then be leveraged for algorithm design, content generation, or even hardware development, essentially contributing to create better virtual experiences for users, and helping unleash the true potential of this medium.

%% file: main.bbl

\begin{thebibliography}{242}


\ifx \showCODEN    \undefined \def \showCODEN     #1{\unskip}     \fi
\ifx \showDOI      \undefined \def \showDOI       #1{#1}\fi
\ifx \showISBNx    \undefined \def \showISBNx     #1{\unskip}     \fi
\ifx \showISBNxiii \undefined \def \showISBNxiii  #1{\unskip}     \fi
\ifx \showISSN     \undefined \def \showISSN      #1{\unskip}     \fi
\ifx \showLCCN     \undefined \def \showLCCN      #1{\unskip}     \fi
\ifx \shownote     \undefined \def \shownote      #1{#1}          \fi
\ifx \showarticletitle \undefined \def \showarticletitle #1{#1}   \fi
\ifx \showURL      \undefined \def \showURL       {\relax}        \fi
\providecommand\bibfield[2]{#2}
\providecommand\bibinfo[2]{#2}
\providecommand\natexlab[1]{#1}
\providecommand\showeprint[2][]{arXiv:#2}

\bibitem[\protect\citeauthoryear{{Akhtar} and {Falk}}{{Akhtar} and
  {Falk}}{2017}]%
        {akhtar2017}
\bibfield{author}{\bibinfo{person}{Z. {Akhtar}} {and} \bibinfo{person}{T.~H.
  {Falk}}.} \bibinfo{year}{2017}\natexlab{}.
\newblock \showarticletitle{Audio-Visual Multimedia Quality Assessment: A
  Comprehensive Survey}.
\newblock \bibinfo{journal}{\emph{IEEE Access}}  \bibinfo{volume}{5}
  (\bibinfo{year}{2017}), \bibinfo{pages}{21090--21117}.
\newblock


\bibitem[\protect\citeauthoryear{Alais and Burr}{Alais and Burr}{2004}]%
        {alais2004ventriloquist}
\bibfield{author}{\bibinfo{person}{D. Alais} {and} \bibinfo{person}{D. Burr}.}
  \bibinfo{year}{2004}\natexlab{}.
\newblock \showarticletitle{The ventriloquist effect results from near-optimal
  bimodal integration}.
\newblock \bibinfo{journal}{\emph{CurrentBiology}} (\bibinfo{year}{2004}).
\newblock


\bibitem[\protect\citeauthoryear{Ali, Ullah, Rabbi, and Alam}{Ali
  et~al\mbox{.}}{2014}]%
        {ali2014effect}
\bibfield{author}{\bibinfo{person}{Numan Ali}, \bibinfo{person}{Sehat Ullah},
  \bibinfo{person}{Ihsan Rabbi}, {and} \bibinfo{person}{Aftab Alam}.}
  \bibinfo{year}{2014}\natexlab{}.
\newblock \showarticletitle{The effect of multimodal virtual chemistry
  laboratory on students’ learning improvement}. In
  \bibinfo{booktitle}{\emph{International Conference on Augmented and Virtual
  Reality}}. Springer, \bibinfo{pages}{65--76}.
\newblock


\bibitem[\protect\citeauthoryear{Alves~Fernandes et~al\mbox{.}}{Alves~Fernandes
  et~al\mbox{.}}{2016}]%
        {alves2016exploring}
\bibfield{author}{\bibinfo{person}{Luis~Miguel Alves~Fernandes}
  {et~al\mbox{.}}} \bibinfo{year}{2016}\natexlab{}.
\newblock \showarticletitle{Exploring educational immersive videogames: an
  empirical study with a 3D multimodal interaction prototype}.
\newblock \bibinfo{journal}{\emph{Behaviour \& Information Technology}}
  \bibinfo{volume}{35}, \bibinfo{number}{11} (\bibinfo{year}{2016}),
  \bibinfo{pages}{907--918}.
\newblock


\bibitem[\protect\citeauthoryear{Ammi and Katz}{Ammi and Katz}{2014}]%
        {ammi2014intermodal}
\bibfield{author}{\bibinfo{person}{Mehdi Ammi} {and} \bibinfo{person}{Brian~FG
  Katz}.} \bibinfo{year}{2014}\natexlab{}.
\newblock \showarticletitle{Intermodal audio-haptic metaphor: improvement of
  target search in abstract environments}.
\newblock \bibinfo{journal}{\emph{International journal of human-computer
  interaction}} \bibinfo{volume}{30}, \bibinfo{number}{11}
  (\bibinfo{year}{2014}), \bibinfo{pages}{921--933}.
\newblock


\bibitem[\protect\citeauthoryear{Andreasen, Geronazzo, Nilsson, Zovnercuka,
  Konovalov, and Serafin}{Andreasen et~al\mbox{.}}{2019}]%
        {andreasen2019auditory}
\bibfield{author}{\bibinfo{person}{A. Andreasen}, \bibinfo{person}{M.
  Geronazzo}, \bibinfo{person}{N. Nilsson}, \bibinfo{person}{J. Zovnercuka},
  \bibinfo{person}{K. Konovalov}, {and} \bibinfo{person}{S. Serafin}.}
  \bibinfo{year}{2019}\natexlab{}.
\newblock \showarticletitle{Auditory feedback for navigation with echoes in
  virtual environments: training procedure and orientation strategies}.
\newblock \bibinfo{journal}{\emph{IEEE Trans. on Visualization and Computer
  Graphics}} \bibinfo{volume}{25}, \bibinfo{number}{5} (\bibinfo{year}{2019}),
  \bibinfo{pages}{1876--1886}.
\newblock


\bibitem[\protect\citeauthoryear{Armbr{\"u}ster, Wolter, Kuhlen, Spijkers, and
  Fimm}{Armbr{\"u}ster et~al\mbox{.}}{2008}]%
        {armbruster2008depth}
\bibfield{author}{\bibinfo{person}{Claudia Armbr{\"u}ster},
  \bibinfo{person}{Marc Wolter}, \bibinfo{person}{Torsten Kuhlen},
  \bibinfo{person}{Will Spijkers}, {and} \bibinfo{person}{Bruno Fimm}.}
  \bibinfo{year}{2008}\natexlab{}.
\newblock \showarticletitle{Depth perception in virtual reality: distance
  estimations in peri-and extrapersonal space}.
\newblock \bibinfo{journal}{\emph{Cyberpsychology \& Behavior}}
  \bibinfo{volume}{11}, \bibinfo{number}{1} (\bibinfo{year}{2008}),
  \bibinfo{pages}{9--15}.
\newblock


\bibitem[\protect\citeauthoryear{Arnold}{Arnold}{2017}]%
        {arnold2017you}
\bibfield{author}{\bibinfo{person}{Peter Arnold}.}
  \bibinfo{year}{2017}\natexlab{}.
\newblock \showarticletitle{You better eat to survive! exploring edible
  interactions in a virtual reality game}. In
  \bibinfo{booktitle}{\emph{Proceedings of the 2017 CHI Conference Extended
  Abstracts on Human Factors in Computing Systems}}. \bibinfo{pages}{206--209}.
\newblock


\bibitem[\protect\citeauthoryear{Arons}{Arons}{1992}]%
        {arons1992review}
\bibfield{author}{\bibinfo{person}{Barry Arons}.}
  \bibinfo{year}{1992}\natexlab{}.
\newblock \showarticletitle{A review of the cocktail party effect}.
\newblock \bibinfo{journal}{\emph{Journal of the American Voice I/O Society}}
  \bibinfo{volume}{12}, \bibinfo{number}{7} (\bibinfo{year}{1992}).
\newblock


\bibitem[\protect\citeauthoryear{Atrey, Hossain, El~Saddik, and
  Kankanhalli}{Atrey et~al\mbox{.}}{2010}]%
        {atrey2010multimodal}
\bibfield{author}{\bibinfo{person}{Pradeep~K Atrey}, \bibinfo{person}{M~Anwar
  Hossain}, \bibinfo{person}{Abdulmotaleb El~Saddik}, {and}
  \bibinfo{person}{Mohan~S Kankanhalli}.} \bibinfo{year}{2010}\natexlab{}.
\newblock \showarticletitle{Multimodal fusion for multimedia analysis: a
  survey}.
\newblock \bibinfo{journal}{\emph{Multimedia Systems}} \bibinfo{volume}{16},
  \bibinfo{number}{6} (\bibinfo{year}{2010}), \bibinfo{pages}{345--379}.
\newblock


\bibitem[\protect\citeauthoryear{Bailey, Mullaney, Gibney, and Kwakye}{Bailey
  et~al\mbox{.}}{2018}]%
        {diggs2018}
\bibfield{author}{\bibinfo{person}{Hudson~Diggs Bailey},
  \bibinfo{person}{Aidan~B. Mullaney}, \bibinfo{person}{Kyla~D. Gibney}, {and}
  \bibinfo{person}{Leslie~Dowell Kwakye}.} \bibinfo{year}{2018}\natexlab{}.
\newblock \showarticletitle{Audiovisual Integration Varies With Target and
  Environment Richness in Immersive Virtual Reality}.
\newblock \bibinfo{journal}{\emph{Multisensory Research}} \bibinfo{volume}{31},
  \bibinfo{number}{7} (\bibinfo{year}{2018}).
\newblock


\bibitem[\protect\citeauthoryear{Bala, Masu, Nisi, and Nunes}{Bala
  et~al\mbox{.}}{2018}]%
        {bala2018cue}
\bibfield{author}{\bibinfo{person}{Paulo Bala}, \bibinfo{person}{Raul Masu},
  \bibinfo{person}{Valentina Nisi}, {and} \bibinfo{person}{Nuno Nunes}.}
  \bibinfo{year}{2018}\natexlab{}.
\newblock \showarticletitle{Cue Control: Interactive Sound Spatialization for
  360º Videos}. In \bibinfo{booktitle}{\emph{International Conference on
  Interactive Digital Storytelling}}. Springer, \bibinfo{pages}{333--337}.
\newblock


\bibitem[\protect\citeauthoryear{Bala, Masu, Nisi, and Nunes}{Bala
  et~al\mbox{.}}{2019}]%
        {bala2019elephant}
\bibfield{author}{\bibinfo{person}{Paulo Bala}, \bibinfo{person}{Raul Masu},
  \bibinfo{person}{Valentina Nisi}, {and} \bibinfo{person}{Nuno Nunes}.}
  \bibinfo{year}{2019}\natexlab{}.
\newblock \showarticletitle{" When the Elephant Trumps" A Comparative Study on
  Spatial Audio for Orientation in 360{\textordmasculine} Videos}. In
  \bibinfo{booktitle}{\emph{Proc. of Conference on Human Factors in Computing
  Systems}}. \bibinfo{pages}{1--13}.
\newblock


\bibitem[\protect\citeauthoryear{Ba{\~n}os, Botella, Alca{\~n}iz, Lia{\~n}o,
  Guerrero, and Rey}{Ba{\~n}os et~al\mbox{.}}{2004}]%
        {banos2004immersion}
\bibfield{author}{\bibinfo{person}{Rosa~Mar{\'\i}a Ba{\~n}os},
  \bibinfo{person}{Cristina Botella}, \bibinfo{person}{Mariano Alca{\~n}iz},
  \bibinfo{person}{V{\'\i}ctor Lia{\~n}o}, \bibinfo{person}{Bel{\'e}n
  Guerrero}, {and} \bibinfo{person}{Beatriz Rey}.}
  \bibinfo{year}{2004}\natexlab{}.
\newblock \showarticletitle{Immersion and emotion: their impact on the sense of
  presence}.
\newblock \bibinfo{journal}{\emph{Cyberpsychology \& behavior}}
  \bibinfo{volume}{7}, \bibinfo{number}{6} (\bibinfo{year}{2004}),
  \bibinfo{pages}{734--741}.
\newblock


\bibitem[\protect\citeauthoryear{Barghout, Cha, El~Saddik, Kammerl, and
  Steinbach}{Barghout et~al\mbox{.}}{2009}]%
        {barghout2009}
\bibfield{author}{\bibinfo{person}{A. Barghout}, \bibinfo{person}{J. Cha},
  \bibinfo{person}{A. El~Saddik}, \bibinfo{person}{J. Kammerl}, {and}
  \bibinfo{person}{E. Steinbach}.} \bibinfo{year}{2009}\natexlab{}.
\newblock \showarticletitle{Spatial resolution of vibrotactile perception on
  the human forearm when exploiting funneling illusion}. In
  \bibinfo{booktitle}{\emph{IEEE Intern. Works. on Haptic Audiovis. Envs. and
  Games}}.
\newblock


\bibitem[\protect\citeauthoryear{Blanke, Slater, and Serino}{Blanke
  et~al\mbox{.}}{2015}]%
        {blanke2015behavioral}
\bibfield{author}{\bibinfo{person}{Olaf Blanke}, \bibinfo{person}{Mel Slater},
  {and} \bibinfo{person}{Andrea Serino}.} \bibinfo{year}{2015}\natexlab{}.
\newblock \showarticletitle{Behavioral, neural, and computational principles of
  bodily self-consciousness}.
\newblock \bibinfo{journal}{\emph{Neuron}} \bibinfo{volume}{88},
  \bibinfo{number}{1} (\bibinfo{year}{2015}), \bibinfo{pages}{145--166}.
\newblock


\bibitem[\protect\citeauthoryear{Bohil, Alicea, and Biocca}{Bohil
  et~al\mbox{.}}{2011}]%
        {bohil2011virtual}
\bibfield{author}{\bibinfo{person}{Corey~J Bohil}, \bibinfo{person}{Bradly
  Alicea}, {and} \bibinfo{person}{Frank~A Biocca}.}
  \bibinfo{year}{2011}\natexlab{}.
\newblock \showarticletitle{Virtual reality in neuroscience research and
  therapy}.
\newblock \bibinfo{journal}{\emph{Nature reviews neuroscience}}
  \bibinfo{volume}{12}, \bibinfo{number}{12} (\bibinfo{year}{2011}),
  \bibinfo{pages}{752--762}.
\newblock


\bibitem[\protect\citeauthoryear{Bolte and Lappe}{Bolte and Lappe}{2015}]%
        {bolte2015subliminal}
\bibfield{author}{\bibinfo{person}{Benjamin Bolte} {and}
  \bibinfo{person}{Markus Lappe}.} \bibinfo{year}{2015}\natexlab{}.
\newblock \showarticletitle{Subliminal reorientation and repositioning in
  immersive virtual environments using saccadic suppression}.
\newblock \bibinfo{journal}{\emph{IEEE Trans. on Visualization and Computer
  Graphics}} \bibinfo{volume}{21}, \bibinfo{number}{4} (\bibinfo{year}{2015}),
  \bibinfo{pages}{545--552}.
\newblock


\bibitem[\protect\citeauthoryear{Boud, Baber, and Steiner}{Boud
  et~al\mbox{.}}{2000}]%
        {boud2000virtual}
\bibfield{author}{\bibinfo{person}{AC Boud}, \bibinfo{person}{Chris Baber},
  {and} \bibinfo{person}{SJ Steiner}.} \bibinfo{year}{2000}\natexlab{}.
\newblock \showarticletitle{Virtual reality: A tool for assembly?}
\newblock \bibinfo{journal}{\emph{Presence: Teleoperators \& Virtual
  Environments}} \bibinfo{volume}{9}, \bibinfo{number}{5}
  (\bibinfo{year}{2000}), \bibinfo{pages}{486--496}.
\newblock


\bibitem[\protect\citeauthoryear{Bouguila, Ishii, and Sato}{Bouguila
  et~al\mbox{.}}{2000}]%
        {bouguila2000}
\bibfield{author}{\bibinfo{person}{Laroussi Bouguila},
  \bibinfo{person}{Masahiro Ishii}, {and} \bibinfo{person}{Makoto Sato}.}
  \bibinfo{year}{2000}\natexlab{}.
\newblock \showarticletitle{Effect of coupling haptics and stereopsis on depth
  perception in virtual environment}. In \bibinfo{booktitle}{\emph{Proc. of the
  1st Workshop on Haptic Human Computer Interaction}}. \bibinfo{pages}{54--62}.
\newblock


\bibitem[\protect\citeauthoryear{Brown, Sheikh, Evans, and Watson}{Brown
  et~al\mbox{.}}{2016}]%
        {brown2016directing}
\bibfield{author}{\bibinfo{person}{A Brown}, \bibinfo{person}{A Sheikh},
  \bibinfo{person}{M Evans}, {and} \bibinfo{person}{Z Watson}.}
  \bibinfo{year}{2016}\natexlab{}.
\newblock \showarticletitle{Directing attention in 360-degree video}. In
  \bibinfo{booktitle}{\emph{IBC 2016 Conference}}.
\newblock


\bibitem[\protect\citeauthoryear{{Burns}, {Razzaque}, {Panter}, {Whitton},
  {McCallus}, and {Brooks}}{{Burns} et~al\mbox{.}}{2005}]%
        {burns2005}
\bibfield{author}{\bibinfo{person}{E. {Burns}}, \bibinfo{person}{S.
  {Razzaque}}, \bibinfo{person}{A.~T. {Panter}}, \bibinfo{person}{M.~C.
  {Whitton}}, \bibinfo{person}{M.~R. {McCallus}}, {and} \bibinfo{person}{F.~P.
  {Brooks}}.} \bibinfo{year}{2005}\natexlab{}.
\newblock \showarticletitle{The hand is slower than the eye: a quantitative
  exploration of visual dominance over proprioception}. In
  \bibinfo{booktitle}{\emph{IEEE Proc. Virtual Reality}}.
  \bibinfo{pages}{3--10}.
\newblock


\bibitem[\protect\citeauthoryear{{\c{C}}amc{\i}}{{\c{C}}amc{\i}}{2019}]%
        {ccamci2019exploring}
\bibfield{author}{\bibinfo{person}{An{\i}l {\c{C}}amc{\i}}.}
  \bibinfo{year}{2019}\natexlab{}.
\newblock \showarticletitle{Exploring the Effects of Diegetic and Non-diegetic
  Audiovisual Cues on Decision-making in Virtual Reality}. In
  \bibinfo{booktitle}{\emph{SMC 2019. Proceedings of the 16th Sound and Music
  Computing Conference}}. \bibinfo{pages}{28--31}.
\newblock


\bibitem[\protect\citeauthoryear{Campos, Butler, and B{\"u}lthoff}{Campos
  et~al\mbox{.}}{2012}]%
        {campos2012multisensory}
\bibfield{author}{\bibinfo{person}{Jennifer~L Campos}, \bibinfo{person}{John~S
  Butler}, {and} \bibinfo{person}{Heinrich~H B{\"u}lthoff}.}
  \bibinfo{year}{2012}\natexlab{}.
\newblock \showarticletitle{Multisensory integration in the estimation of
  walked distances}.
\newblock \bibinfo{journal}{\emph{Experimental brain research}}
  \bibinfo{volume}{218}, \bibinfo{number}{4} (\bibinfo{year}{2012}),
  \bibinfo{pages}{551--565}.
\newblock


\bibitem[\protect\citeauthoryear{Carlon}{Carlon}{2018}]%
        {carlon2018}
\bibfield{author}{\bibinfo{person}{Alexsis~Danae Carlon}.}
  \bibinfo{year}{2018}\natexlab{}.
\newblock \emph{\bibinfo{title}{Virtual Reality's Utility for Examining the
  Multimodal Perception of Heaviness}}.
\newblock \bibinfo{thesistype}{Ph.D. Dissertation}. \bibinfo{school}{California
  State University, Fresno}.
\newblock


\bibitem[\protect\citeauthoryear{Chalmers, Debattista, and
  Ramic-Brkic}{Chalmers et~al\mbox{.}}{2009}]%
        {chalmers2009towards}
\bibfield{author}{\bibinfo{person}{Alan Chalmers}, \bibinfo{person}{Kurt
  Debattista}, {and} \bibinfo{person}{Belma Ramic-Brkic}.}
  \bibinfo{year}{2009}\natexlab{}.
\newblock \showarticletitle{Towards high-fidelity multi-sensory virtual
  environments}.
\newblock \bibinfo{journal}{\emph{The Visual Computer}} \bibinfo{volume}{25},
  \bibinfo{number}{12} (\bibinfo{year}{2009}), \bibinfo{pages}{1101}.
\newblock


\bibitem[\protect\citeauthoryear{Chao, Ozcinar, Wang, Zerman, Zhang,
  Hamidouche, Deforges, and Smolic}{Chao et~al\mbox{.}}{2020}]%
        {chao2020audio}
\bibfield{author}{\bibinfo{person}{F. Chao}, \bibinfo{person}{C. Ozcinar},
  \bibinfo{person}{C. Wang}, \bibinfo{person}{E. Zerman}, \bibinfo{person}{L.
  Zhang}, \bibinfo{person}{W. Hamidouche}, \bibinfo{person}{O. Deforges}, {and}
  \bibinfo{person}{A. Smolic}.} \bibinfo{year}{2020}\natexlab{}.
\newblock \showarticletitle{Audio-Visual Perception of Omnidirectional Video
  for Virtual Reality Applications}. In \bibinfo{booktitle}{\emph{IEEE Inter.
  Conf. on Mult. \& Expo Workshops}}.
\newblock


\bibitem[\protect\citeauthoryear{Chauvel, Wulf, and Maquestiaux}{Chauvel
  et~al\mbox{.}}{2015}]%
        {chauvel2015}
\bibfield{author}{\bibinfo{person}{Guillaume Chauvel},
  \bibinfo{person}{Gabriele Wulf}, {and} \bibinfo{person}{Fran{\c{c}}ois
  Maquestiaux}.} \bibinfo{year}{2015}\natexlab{}.
\newblock \showarticletitle{Visual illusions can facilitate sport skill
  learning}.
\newblock \bibinfo{journal}{\emph{Psychonomic bulletin \& review}}
  \bibinfo{volume}{22}, \bibinfo{number}{3} (\bibinfo{year}{2015}),
  \bibinfo{pages}{717--721}.
\newblock


\bibitem[\protect\citeauthoryear{Checa and Bustillo}{Checa and
  Bustillo}{2019}]%
        {checa2019review}
\bibfield{author}{\bibinfo{person}{David Checa} {and} \bibinfo{person}{Andres
  Bustillo}.} \bibinfo{year}{2019}\natexlab{}.
\newblock \showarticletitle{A review of immersive virtual reality serious games
  to enhance learning and training}.
\newblock \bibinfo{journal}{\emph{Multimedia Tools and Applications}}
  (\bibinfo{year}{2019}), \bibinfo{pages}{1--27}.
\newblock


\bibitem[\protect\citeauthoryear{Cheng and Liu}{Cheng and Liu}{2019}]%
        {cheng2019haptic}
\bibfield{author}{\bibinfo{person}{Haonan Cheng} {and}
  \bibinfo{person}{Shiguang Liu}.} \bibinfo{year}{2019}\natexlab{}.
\newblock \showarticletitle{Haptic force guided sound synthesis in multisensory
  virtual reality (VR) simulation for rigid-fluid interaction}. In
  \bibinfo{booktitle}{\emph{2019 IEEE Conference on Virtual Reality and 3D User
  Interfaces (VR)}}. IEEE.
\newblock


\bibitem[\protect\citeauthoryear{Christopoulos and Gaitatzes}{Christopoulos and
  Gaitatzes}{2009}]%
        {christopoulos2009multimodal}
\bibfield{author}{\bibinfo{person}{Dimitrios Christopoulos} {and}
  \bibinfo{person}{Athanasios Gaitatzes}.} \bibinfo{year}{2009}\natexlab{}.
\newblock \showarticletitle{Multimodal interfaces for educational virtual
  environments}. In \bibinfo{booktitle}{\emph{2009 13th Panhellenic Conference
  on Informatics}}. IEEE, \bibinfo{pages}{197--201}.
\newblock


\bibitem[\protect\citeauthoryear{Cipresso, Albani, et~al\mbox{.}}{Cipresso
  et~al\mbox{.}}{2014}]%
        {cipresso2014virtual}
\bibfield{author}{\bibinfo{person}{P. Cipresso}, \bibinfo{person}{G. Albani},
  {et~al\mbox{.}}} \bibinfo{year}{2014}\natexlab{}.
\newblock \showarticletitle{Virtual multiple errands test (VMET): a virtual
  reality-based tool to detect early executive functions deficit in
  Parkinson’s disease}.
\newblock \bibinfo{journal}{\emph{Frontiers in Behavioral Neuroscience}}
  \bibinfo{volume}{8} (\bibinfo{year}{2014}), \bibinfo{pages}{405}.
\newblock


\bibitem[\protect\citeauthoryear{Cipresso, La~Paglia, La~Cascia, Riva, Albani,
  and La~Barbera}{Cipresso et~al\mbox{.}}{2013}]%
        {cipresso2013break}
\bibfield{author}{\bibinfo{person}{P. Cipresso}, \bibinfo{person}{F.
  La~Paglia}, \bibinfo{person}{C. La~Cascia}, \bibinfo{person}{G. Riva},
  \bibinfo{person}{G. Albani}, {and} \bibinfo{person}{D. La~Barbera}.}
  \bibinfo{year}{2013}\natexlab{}.
\newblock \showarticletitle{Break in volition: A virtual reality study in
  patients with obsessive-compulsive disorder}.
\newblock \bibinfo{journal}{\emph{Experimental brain research}}
  \bibinfo{volume}{229}, \bibinfo{number}{3} (\bibinfo{year}{2013}),
  \bibinfo{pages}{443--449}.
\newblock


\bibitem[\protect\citeauthoryear{Cole and Montero}{Cole and Montero}{2007}]%
        {cole2007affective}
\bibfield{author}{\bibinfo{person}{Jonathan Cole} {and}
  \bibinfo{person}{Barbara Montero}.} \bibinfo{year}{2007}\natexlab{}.
\newblock \showarticletitle{Affective proprioception}.
\newblock \bibinfo{journal}{\emph{Janus Head}} \bibinfo{volume}{9},
  \bibinfo{number}{2} (\bibinfo{year}{2007}), \bibinfo{pages}{299--317}.
\newblock


\bibitem[\protect\citeauthoryear{Colley, V{\"a}yrynen, and
  H{\"a}kkil{\"a}}{Colley et~al\mbox{.}}{2015}]%
        {colley2015skiing}
\bibfield{author}{\bibinfo{person}{Ashley Colley}, \bibinfo{person}{Jani
  V{\"a}yrynen}, {and} \bibinfo{person}{Jonna H{\"a}kkil{\"a}}.}
  \bibinfo{year}{2015}\natexlab{}.
\newblock \showarticletitle{Skiing in a blended virtuality: an in-the-wild
  experiment}. In \bibinfo{booktitle}{\emph{Proceedings of the 19th
  International Academic Mindtrek Conference}}. \bibinfo{pages}{89--91}.
\newblock


\bibitem[\protect\citeauthoryear{Covaci, Ghinea, Lin, Huang, and Shih}{Covaci
  et~al\mbox{.}}{2018}]%
        {covaci2018multisensory}
\bibfield{author}{\bibinfo{person}{A. Covaci}, \bibinfo{person}{G. Ghinea},
  \bibinfo{person}{C. Lin}, \bibinfo{person}{S. Huang}, {and}
  \bibinfo{person}{J. Shih}.} \bibinfo{year}{2018}\natexlab{}.
\newblock \showarticletitle{Multisensory games-based learning-lessons learnt
  from olfactory enhancement of a digital board game}.
\newblock \bibinfo{journal}{\emph{Multimedia Tools and Applications}}
  \bibinfo{volume}{77}, \bibinfo{number}{16} (\bibinfo{year}{2018}).
\newblock


\bibitem[\protect\citeauthoryear{Crison, Lecuyer, d'Huart, Burkhardt, Michel,
  and Dautin}{Crison et~al\mbox{.}}{2005}]%
        {crison2005virtual}
\bibfield{author}{\bibinfo{person}{F. Crison}, \bibinfo{person}{A. Lecuyer},
  \bibinfo{person}{D. d'Huart}, \bibinfo{person}{J. Burkhardt},
  \bibinfo{person}{G. Michel}, {and} \bibinfo{person}{J. Dautin}.}
  \bibinfo{year}{2005}\natexlab{}.
\newblock \showarticletitle{Virtual technical trainer: Learning how to use
  milling machines with multi-sensory feedback in virtual reality}. In
  \bibinfo{booktitle}{\emph{IEEE Proc. Virtual Reality}}.
  \bibinfo{pages}{139--145}.
\newblock


\bibitem[\protect\citeauthoryear{De~Coensel and Botteldooren}{De~Coensel and
  Botteldooren}{2010}]%
        {de2010model}
\bibfield{author}{\bibinfo{person}{Bert De~Coensel} {and} \bibinfo{person}{Dick
  Botteldooren}.} \bibinfo{year}{2010}\natexlab{}.
\newblock \showarticletitle{A model of saliency-based auditory attention to
  environmental sound}. In \bibinfo{booktitle}{\emph{20th International
  Congress on Acoustics (ICA-2010)}}. \bibinfo{pages}{1--8}.
\newblock


\bibitem[\protect\citeauthoryear{De~Coensel, Botteldooren, Berglund, and
  Nilsson}{De~Coensel et~al\mbox{.}}{2009}]%
        {de2009computational}
\bibfield{author}{\bibinfo{person}{Bert De~Coensel}, \bibinfo{person}{Dick
  Botteldooren}, \bibinfo{person}{Birgitta Berglund}, {and}
  \bibinfo{person}{Mats~E Nilsson}.} \bibinfo{year}{2009}\natexlab{}.
\newblock \showarticletitle{A computational model for auditory saliency of
  environmental sound.}
\newblock \bibinfo{journal}{\emph{The Journal of the Acoustical Society of
  America}} \bibinfo{volume}{125}, \bibinfo{number}{4} (\bibinfo{year}{2009}),
  \bibinfo{pages}{2528--2528}.
\newblock


\bibitem[\protect\citeauthoryear{De~Sa and Zachmann}{De~Sa and
  Zachmann}{1999}]%
        {de1999virtual}
\bibfield{author}{\bibinfo{person}{Antonino~Gomes De~Sa} {and}
  \bibinfo{person}{Gabriel Zachmann}.} \bibinfo{year}{1999}\natexlab{}.
\newblock \showarticletitle{Virtual reality as a tool for verification of
  assembly and maintenance processes}.
\newblock \bibinfo{journal}{\emph{Computers \& Graphics}} \bibinfo{volume}{23},
  \bibinfo{number}{3} (\bibinfo{year}{1999}), \bibinfo{pages}{389--403}.
\newblock


\bibitem[\protect\citeauthoryear{Deng, Kirkby, Chang, and Zhang}{Deng
  et~al\mbox{.}}{2014}]%
        {deng2014multimodality}
\bibfield{author}{\bibinfo{person}{Shujie Deng}, \bibinfo{person}{Julie~A
  Kirkby}, \bibinfo{person}{Jian Chang}, {and} \bibinfo{person}{Jian~J Zhang}.}
  \bibinfo{year}{2014}\natexlab{}.
\newblock \showarticletitle{Multimodality with eye tracking and haptics: a new
  horizon for serious games?}
\newblock \bibinfo{journal}{\emph{International Journal of Serious Games}}
  \bibinfo{volume}{1}, \bibinfo{number}{4} (\bibinfo{year}{2014}),
  \bibinfo{pages}{17--34}.
\newblock


\bibitem[\protect\citeauthoryear{Dichgans and Brandt}{Dichgans and
  Brandt}{1978}]%
        {dichgans1978visual}
\bibfield{author}{\bibinfo{person}{Johannes Dichgans} {and}
  \bibinfo{person}{Thomas Brandt}.} \bibinfo{year}{1978}\natexlab{}.
\newblock \showarticletitle{Visual-vestibular interaction: Effects on
  self-motion perception and postural control}.
\newblock In \bibinfo{booktitle}{\emph{Perception}}.
  \bibinfo{publisher}{Springer}, \bibinfo{pages}{755--804}.
\newblock


\bibitem[\protect\citeauthoryear{Duangudom and Anderson}{Duangudom and
  Anderson}{2007}]%
        {duangudom2007using}
\bibfield{author}{\bibinfo{person}{Varinthira Duangudom} {and}
  \bibinfo{person}{David~V Anderson}.} \bibinfo{year}{2007}\natexlab{}.
\newblock \showarticletitle{Using auditory saliency to understand complex
  auditory scenes}. In \bibinfo{booktitle}{\emph{2007 15th European Signal
  Processing Conference}}. IEEE, \bibinfo{pages}{1206--1210}.
\newblock


\bibitem[\protect\citeauthoryear{Dumas, Lalanne, and Oviatt}{Dumas
  et~al\mbox{.}}{2009}]%
        {dumas2009multimodal}
\bibfield{author}{\bibinfo{person}{Bruno Dumas}, \bibinfo{person}{Denis
  Lalanne}, {and} \bibinfo{person}{Sharon Oviatt}.}
  \bibinfo{year}{2009}\natexlab{}.
\newblock \showarticletitle{Multimodal interfaces: A survey of principles,
  models and frameworks}.
\newblock In \bibinfo{booktitle}{\emph{Human machine interaction}}.
  \bibinfo{publisher}{Springer}, \bibinfo{pages}{3--26}.
\newblock


\bibitem[\protect\citeauthoryear{Eg and Behne}{Eg and Behne}{2015}]%
        {eg2015}
\bibfield{author}{\bibinfo{person}{Ragnhild Eg} {and} \bibinfo{person}{Dawn~M
  Behne}.} \bibinfo{year}{2015}\natexlab{}.
\newblock \showarticletitle{Perceived synchrony for realistic and dynamic
  audiovisual events}.
\newblock \bibinfo{journal}{\emph{Frontiers in psychology}}
  \bibinfo{volume}{6} (\bibinfo{year}{2015}), \bibinfo{pages}{736}.
\newblock


\bibitem[\protect\citeauthoryear{Elbamby, Perfecto, Bennis, and
  Doppler}{Elbamby et~al\mbox{.}}{2018}]%
        {elbamby2018toward}
\bibfield{author}{\bibinfo{person}{Mohammed~S Elbamby},
  \bibinfo{person}{Cristina Perfecto}, \bibinfo{person}{Mehdi Bennis}, {and}
  \bibinfo{person}{Klaus Doppler}.} \bibinfo{year}{2018}\natexlab{}.
\newblock \showarticletitle{Toward low-latency and ultra-reliable virtual
  reality}.
\newblock \bibinfo{journal}{\emph{IEEE Network}} \bibinfo{volume}{32},
  \bibinfo{number}{2} (\bibinfo{year}{2018}), \bibinfo{pages}{78--84}.
\newblock


\bibitem[\protect\citeauthoryear{Emerge}{Emerge}{2021}]%
        {Emergeio}
\bibfield{author}{\bibinfo{person}{Emerge}.} \bibinfo{year}{2021}\natexlab{}.
\newblock \bibinfo{booktitle}{\emph{Bringing touch and emotion to virtual
  experiences}}.
\newblock
\urldef\tempurl%
\url{https://emerge.io/}
\showURL{%
\tempurl}
\newblock
\shownote{Last accessed on 2021-11-02.}


\bibitem[\protect\citeauthoryear{Evangelopoulos, Zlatintsi, Potamianos,
  Maragos, Rapantzikos, Skoumas, and Avrithis}{Evangelopoulos
  et~al\mbox{.}}{2013}]%
        {evangelopoulos2013multimodal}
\bibfield{author}{\bibinfo{person}{G. Evangelopoulos}, \bibinfo{person}{A.
  Zlatintsi}, \bibinfo{person}{A. Potamianos}, \bibinfo{person}{P. Maragos},
  \bibinfo{person}{K. Rapantzikos}, \bibinfo{person}{G. Skoumas}, {and}
  \bibinfo{person}{Y. Avrithis}.} \bibinfo{year}{2013}\natexlab{}.
\newblock \showarticletitle{Multimodal saliency and fusion for movie
  summarization based on aural, visual, and textual attention}.
\newblock \bibinfo{journal}{\emph{IEEE Trans. on Multimedia}}
  \bibinfo{volume}{15}, \bibinfo{number}{7} (\bibinfo{year}{2013}),
  \bibinfo{pages}{1553--1568}.
\newblock


\bibitem[\protect\citeauthoryear{Feigl, K{\~o}re, Mutschler, and
  Philippsen}{Feigl et~al\mbox{.}}{2017}]%
        {feigl2017acoustical}
\bibfield{author}{\bibinfo{person}{Tobias Feigl}, \bibinfo{person}{Eliise
  K{\~o}re}, \bibinfo{person}{Christopher Mutschler}, {and}
  \bibinfo{person}{Michael Philippsen}.} \bibinfo{year}{2017}\natexlab{}.
\newblock \showarticletitle{Acoustical manipulation for redirected walking}. In
  \bibinfo{booktitle}{\emph{Proceedings of the 23rd ACM Symposium on Virtual
  Reality Software and Technology}}. \bibinfo{pages}{1--2}.
\newblock


\bibitem[\protect\citeauthoryear{Feng, Dey, and Lindeman}{Feng
  et~al\mbox{.}}{2016}]%
        {feng2016effect}
\bibfield{author}{\bibinfo{person}{Mi Feng}, \bibinfo{person}{Arindam Dey},
  {and} \bibinfo{person}{Robert~W Lindeman}.} \bibinfo{year}{2016}\natexlab{}.
\newblock \showarticletitle{The effect of multi-sensory cues on performance and
  experience during walking in immersive virtual environments}. In
  \bibinfo{booktitle}{\emph{2016 IEEE Virtual Reality (VR)}}. IEEE,
  \bibinfo{pages}{173--174}.
\newblock


\bibitem[\protect\citeauthoryear{Fernandes et~al\mbox{.}}{Fernandes
  et~al\mbox{.}}{2015}]%
        {fernandes2015bringing}
\bibfield{author}{\bibinfo{person}{Lu{\'\i}s Fernandes} {et~al\mbox{.}}}
  \bibinfo{year}{2015}\natexlab{}.
\newblock \showarticletitle{Bringing user experience empirical data to
  gesture-control and somatic interaction in virtual reality videogames: an
  exploratory study with a multimodal interaction prototype}. In
  \bibinfo{booktitle}{\emph{SciTecIn15-Confer{\^e}ncia Ci{\^e}ncias E
  Tecnologias Da Intera{\c{c}}{\~a}o 2015}}.
\newblock


\bibitem[\protect\citeauthoryear{Feuchtner and M{\"u}ller}{Feuchtner and
  M{\"u}ller}{2017}]%
        {feuchtner2017extending}
\bibfield{author}{\bibinfo{person}{Tiare Feuchtner} {and}
  \bibinfo{person}{J{\"o}rg M{\"u}ller}.} \bibinfo{year}{2017}\natexlab{}.
\newblock \showarticletitle{Extending the body for interaction with reality}.
  In \bibinfo{booktitle}{\emph{Proceedings of the Conference on Human Factors
  in Computing Systems}}. \bibinfo{pages}{5145--5157}.
\newblock


\bibitem[\protect\citeauthoryear{Feuchtner and M{\"u}ller}{Feuchtner and
  M{\"u}ller}{2018}]%
        {feuchtner2018ownershift}
\bibfield{author}{\bibinfo{person}{Tiare Feuchtner} {and}
  \bibinfo{person}{J{\"o}rg M{\"u}ller}.} \bibinfo{year}{2018}\natexlab{}.
\newblock \showarticletitle{Ownershift: Facilitating overhead interaction in
  virtual reality with an ownership-preserving hand space shift}. In
  \bibinfo{booktitle}{\emph{Proc. of the 31st ACM Symposium on User Interface
  Software and Technology}}.
\newblock


\bibitem[\protect\citeauthoryear{Fordell, Bodin, Eklund, and Malm}{Fordell
  et~al\mbox{.}}{2016}]%
        {fordell2016rehatt}
\bibfield{author}{\bibinfo{person}{Helena Fordell}, \bibinfo{person}{Kenneth
  Bodin}, \bibinfo{person}{Anders Eklund}, {and} \bibinfo{person}{Jan Malm}.}
  \bibinfo{year}{2016}\natexlab{}.
\newblock \showarticletitle{RehAtt--scanning training for neglect enhanced by
  multi-sensory stimulation in Virtual Reality}.
\newblock \bibinfo{journal}{\emph{Topics in stroke rehabilitation}}
  \bibinfo{volume}{23}, \bibinfo{number}{3} (\bibinfo{year}{2016}),
  \bibinfo{pages}{191--199}.
\newblock


\bibitem[\protect\citeauthoryear{Freina and Ott}{Freina and Ott}{2015}]%
        {freina2015literature}
\bibfield{author}{\bibinfo{person}{Laura Freina} {and} \bibinfo{person}{Michela
  Ott}.} \bibinfo{year}{2015}\natexlab{}.
\newblock \showarticletitle{A literature review on immersive virtual reality in
  education: state of the art and perspectives}. In
  \bibinfo{booktitle}{\emph{The international scientific conference elearning
  and software for education}}, Vol.~\bibinfo{volume}{1}.
  \bibinfo{pages}{10--1007}.
\newblock


\bibitem[\protect\citeauthoryear{Gallace, Ngo, Sulaitis, and Spence}{Gallace
  et~al\mbox{.}}{2012}]%
        {gallace2012multisensory}
\bibfield{author}{\bibinfo{person}{Alberto Gallace}, \bibinfo{person}{Mary~K
  Ngo}, \bibinfo{person}{John Sulaitis}, {and} \bibinfo{person}{Charles
  Spence}.} \bibinfo{year}{2012}\natexlab{}.
\newblock \showarticletitle{Multisensory presence in virtual reality:
  possibilities \& limitations}.
\newblock In \bibinfo{booktitle}{\emph{Multiple sensorial media advances and
  applications}}. \bibinfo{publisher}{IGI Global}, \bibinfo{pages}{1--38}.
\newblock


\bibitem[\protect\citeauthoryear{Gao, Chen, Al-Halah, Schissler, and
  Grauman}{Gao et~al\mbox{.}}{2020}]%
        {gao2020visualechoes}
\bibfield{author}{\bibinfo{person}{Ruohan Gao}, \bibinfo{person}{Changan Chen},
  \bibinfo{person}{Ziad Al-Halah}, \bibinfo{person}{Carl Schissler}, {and}
  \bibinfo{person}{Kristen Grauman}.} \bibinfo{year}{2020}\natexlab{}.
\newblock \showarticletitle{VisualEchoes: Spatial Image Representation Learning
  through Echolocation}.
\newblock \bibinfo{journal}{\emph{arXiv preprint arXiv:2005.01616}}
  (\bibinfo{year}{2020}).
\newblock


\bibitem[\protect\citeauthoryear{Gao and Grauman}{Gao and Grauman}{2019}]%
        {gao20192}
\bibfield{author}{\bibinfo{person}{R. Gao} {and} \bibinfo{person}{K. Grauman}.}
  \bibinfo{year}{2019}\natexlab{}.
\newblock \showarticletitle{2.5D visual sound}. In
  \bibinfo{booktitle}{\emph{Proc. of IEEE Conf. on Comp. Vision and Pattern
  Recogn.}}
\newblock


\bibitem[\protect\citeauthoryear{Gon{\c{c}}alves, Melo, Vasconcelos-Raposo, and
  Bessa}{Gon{\c{c}}alves et~al\mbox{.}}{2019}]%
        {gonccalves2019impact}
\bibfield{author}{\bibinfo{person}{Guilherme Gon{\c{c}}alves},
  \bibinfo{person}{Miguel Melo}, \bibinfo{person}{Jos{\'e} Vasconcelos-Raposo},
  {and} \bibinfo{person}{Maximino Bessa}.} \bibinfo{year}{2019}\natexlab{}.
\newblock \showarticletitle{Impact of different sensory stimuli on presence in
  credible virtual environments}.
\newblock \bibinfo{journal}{\emph{IEEE Trans. on Vis. and Computer Graphics}}
  \bibinfo{volume}{26}, \bibinfo{number}{11} (\bibinfo{year}{2019}).
\newblock


\bibitem[\protect\citeauthoryear{Gopher}{Gopher}{2012}]%
        {gopher2012skill}
\bibfield{author}{\bibinfo{person}{Daniel Gopher}.}
  \bibinfo{year}{2012}\natexlab{}.
\newblock \showarticletitle{Skill training in multimodal virtual environments}.
\newblock \bibinfo{journal}{\emph{Work}} \bibinfo{volume}{41},
  \bibinfo{number}{Supplement 1} (\bibinfo{year}{2012}),
  \bibinfo{pages}{2284--2287}.
\newblock


\bibitem[\protect\citeauthoryear{Gospodarek, Genovese, Dembeck, Brenner,
  Roginska, and Perlin}{Gospodarek et~al\mbox{.}}{2019}]%
        {gospodarek2019sound}
\bibfield{author}{\bibinfo{person}{M. Gospodarek}, \bibinfo{person}{A.
  Genovese}, \bibinfo{person}{D. Dembeck}, \bibinfo{person}{C. Brenner},
  \bibinfo{person}{Ag. Roginska}, {and} \bibinfo{person}{K. Perlin}.}
  \bibinfo{year}{2019}\natexlab{}.
\newblock \showarticletitle{Sound design and reproduction techniques for
  co-located narrative VR experiences}. In \bibinfo{booktitle}{\emph{Audio
  Engineering Society Convention 147}}.
\newblock


\bibitem[\protect\citeauthoryear{Groehn, Lokki, Savioja, and Takala}{Groehn
  et~al\mbox{.}}{2001}]%
        {groehn2001some}
\bibfield{author}{\bibinfo{person}{Matti Groehn}, \bibinfo{person}{Tapio
  Lokki}, \bibinfo{person}{Lauri Savioja}, {and} \bibinfo{person}{Tapio
  Takala}.} \bibinfo{year}{2001}\natexlab{}.
\newblock \showarticletitle{Some aspects of role of audio in immersive
  visualization}. In \bibinfo{booktitle}{\emph{Visual Data Exploration and
  Analysis VIII}}, Vol.~\bibinfo{volume}{4302}. International Society for
  Optics and Photonics.
\newblock


\bibitem[\protect\citeauthoryear{Gugenheimer, Wolf, Haas, Krebs, and
  Rukzio}{Gugenheimer et~al\mbox{.}}{2016}]%
        {gugenheimer2016swivrchair}
\bibfield{author}{\bibinfo{person}{J. Gugenheimer}, \bibinfo{person}{D. Wolf},
  \bibinfo{person}{G. Haas}, \bibinfo{person}{S. Krebs}, {and}
  \bibinfo{person}{E. Rukzio}.} \bibinfo{year}{2016}\natexlab{}.
\newblock \showarticletitle{Swivrchair: A motorized swivel chair to nudge
  users' orientation for 360 degree storytelling in virtual reality}. In
  \bibinfo{booktitle}{\emph{Proc. of the Conf. on Human Factors in Comp.
  Systems}}.
\newblock


\bibitem[\protect\citeauthoryear{G{\"u}rk{\"o}k and Nijholt}{G{\"u}rk{\"o}k and
  Nijholt}{2012}]%
        {gurkok2012brain}
\bibfield{author}{\bibinfo{person}{Hayrettin G{\"u}rk{\"o}k} {and}
  \bibinfo{person}{Anton Nijholt}.} \bibinfo{year}{2012}\natexlab{}.
\newblock \showarticletitle{Brain--computer interfaces for multimodal
  interaction: A survey and principles}.
\newblock \bibinfo{journal}{\emph{International Journal of Human-Computer
  Interaction}} \bibinfo{volume}{28}, \bibinfo{number}{5}
  (\bibinfo{year}{2012}), \bibinfo{pages}{292--307}.
\newblock


\bibitem[\protect\citeauthoryear{Ha, Kim, Park, Jun, and Rho}{Ha
  et~al\mbox{.}}{2009}]%
        {ha2009virtual}
\bibfield{author}{\bibinfo{person}{Sungdo Ha}, \bibinfo{person}{Laehyun Kim},
  \bibinfo{person}{Sehyung Park}, \bibinfo{person}{Cha-soo Jun}, {and}
  \bibinfo{person}{H Rho}.} \bibinfo{year}{2009}\natexlab{}.
\newblock \showarticletitle{Virtual prototyping enhanced by a haptic
  interface}.
\newblock \bibinfo{journal}{\emph{CIRP annals}} \bibinfo{volume}{58},
  \bibinfo{number}{1} (\bibinfo{year}{2009}), \bibinfo{pages}{135--138}.
\newblock


\bibitem[\protect\citeauthoryear{Harders, Bianchi, and Knoerlein}{Harders
  et~al\mbox{.}}{2007}]%
        {harders2007multimodal}
\bibfield{author}{\bibinfo{person}{Matthias Harders}, \bibinfo{person}{Gerald
  Bianchi}, {and} \bibinfo{person}{Benjamin Knoerlein}.}
  \bibinfo{year}{2007}\natexlab{}.
\newblock \showarticletitle{Multimodal augmented reality in medicine}. In
  \bibinfo{booktitle}{\emph{International Conference on Universal Access in
  Human-Computer Interaction}}. Springer, \bibinfo{pages}{652--658}.
\newblock


\bibitem[\protect\citeauthoryear{Hayashi, Fujita, Takashima, Lindernan, and
  Kitarnura}{Hayashi et~al\mbox{.}}{2019}]%
        {hayashi2019redirected}
\bibfield{author}{\bibinfo{person}{O. Hayashi}, \bibinfo{person}{K. Fujita},
  \bibinfo{person}{K. Takashima}, \bibinfo{person}{R. Lindernan}, {and}
  \bibinfo{person}{Y. Kitarnura}.} \bibinfo{year}{2019}\natexlab{}.
\newblock \showarticletitle{Redirected Jumping: Imperceptibly Manipulating Jump
  Motions in Virtual Reality}. In \bibinfo{booktitle}{\emph{IEEE Conf. on
  Virtual Reality and 3D User Interfaces}}.
\newblock


\bibitem[\protect\citeauthoryear{Hecht, Reiner, and Halevy}{Hecht
  et~al\mbox{.}}{2006}]%
        {hecht2006multimodal}
\bibfield{author}{\bibinfo{person}{David Hecht}, \bibinfo{person}{Miriam
  Reiner}, {and} \bibinfo{person}{Gad Halevy}.}
  \bibinfo{year}{2006}\natexlab{}.
\newblock \showarticletitle{Multimodal virtual environments: response times,
  attention, and presence}.
\newblock \bibinfo{journal}{\emph{Presence: Teleoperators and virtual
  environments}} \bibinfo{volume}{15}, \bibinfo{number}{5}
  (\bibinfo{year}{2006}), \bibinfo{pages}{515--523}.
\newblock


\bibitem[\protect\citeauthoryear{Herrera and McMahan}{Herrera and
  McMahan}{2014}]%
        {herrera2014development}
\bibfield{author}{\bibinfo{person}{Nicolas~S Herrera} {and}
  \bibinfo{person}{Ryan~P McMahan}.} \bibinfo{year}{2014}\natexlab{}.
\newblock \showarticletitle{Development of a simple and low-cost olfactory
  display for immersive media experiences}. In
  \bibinfo{booktitle}{\emph{Proceedings of the ACM International Workshop on
  Immersive Media Experiences}}. \bibinfo{pages}{1--6}.
\newblock


\bibitem[\protect\citeauthoryear{Hidaka and Ide}{Hidaka and Ide}{2015}]%
        {hidaka2015sound}
\bibfield{author}{\bibinfo{person}{Souta Hidaka} {and}
  \bibinfo{person}{Masakazu Ide}.} \bibinfo{year}{2015}\natexlab{}.
\newblock \showarticletitle{Sound can suppress visual perception}.
\newblock \bibinfo{journal}{\emph{Scientific Reports}} \bibinfo{volume}{5},
  \bibinfo{number}{1} (\bibinfo{year}{2015}), \bibinfo{pages}{1--9}.
\newblock


\bibitem[\protect\citeauthoryear{Hoffman}{Hoffman}{1998}]%
        {hoffman1998physically_a}
\bibfield{author}{\bibinfo{person}{Hunter~G Hoffman}.}
  \bibinfo{year}{1998}\natexlab{}.
\newblock \showarticletitle{Physically touching virtual objects using tactile
  augmentation enhances the realism of virtual environments}. In
  \bibinfo{booktitle}{\emph{Proceedings. IEEE 1998 Virtual Reality Annual
  International Symposium}}. IEEE, \bibinfo{pages}{59--63}.
\newblock


\bibitem[\protect\citeauthoryear{Hoffman, Garcia-Palacios, Carlin, Furness~Iii,
  and Botella-Arbona}{Hoffman et~al\mbox{.}}{2003}]%
        {hoffman2003interfaces}
\bibfield{author}{\bibinfo{person}{Hunter~G Hoffman}, \bibinfo{person}{Azucena
  Garcia-Palacios}, \bibinfo{person}{Albert Carlin}, \bibinfo{person}{Thomas~A
  Furness~Iii}, {and} \bibinfo{person}{Cristina Botella-Arbona}.}
  \bibinfo{year}{2003}\natexlab{}.
\newblock \showarticletitle{Interfaces that heal: coupling real and virtual
  objects to treat spider phobia}.
\newblock \bibinfo{journal}{\emph{Intern. Journal of HCI}}
  \bibinfo{volume}{16}, \bibinfo{number}{2} (\bibinfo{year}{2003}).
\newblock


\bibitem[\protect\citeauthoryear{Hoffman, Hollander, Schroder, Rousseau, and
  Furness}{Hoffman et~al\mbox{.}}{1998}]%
        {hoffman1998physically_b}
\bibfield{author}{\bibinfo{person}{Hunter~G Hoffman}, \bibinfo{person}{Ari
  Hollander}, \bibinfo{person}{Konrad Schroder}, \bibinfo{person}{Scott
  Rousseau}, {and} \bibinfo{person}{Tom Furness}.}
  \bibinfo{year}{1998}\natexlab{}.
\newblock \showarticletitle{Physically touching and tasting virtual objects
  enhances the realism of virtual experiences}.
\newblock \bibinfo{journal}{\emph{Virtual Reality}} \bibinfo{volume}{3},
  \bibinfo{number}{4} (\bibinfo{year}{1998}), \bibinfo{pages}{226--234}.
\newblock


\bibitem[\protect\citeauthoryear{Howard, Jenkin, and Hu}{Howard
  et~al\mbox{.}}{2000}]%
        {howard2000visually}
\bibfield{author}{\bibinfo{person}{IP Howard}, \bibinfo{person}{HL Jenkin},
  {and} \bibinfo{person}{G Hu}.} \bibinfo{year}{2000}\natexlab{}.
\newblock \showarticletitle{Visually-induced reorientation illusions as a
  function of age.}
\newblock \bibinfo{journal}{\emph{Aviation, space, and environmental medicine}}
  \bibinfo{volume}{71}, \bibinfo{number}{9 Suppl} (\bibinfo{year}{2000}),
  \bibinfo{pages}{A87--91}.
\newblock


\bibitem[\protect\citeauthoryear{Hu, Sun, Didyk, Wei, and Kaufman}{Hu
  et~al\mbox{.}}{2019}]%
        {hu2019reducing}
\bibfield{author}{\bibinfo{person}{Ping Hu}, \bibinfo{person}{Qi Sun},
  \bibinfo{person}{Piotr Didyk}, \bibinfo{person}{Li-Yi Wei}, {and}
  \bibinfo{person}{Arie~E Kaufman}.} \bibinfo{year}{2019}\natexlab{}.
\newblock \showarticletitle{Reducing simulator sickness with perceptual camera
  control}.
\newblock \bibinfo{journal}{\emph{ACM Trans. on Graphics (TOG)}}
  \bibinfo{volume}{38}, \bibinfo{number}{6} (\bibinfo{year}{2019}),
  \bibinfo{pages}{1--12}.
\newblock


\bibitem[\protect\citeauthoryear{Hu, Li, Zhang, Yi, Wang, and Manocha}{Hu
  et~al\mbox{.}}{2020}]%
        {hu2020dgaze}
\bibfield{author}{\bibinfo{person}{Zhiming Hu}, \bibinfo{person}{Sheng Li},
  \bibinfo{person}{Congyi Zhang}, \bibinfo{person}{Kangrui Yi},
  \bibinfo{person}{Guoping Wang}, {and} \bibinfo{person}{Dinesh Manocha}.}
  \bibinfo{year}{2020}\natexlab{}.
\newblock \showarticletitle{Dgaze: Cnn-based gaze prediction in dynamic
  scenes}.
\newblock \bibinfo{journal}{\emph{IEEE Trans. on Visualization and Computer
  Graphics}} \bibinfo{volume}{26}, \bibinfo{number}{5} (\bibinfo{year}{2020}),
  \bibinfo{pages}{1902--1911}.
\newblock


\bibitem[\protect\citeauthoryear{Huang, Solah, Li, and Yu}{Huang
  et~al\mbox{.}}{2019}]%
        {huang2019audible}
\bibfield{author}{\bibinfo{person}{Haikun Huang}, \bibinfo{person}{Michael
  Solah}, \bibinfo{person}{Dingzeyu Li}, {and} \bibinfo{person}{Lap-Fai Yu}.}
  \bibinfo{year}{2019}\natexlab{}.
\newblock \showarticletitle{Audible panorama: Automatic spatial audio
  generation for panorama imagery}. In \bibinfo{booktitle}{\emph{Proceedings of
  the 2019 CHI conference on human factors in computing systems}}.
\newblock


\bibitem[\protect\citeauthoryear{Hutchins, Adcock, Stevenson, Gunn, and
  Krumpholz}{Hutchins et~al\mbox{.}}{2005}]%
        {hutchins2005design}
\bibfield{author}{\bibinfo{person}{Matthew Hutchins}, \bibinfo{person}{Matt
  Adcock}, \bibinfo{person}{Duncan Stevenson}, \bibinfo{person}{Chris Gunn},
  {and} \bibinfo{person}{Alexander Krumpholz}.}
  \bibinfo{year}{2005}\natexlab{}.
\newblock \showarticletitle{The design of perceptual representations for
  practical networked multimodal virtual training environments}. In
  \bibinfo{booktitle}{\emph{the Proceedings of the 11th International
  Conference on Human-Computer Interaction: HCI International'05}}.
\newblock


\bibitem[\protect\citeauthoryear{Iachini, Maffei, Ruotolo, Senese, Ruggiero,
  Masullo, and Alekseeva}{Iachini et~al\mbox{.}}{2012}]%
        {iachini2012multisensory}
\bibfield{author}{\bibinfo{person}{T. Iachini}, \bibinfo{person}{L. Maffei},
  \bibinfo{person}{F. Ruotolo}, \bibinfo{person}{V. Senese},
  \bibinfo{person}{G. Ruggiero}, \bibinfo{person}{M. Masullo}, {and}
  \bibinfo{person}{N. Alekseeva}.} \bibinfo{year}{2012}\natexlab{}.
\newblock \showarticletitle{Multisensory assessment of acoustic comfort aboard
  metros: a virtual reality study}.
\newblock \bibinfo{journal}{\emph{Applied Cognitive Psychology}}
  \bibinfo{volume}{26}, \bibinfo{number}{5} (\bibinfo{year}{2012}).
\newblock


\bibitem[\protect\citeauthoryear{Ichimura, Nakajima, and Juzoji}{Ichimura
  et~al\mbox{.}}{2001}]%
        {ichimura2001investigation}
\bibfield{author}{\bibinfo{person}{Atsushi Ichimura}, \bibinfo{person}{Isao
  Nakajima}, {and} \bibinfo{person}{Hiroshi Juzoji}.}
  \bibinfo{year}{2001}\natexlab{}.
\newblock \showarticletitle{Investigation and analysis of a reported incident
  resulting in an actual airline hijacking due to a fanatical and engrossed VR
  state}.
\newblock \bibinfo{journal}{\emph{CyberPsychology \& Behavior}}
  \bibinfo{volume}{4}, \bibinfo{number}{3} (\bibinfo{year}{2001}).
\newblock


\bibitem[\protect\citeauthoryear{Ide and Hidaka}{Ide and Hidaka}{2013}]%
        {ide2013}
\bibfield{author}{\bibinfo{person}{Masakazu Ide} {and} \bibinfo{person}{Souta
  Hidaka}.} \bibinfo{year}{2013}\natexlab{}.
\newblock \showarticletitle{Tactile stimulation can suppress visual
  perception}.
\newblock \bibinfo{journal}{\emph{Scientific reports}}  \bibinfo{volume}{3}
  (\bibinfo{year}{2013}).
\newblock


\bibitem[\protect\citeauthoryear{Ito}{Ito}{2012}]%
        {ito2012}
\bibfield{author}{\bibinfo{person}{Hiroyuki Ito}.}
  \bibinfo{year}{2012}\natexlab{}.
\newblock \showarticletitle{Cortical shape adaptation transforms a circle into
  a hexagon: A novel afterimage illusion}.
\newblock \bibinfo{journal}{\emph{Psychological Science}} \bibinfo{volume}{23},
  \bibinfo{number}{2} (\bibinfo{year}{2012}), \bibinfo{pages}{126--132}.
\newblock


\bibitem[\protect\citeauthoryear{Jaimes and Sebe}{Jaimes and Sebe}{2007}]%
        {jaimes2007multimodal}
\bibfield{author}{\bibinfo{person}{Alejandro Jaimes} {and}
  \bibinfo{person}{Nicu Sebe}.} \bibinfo{year}{2007}\natexlab{}.
\newblock \showarticletitle{Multimodal human--computer interaction: A survey}.
\newblock \bibinfo{journal}{\emph{Computer vision and image understanding}}
  \bibinfo{volume}{108}, \bibinfo{number}{1-2} (\bibinfo{year}{2007}),
  \bibinfo{pages}{116--134}.
\newblock


\bibitem[\protect\citeauthoryear{Je, Kim, Lee, Lee, Yang, Lopes, and
  Bianchi}{Je et~al\mbox{.}}{2019}]%
        {je2019aero}
\bibfield{author}{\bibinfo{person}{Seungwoo Je}, \bibinfo{person}{Myung~Jin
  Kim}, \bibinfo{person}{Woojin Lee}, \bibinfo{person}{Byungjoo Lee},
  \bibinfo{person}{Xing-Dong Yang}, \bibinfo{person}{Pedro Lopes}, {and}
  \bibinfo{person}{Andrea Bianchi}.} \bibinfo{year}{2019}\natexlab{}.
\newblock \showarticletitle{Aero-plane: A Handheld Force-Feedback Device that
  Renders Weight Motion Illusion on a Virtual 2D Plane}. In
  \bibinfo{booktitle}{\emph{Proceedings of the ACM Symposium on User Interface
  Software and Technology}}. \bibinfo{pages}{763--775}.
\newblock


\bibitem[\protect\citeauthoryear{Jensen and Konradsen}{Jensen and
  Konradsen}{2018}]%
        {jensen2018review}
\bibfield{author}{\bibinfo{person}{Lasse Jensen} {and}
  \bibinfo{person}{Flemming Konradsen}.} \bibinfo{year}{2018}\natexlab{}.
\newblock \showarticletitle{A review of the use of virtual reality head-mounted
  displays in education and training}.
\newblock \bibinfo{journal}{\emph{Education and Information Technologies}}
  \bibinfo{volume}{23}, \bibinfo{number}{4} (\bibinfo{year}{2018}),
  \bibinfo{pages}{1515--1529}.
\newblock


\bibitem[\protect\citeauthoryear{Jung, Wood, Hoermann, Abhayawardhana, and
  Lindeman}{Jung et~al\mbox{.}}{2020}]%
        {jung2020impact}
\bibfield{author}{\bibinfo{person}{S. Jung}, \bibinfo{person}{A. Wood},
  \bibinfo{person}{S. Hoermann}, \bibinfo{person}{P. Abhayawardhana}, {and}
  \bibinfo{person}{R. Lindeman}.} \bibinfo{year}{2020}\natexlab{}.
\newblock \showarticletitle{The Impact of Multi-sensory Stimuli on Confidence
  Levels for Perceptual-cognitive Tasks in VR}. In
  \bibinfo{booktitle}{\emph{IEEE Conf. on Virtual Reality and 3D User
  Interfaces}}.
\newblock


\bibitem[\protect\citeauthoryear{Kayser, Petkov, Lippert, and
  Logothetis}{Kayser et~al\mbox{.}}{2005}]%
        {kayser2005mechanisms}
\bibfield{author}{\bibinfo{person}{Christoph Kayser},
  \bibinfo{person}{Christopher~I Petkov}, \bibinfo{person}{Michael Lippert},
  {and} \bibinfo{person}{Nikos~K Logothetis}.} \bibinfo{year}{2005}\natexlab{}.
\newblock \showarticletitle{Mechanisms for allocating auditory attention: an
  auditory saliency map}.
\newblock \bibinfo{journal}{\emph{Current Biology}} \bibinfo{volume}{15},
  \bibinfo{number}{21} (\bibinfo{year}{2005}), \bibinfo{pages}{1943--1947}.
\newblock


\bibitem[\protect\citeauthoryear{Keshavarz, Hettinger, Vena, and
  Campos}{Keshavarz et~al\mbox{.}}{2014}]%
        {keshavarz2014combined}
\bibfield{author}{\bibinfo{person}{Behrang Keshavarz},
  \bibinfo{person}{Lawrence~J Hettinger}, \bibinfo{person}{Daniel Vena}, {and}
  \bibinfo{person}{Jennifer~L Campos}.} \bibinfo{year}{2014}\natexlab{}.
\newblock \showarticletitle{Combined effects of auditory and visual cues on the
  perception of vection}.
\newblock \bibinfo{journal}{\emph{Experimental brain research}}
  \bibinfo{volume}{232}, \bibinfo{number}{3} (\bibinfo{year}{2014}),
  \bibinfo{pages}{827--836}.
\newblock


\bibitem[\protect\citeauthoryear{Kilteni, Groten, and Slater}{Kilteni
  et~al\mbox{.}}{2012}]%
        {kilteni2012sense}
\bibfield{author}{\bibinfo{person}{Konstantina Kilteni},
  \bibinfo{person}{Raphaela Groten}, {and} \bibinfo{person}{Mel Slater}.}
  \bibinfo{year}{2012}\natexlab{}.
\newblock \showarticletitle{The sense of embodiment in virtual reality}.
\newblock \bibinfo{journal}{\emph{Presence: Teleoperators and Virtual
  Environments}} \bibinfo{volume}{21}, \bibinfo{number}{4}
  (\bibinfo{year}{2012}), \bibinfo{pages}{373--387}.
\newblock


\bibitem[\protect\citeauthoryear{Kimura and Sato}{Kimura and Sato}{2020}]%
        {kimura2020auditory}
\bibfield{author}{\bibinfo{person}{Zentaro Kimura} {and} \bibinfo{person}{Mie
  Sato}.} \bibinfo{year}{2020}\natexlab{}.
\newblock \showarticletitle{Auditory Stimulation on Touching a Virtual Object
  Outside a User’s Field of View}. In \bibinfo{booktitle}{\emph{2020 IEEE
  Conference on Virtual Reality and 3D User Interfaces Abstracts and Workshops
  (VRW)}}. IEEE.
\newblock


\bibitem[\protect\citeauthoryear{Koelewijn, Bronkhorst, and Theeuwes}{Koelewijn
  et~al\mbox{.}}{2010}]%
        {koelewijn2010attention}
\bibfield{author}{\bibinfo{person}{Thomas Koelewijn}, \bibinfo{person}{Adelbert
  Bronkhorst}, {and} \bibinfo{person}{Jan Theeuwes}.}
  \bibinfo{year}{2010}\natexlab{}.
\newblock \showarticletitle{Attention and the multiple stages of multisensory
  integration: A review of audiovisual studies}.
\newblock \bibinfo{journal}{\emph{Acta psychologica}} \bibinfo{volume}{134},
  \bibinfo{number}{3} (\bibinfo{year}{2010}), \bibinfo{pages}{372--384}.
\newblock


\bibitem[\protect\citeauthoryear{Kouakoua, Duclos, Aissaoui, Nadeau, and
  Labbe}{Kouakoua et~al\mbox{.}}{2020}]%
        {kouakoua2020rhythmic}
\bibfield{author}{\bibinfo{person}{Kean Kouakoua}, \bibinfo{person}{Cyril
  Duclos}, \bibinfo{person}{Rachid Aissaoui}, \bibinfo{person}{Sylvie Nadeau},
  {and} \bibinfo{person}{David~R Labbe}.} \bibinfo{year}{2020}\natexlab{}.
\newblock \showarticletitle{Rhythmic proprioceptive stimulation improves
  embodiment in a walking avatar when added to visual stimulation}. In
  \bibinfo{booktitle}{\emph{2020 IEEE Conference on Virtual Reality and 3D User
  Interfaces Abstracts and Workshops (VRW)}}. IEEE, \bibinfo{pages}{573--574}.
\newblock


\bibitem[\protect\citeauthoryear{Kruijff, Marquardt, Trepkowski, Lindeman,
  Hinkenjann, Maiero, and Riecke}{Kruijff et~al\mbox{.}}{2016}]%
        {kruijff2016your}
\bibfield{author}{\bibinfo{person}{E. Kruijff}, \bibinfo{person}{A. Marquardt},
  \bibinfo{person}{C. Trepkowski}, \bibinfo{person}{R. Lindeman},
  \bibinfo{person}{A. Hinkenjann}, \bibinfo{person}{J. Maiero}, {and}
  \bibinfo{person}{B. Riecke}.} \bibinfo{year}{2016}\natexlab{}.
\newblock \showarticletitle{On your feet! Enhancing vection in leaning-based
  interfaces through multisensory stimuli}. In \bibinfo{booktitle}{\emph{Proc.
  Symp. on Spatial User Interact.}}
\newblock


\bibitem[\protect\citeauthoryear{Kruijff, Marquardt, Trepkowski, Schild, and
  Hinkenjann}{Kruijff et~al\mbox{.}}{2015}]%
        {kruijff2015enhancing}
\bibfield{author}{\bibinfo{person}{E. Kruijff}, \bibinfo{person}{A. Marquardt},
  \bibinfo{person}{C. Trepkowski}, \bibinfo{person}{J. Schild}, {and}
  \bibinfo{person}{A. Hinkenjann}.} \bibinfo{year}{2015}\natexlab{}.
\newblock \showarticletitle{Enhancing user engagement in immersive games
  through multisensory cues}. In \bibinfo{booktitle}{\emph{International
  Conference on Games and Virtual Worlds for Serious Applications}}.
\newblock


\bibitem[\protect\citeauthoryear{Kyt{\"o}, Kusumoto, and Oittinen}{Kyt{\"o}
  et~al\mbox{.}}{2015}]%
        {kyto2015ventriloquist}
\bibfield{author}{\bibinfo{person}{Mikko Kyt{\"o}}, \bibinfo{person}{Kenta
  Kusumoto}, {and} \bibinfo{person}{Pirkko Oittinen}.}
  \bibinfo{year}{2015}\natexlab{}.
\newblock \showarticletitle{The ventriloquist effect in augmented reality}. In
  \bibinfo{booktitle}{\emph{2015 IEEE International Symposium on Mixed and
  Augmented Reality}}. IEEE, \bibinfo{pages}{49--53}.
\newblock


\bibitem[\protect\citeauthoryear{Lalanne and Lorenceau}{Lalanne and
  Lorenceau}{2004}]%
        {lalanne2004crossmodal}
\bibfield{author}{\bibinfo{person}{Christophe Lalanne} {and}
  \bibinfo{person}{Jean Lorenceau}.} \bibinfo{year}{2004}\natexlab{}.
\newblock \showarticletitle{Crossmodal integration for perception and action}.
\newblock \bibinfo{journal}{\emph{Journal of Physiology-Paris}}
  \bibinfo{volume}{98}, \bibinfo{number}{1-3} (\bibinfo{year}{2004}),
  \bibinfo{pages}{265--279}.
\newblock


\bibitem[\protect\citeauthoryear{Langbehn, Steinicke, Lappe, Welch, and
  Bruder}{Langbehn et~al\mbox{.}}{2018}]%
        {langbehn2018}
\bibfield{author}{\bibinfo{person}{E. Langbehn}, \bibinfo{person}{F.
  Steinicke}, \bibinfo{person}{M. Lappe}, \bibinfo{person}{G. Welch}, {and}
  \bibinfo{person}{G. Bruder}.} \bibinfo{year}{2018}\natexlab{}.
\newblock \showarticletitle{In the blink of an eye: leveraging blink-induced
  suppression for imperceptible position and orientation redirection in virtual
  reality}.
\newblock \bibinfo{journal}{\emph{ACM Trans. on Graph.}}
  (\bibinfo{year}{2018}).
\newblock


\bibitem[\protect\citeauthoryear{Lathan, Tracey, Sebrechts, Clawson, and
  Higgins}{Lathan et~al\mbox{.}}{2002}]%
        {lathan2002using}
\bibfield{author}{\bibinfo{person}{C. Lathan}, \bibinfo{person}{M. Tracey},
  \bibinfo{person}{M. Sebrechts}, \bibinfo{person}{D. Clawson}, {and}
  \bibinfo{person}{G. Higgins}.} \bibinfo{year}{2002}\natexlab{}.
\newblock \showarticletitle{Using virtual environments as training simulators:
  Measuring transfer}.
\newblock \bibinfo{journal}{\emph{Handbook of Virtual Environments: Design,
  Implementation, and Applications}} (\bibinfo{year}{2002}).
\newblock


\bibitem[\protect\citeauthoryear{Laurienti, Burdette, Wallace, Yen, Field, and
  Stein}{Laurienti et~al\mbox{.}}{2002}]%
        {laurienti2002}
\bibfield{author}{\bibinfo{person}{Paul~J Laurienti},
  \bibinfo{person}{Jonathan~H Burdette}, \bibinfo{person}{Mark~T Wallace},
  \bibinfo{person}{Yi-Fen Yen}, \bibinfo{person}{Aaron~S Field}, {and}
  \bibinfo{person}{Barry~E Stein}.} \bibinfo{year}{2002}\natexlab{}.
\newblock \showarticletitle{Deactivation of sensory-specific cortex by
  cross-modal stimuli}.
\newblock \bibinfo{journal}{\emph{Journal of cognitive neuroscience}}
  \bibinfo{volume}{14}, \bibinfo{number}{3} (\bibinfo{year}{2002}).
\newblock


\bibitem[\protect\citeauthoryear{LaViola~Jr}{LaViola~Jr}{2000}]%
        {laviola2000discussion}
\bibfield{author}{\bibinfo{person}{Joseph~J LaViola~Jr}.}
  \bibinfo{year}{2000}\natexlab{}.
\newblock \showarticletitle{A discussion of cybersickness in virtual
  environments}.
\newblock \bibinfo{journal}{\emph{ACM Sigchi Bulletin}} \bibinfo{volume}{32},
  \bibinfo{number}{1} (\bibinfo{year}{2000}).
\newblock


\bibitem[\protect\citeauthoryear{Lawson, Roper, and Abdullah}{Lawson
  et~al\mbox{.}}{2016}]%
        {glyn2016}
\bibfield{author}{\bibinfo{person}{G. Lawson}, \bibinfo{person}{T. Roper},
  {and} \bibinfo{person}{C. Abdullah}.} \bibinfo{year}{2016}\natexlab{}.
\newblock \showarticletitle{Multimodal “Sensory Illusions” for Improving
  Spatial Awareness in Virtual Environments}. In
  \bibinfo{booktitle}{\emph{Proc. of the European Conference on Cognitive
  Ergonomics}}.
\newblock


\bibitem[\protect\citeauthoryear{Laycock and Day}{Laycock and Day}{2007}]%
        {laycock2007survey}
\bibfield{author}{\bibinfo{person}{Stephen~D Laycock} {and} \bibinfo{person}{AM
  Day}.} \bibinfo{year}{2007}\natexlab{}.
\newblock \showarticletitle{A survey of haptic rendering techniques}. In
  \bibinfo{booktitle}{\emph{Computer Graphics Forum}},
  Vol.~\bibinfo{volume}{26}. Wiley Online Library, \bibinfo{pages}{50--65}.
\newblock


\bibitem[\protect\citeauthoryear{Li, Yu, Shi, Shi, Tian, Yang, Wang, and
  Jiang}{Li et~al\mbox{.}}{2017}]%
        {li2017application}
\bibfield{author}{\bibinfo{person}{Lan Li}, \bibinfo{person}{Fei Yu},
  \bibinfo{person}{Dongquan Shi}, \bibinfo{person}{Jianping Shi},
  \bibinfo{person}{Zongjun Tian}, \bibinfo{person}{Jiquan Yang},
  \bibinfo{person}{Xingsong Wang}, {and} \bibinfo{person}{Qing Jiang}.}
  \bibinfo{year}{2017}\natexlab{}.
\newblock \showarticletitle{Application of virtual reality technology in
  clinical medicine}.
\newblock \bibinfo{journal}{\emph{American journal of translational research}}
  \bibinfo{volume}{9}, \bibinfo{number}{9} (\bibinfo{year}{2017}).
\newblock


\bibitem[\protect\citeauthoryear{Liao, Xie, Li, Li, Su, Jiang, Huang, and
  Shen}{Liao et~al\mbox{.}}{2020}]%
        {liao2020data}
\bibfield{author}{\bibinfo{person}{Haodong Liao}, \bibinfo{person}{Ning Xie},
  \bibinfo{person}{Huiyuan Li}, \bibinfo{person}{Yuhang Li},
  \bibinfo{person}{Jianping Su}, \bibinfo{person}{Feng Jiang},
  \bibinfo{person}{Weipeng Huang}, {and} \bibinfo{person}{Heng~Tao Shen}.}
  \bibinfo{year}{2020}\natexlab{}.
\newblock \showarticletitle{Data-Driven Spatio-Temporal Analysis via
  Multi-Modal Zeitgebers and Cognitive Load in VR}. In
  \bibinfo{booktitle}{\emph{2020 IEEE Conference on Virtual Reality and 3D User
  Interfaces (VR)}}. IEEE, \bibinfo{pages}{473--482}.
\newblock


\bibitem[\protect\citeauthoryear{Lin, Chang, Hu, Cheng, Huang, and Sun}{Lin
  et~al\mbox{.}}{2017a}]%
        {lin2017tell}
\bibfield{author}{\bibinfo{person}{Yen-Chen Lin}, \bibinfo{person}{Yung-Ju
  Chang}, \bibinfo{person}{Hou-Ning Hu}, \bibinfo{person}{Hsien-Tzu Cheng},
  \bibinfo{person}{Chi-Wen Huang}, {and} \bibinfo{person}{Min Sun}.}
  \bibinfo{year}{2017}\natexlab{a}.
\newblock \showarticletitle{Tell me where to look: Investigating ways for
  assisting focus in 360 video}. In \bibinfo{booktitle}{\emph{Proceedings of
  the 2017 CHI Conference on Human Factors in Computing Systems}}.
  \bibinfo{pages}{2535--2545}.
\newblock


\bibitem[\protect\citeauthoryear{Lin, Liao, Teng, Chung, Chan, and Chen}{Lin
  et~al\mbox{.}}{2017b}]%
        {lin2017outside}
\bibfield{author}{\bibinfo{person}{Yung-Ta Lin}, \bibinfo{person}{Yi-Chi Liao},
  \bibinfo{person}{Shan-Yuan Teng}, \bibinfo{person}{Yi-Ju Chung},
  \bibinfo{person}{Liwei Chan}, {and} \bibinfo{person}{Bing-Yu Chen}.}
  \bibinfo{year}{2017}\natexlab{b}.
\newblock \showarticletitle{Outside-in: visualizing out-of-sight
  regions-of-interest in a 360 video using spatial picture-in-picture
  previews}. In \bibinfo{booktitle}{\emph{Proceedings of the 30th Annual ACM
  Symposium on User Interface Software and Technology}}.
  \bibinfo{pages}{255--265}.
\newblock


\bibitem[\protect\citeauthoryear{Long, Seah, Carter, and Subramanian}{Long
  et~al\mbox{.}}{2014}]%
        {long2014rendering}
\bibfield{author}{\bibinfo{person}{Benjamin Long}, \bibinfo{person}{Sue~Ann
  Seah}, \bibinfo{person}{Tom Carter}, {and} \bibinfo{person}{Sriram
  Subramanian}.} \bibinfo{year}{2014}\natexlab{}.
\newblock \showarticletitle{Rendering volumetric haptic shapes in mid-air using
  ultrasound}.
\newblock \bibinfo{journal}{\emph{ACM Trans. on Graphics (TOG)}}
  \bibinfo{volume}{33}, \bibinfo{number}{6} (\bibinfo{year}{2014}),
  \bibinfo{pages}{1--10}.
\newblock


\bibitem[\protect\citeauthoryear{Lopes, You, Ion, and Baudisch}{Lopes
  et~al\mbox{.}}{2018}]%
        {lopes2018adding}
\bibfield{author}{\bibinfo{person}{P. Lopes}, \bibinfo{person}{S. You},
  \bibinfo{person}{A. Ion}, {and} \bibinfo{person}{P. Baudisch}.}
  \bibinfo{year}{2018}\natexlab{}.
\newblock \showarticletitle{Adding force feedback to mixed reality experiences
  and games using electrical muscle stimulation}. In
  \bibinfo{booktitle}{\emph{Proc. of the Conference on Human Factors in
  Computing Systems}}. \bibinfo{pages}{1--13}.
\newblock


\bibitem[\protect\citeauthoryear{Lovelace, Stein, and Wallace}{Lovelace
  et~al\mbox{.}}{2003}]%
        {lovelace2003irrelevant}
\bibfield{author}{\bibinfo{person}{C. Lovelace}, \bibinfo{person}{B. Stein},
  {and} \bibinfo{person}{M. Wallace}.} \bibinfo{year}{2003}\natexlab{}.
\newblock \showarticletitle{An irrelevant light enhances auditory detection in
  humans: a psychophysical analysis of multisensory integration in stimulus
  detection}.
\newblock \bibinfo{journal}{\emph{Cognitive brain research}}
  \bibinfo{volume}{17}, \bibinfo{number}{2} (\bibinfo{year}{2003}),
  \bibinfo{pages}{447--453}.
\newblock


\bibitem[\protect\citeauthoryear{Lu, Li, and Sun}{Lu et~al\mbox{.}}{2010}]%
        {lu2010multimodal}
\bibfield{author}{\bibinfo{person}{Jianfeng Lu}, \bibinfo{person}{Li Li}, {and}
  \bibinfo{person}{Goh~Poh Sun}.} \bibinfo{year}{2010}\natexlab{}.
\newblock \showarticletitle{A multimodal virtual anatomy e-learning tool for
  medical education}. In \bibinfo{booktitle}{\emph{International Conference on
  Technologies for E-Learning and Digital Entertainment}}. Springer,
  \bibinfo{pages}{278--287}.
\newblock


\bibitem[\protect\citeauthoryear{MacDonald, Balakrishnan, Orosz, and
  Karplus}{MacDonald et~al\mbox{.}}{2002}]%
        {macdonald2002intelligibility}
\bibfield{author}{\bibinfo{person}{Justin~A MacDonald}, \bibinfo{person}{JD
  Balakrishnan}, \bibinfo{person}{Michael~D Orosz}, {and}
  \bibinfo{person}{Walter~J Karplus}.} \bibinfo{year}{2002}\natexlab{}.
\newblock \showarticletitle{Intelligibility of speech in a virtual 3-D
  environment}.
\newblock \bibinfo{journal}{\emph{Human Factors}} \bibinfo{volume}{44},
  \bibinfo{number}{2} (\bibinfo{year}{2002}), \bibinfo{pages}{272--286}.
\newblock


\bibitem[\protect\citeauthoryear{Maculewicz, Nilsson, and Serafin}{Maculewicz
  et~al\mbox{.}}{2016}]%
        {maculewicz2016investigation}
\bibfield{author}{\bibinfo{person}{Justyna Maculewicz},
  \bibinfo{person}{Niels~Christian Nilsson}, {and} \bibinfo{person}{Stefania
  Serafin}.} \bibinfo{year}{2016}\natexlab{}.
\newblock \showarticletitle{An investigation of the effect of immersive visual
  and auditory feedback on rhythmic walking interaction}.
\newblock In \bibinfo{booktitle}{\emph{Proceedings of the Audio Mostly 2016}}.
  \bibinfo{pages}{194--201}.
\newblock


\bibitem[\protect\citeauthoryear{Magalh{\~a}es, Jacob, Nilsson, Nordahl, and
  Bernardes}{Magalh{\~a}es et~al\mbox{.}}{2020}]%
        {magalhaes2020physics}
\bibfield{author}{\bibinfo{person}{Eduardo Magalh{\~a}es},
  \bibinfo{person}{Jo{\~a}o Jacob}, \bibinfo{person}{Niels Nilsson},
  \bibinfo{person}{Rolf Nordahl}, {and} \bibinfo{person}{Gilberto Bernardes}.}
  \bibinfo{year}{2020}\natexlab{}.
\newblock \showarticletitle{Physics-based Concatenative Sound Synthesis of
  Photogrammetric models for Aural and Haptic Feedback in Virtual
  Environments}. In \bibinfo{booktitle}{\emph{IEEE Conference on Virtual
  Reality and 3D User Interfaces Abstracts and Workshops (VRW)}}.
  \bibinfo{pages}{376--379}.
\newblock


\bibitem[\protect\citeauthoryear{Maggioni, Cobden, Dmitrenko, and
  Obrist}{Maggioni et~al\mbox{.}}{2018}]%
        {maggioni2018smell}
\bibfield{author}{\bibinfo{person}{Emanuela Maggioni}, \bibinfo{person}{Robert
  Cobden}, \bibinfo{person}{Dmitrijs Dmitrenko}, {and}
  \bibinfo{person}{Marianna Obrist}.} \bibinfo{year}{2018}\natexlab{}.
\newblock \showarticletitle{Smell-O-Message: integration of olfactory
  notifications into a messaging application to improve users' performance}. In
  \bibinfo{booktitle}{\emph{Proceedings of the 20th ACM International
  Conference on Multimodal Interaction}}. \bibinfo{pages}{45--54}.
\newblock


\bibitem[\protect\citeauthoryear{Malpica, Serrano, Allue, Bedia, and
  Masia}{Malpica et~al\mbox{.}}{2020a}]%
        {malpica2020crossmodal}
\bibfield{author}{\bibinfo{person}{S Malpica}, \bibinfo{person}{A Serrano},
  \bibinfo{person}{M Allue}, \bibinfo{person}{MG Bedia}, {and}
  \bibinfo{person}{B Masia}.} \bibinfo{year}{2020}\natexlab{a}.
\newblock \showarticletitle{Crossmodal perception in virtual reality}.
\newblock \bibinfo{journal}{\emph{Multimedia Tools and Applications}}
  \bibinfo{volume}{79}, \bibinfo{number}{5} (\bibinfo{year}{2020}),
  \bibinfo{pages}{3311--3331}.
\newblock


\bibitem[\protect\citeauthoryear{Malpica, Serrano, Gutierrez, and
  Masia}{Malpica et~al\mbox{.}}{2020b}]%
        {malpica2020}
\bibfield{author}{\bibinfo{person}{Sandra Malpica}, \bibinfo{person}{Ana
  Serrano}, \bibinfo{person}{Diego Gutierrez}, {and} \bibinfo{person}{Belen
  Masia}.} \bibinfo{year}{2020}\natexlab{b}.
\newblock \showarticletitle{Auditory stimuli degrade visual performance in
  virtual reality}.
\newblock \bibinfo{journal}{\emph{Scientific Reports}}  \bibinfo{volume}{10}
  (\bibinfo{year}{2020}).
\newblock
\showISSN{2045-2322}


\bibitem[\protect\citeauthoryear{Mara{\~n}es, Gutierrez, and
  Serrano}{Mara{\~n}es et~al\mbox{.}}{2020}]%
        {maranes2020exploring}
\bibfield{author}{\bibinfo{person}{Carlos Mara{\~n}es}, \bibinfo{person}{Diego
  Gutierrez}, {and} \bibinfo{person}{Ana Serrano}.}
  \bibinfo{year}{2020}\natexlab{}.
\newblock \showarticletitle{Exploring the impact of 360° movie cuts in
  users’ attention}. In \bibinfo{booktitle}{\emph{2020 IEEE Conference on
  Virtual Reality and 3D User Interfaces (VR)}}. IEEE, \bibinfo{pages}{73--82}.
\newblock


\bibitem[\protect\citeauthoryear{Marbury}{Marbury}{2021}]%
        {FutureMedicine}
\bibfield{author}{\bibinfo{person}{Donna Marbury}.}
  \bibinfo{year}{2021}\natexlab{}.
\newblock \bibinfo{booktitle}{\emph{What Does the Future Hold for AR and VR in
  Healthcare?}}
\newblock
\urldef\tempurl%
\url{https://healthtechmagazine.net/article/2020/11/what-does-future-hold-ar-and-vr-healthcare}
\showURL{%
\tempurl}
\newblock
\shownote{Last accessed on 2021-11-02.}


\bibitem[\protect\citeauthoryear{Marchal, Gallagher, L{\'e}cuyer, and
  Pacchierotti}{Marchal et~al\mbox{.}}{2020}]%
        {marchal2020can}
\bibfield{author}{\bibinfo{person}{Maud Marchal}, \bibinfo{person}{Gerard
  Gallagher}, \bibinfo{person}{Anatole L{\'e}cuyer}, {and}
  \bibinfo{person}{Claudio Pacchierotti}.} \bibinfo{year}{2020}\natexlab{}.
\newblock \showarticletitle{Can Stiffness Sensations be Rendered in Virtual
  Reality Using Mid-air Ultrasound Haptic Technologies?}. In
  \bibinfo{booktitle}{\emph{International Conference on Human Haptic Sensing
  and Touch Enabled Computer Applications}}. Springer,
  \bibinfo{pages}{297--306}.
\newblock


\bibitem[\protect\citeauthoryear{Martin, Serrano, Bergman, Wetzstein, and
  Masia}{Martin et~al\mbox{.}}{2022}]%
        {martin2021scangan360}
\bibfield{author}{\bibinfo{person}{Daniel Martin}, \bibinfo{person}{Ana
  Serrano}, \bibinfo{person}{Alexander~W. Bergman}, \bibinfo{person}{Gordon
  Wetzstein}, {and} \bibinfo{person}{Belen Masia}.}
  \bibinfo{year}{2022}\natexlab{}.
\newblock \showarticletitle{ScanGAN360: A Generative Model of Realistic
  Scanpaths for 360$^{\circ}$ Images}.
\newblock \bibinfo{journal}{\emph{IEEE Trans. on Vis. and Computer Graphics}}
  (\bibinfo{year}{2022}).
\newblock


\bibitem[\protect\citeauthoryear{Martin, Serrano, and Masia}{Martin
  et~al\mbox{.}}{2020}]%
        {martin20saliency}
\bibfield{author}{\bibinfo{person}{Daniel Martin}, \bibinfo{person}{Ana
  Serrano}, {and} \bibinfo{person}{Belen Masia}.}
  \bibinfo{year}{2020}\natexlab{}.
\newblock \showarticletitle{Panoramic convolutions for $360^{\circ}$
  single-image saliency prediction}. In \bibinfo{booktitle}{\emph{CVPR Workshop
  on Computer Vision for Augmented and Virtual Reality}}.
\newblock


\bibitem[\protect\citeauthoryear{Mart{\'\i}nez, Garc{\'\i}a, Oliver, Molina,
  and Gonz{\'a}lez}{Mart{\'\i}nez et~al\mbox{.}}{2014}]%
        {martinez2014vitaki}
\bibfield{author}{\bibinfo{person}{J. Mart{\'\i}nez}, \bibinfo{person}{A.
  Garc{\'\i}a}, \bibinfo{person}{M. Oliver}, \bibinfo{person}{J. Molina}, {and}
  \bibinfo{person}{P. Gonz{\'a}lez}.} \bibinfo{year}{2014}\natexlab{}.
\newblock \showarticletitle{Vitaki: a vibrotactile prototyping toolkit for
  virtual reality and video games}.
\newblock \bibinfo{journal}{\emph{International Journal of Human-Computer
  Interaction}} \bibinfo{volume}{30}, \bibinfo{number}{11}
  (\bibinfo{year}{2014}).
\newblock


\bibitem[\protect\citeauthoryear{Maselli and Slater}{Maselli and
  Slater}{2013}]%
        {maselli2013building}
\bibfield{author}{\bibinfo{person}{Antonella Maselli} {and}
  \bibinfo{person}{Mel Slater}.} \bibinfo{year}{2013}\natexlab{}.
\newblock \showarticletitle{The building blocks of the full body ownership
  illusion}.
\newblock \bibinfo{journal}{\emph{Frontiers in human neuroscience}}
  \bibinfo{volume}{7} (\bibinfo{year}{2013}), \bibinfo{pages}{83}.
\newblock


\bibitem[\protect\citeauthoryear{{Masia}, {Camon}, {Gutierrez}, and
  {Serrano}}{{Masia} et~al\mbox{.}}{2021}]%
        {masiaSound2021}
\bibfield{author}{\bibinfo{person}{B. {Masia}}, \bibinfo{person}{J. {Camon}},
  \bibinfo{person}{D. {Gutierrez}}, {and} \bibinfo{person}{A. {Serrano}}.}
  \bibinfo{year}{2021}\natexlab{}.
\newblock \showarticletitle{Influence of directional sound cues on users
  exploration across 360 movie cuts}.
\newblock \bibinfo{journal}{\emph{IEEE Computer Graphics and Applications}}
  (\bibinfo{year}{2021}), \bibinfo{pages}{1--1}.
\newblock


\bibitem[\protect\citeauthoryear{Mast and Oman}{Mast and Oman}{2004}]%
        {mast2004top}
\bibfield{author}{\bibinfo{person}{Fred~W Mast} {and}
  \bibinfo{person}{Charles~M Oman}.} \bibinfo{year}{2004}\natexlab{}.
\newblock \showarticletitle{Top-down processing and visual reorientation
  illusions in a virtual reality environment.}
\newblock \bibinfo{journal}{\emph{Swiss Journal of Psychology/Schweizerische
  Zeitschrift f{\"u}r Psychologie}} \bibinfo{volume}{63}, \bibinfo{number}{3}
  (\bibinfo{year}{2004}), \bibinfo{pages}{143}.
\newblock


\bibitem[\protect\citeauthoryear{Matsuda, Nakamura, Amemiya, Ikei, and
  Kitazaki}{Matsuda et~al\mbox{.}}{2020}]%
        {matsuda2020perception}
\bibfield{author}{\bibinfo{person}{Y. Matsuda}, \bibinfo{person}{J. Nakamura},
  \bibinfo{person}{T. Amemiya}, \bibinfo{person}{Y. Ikei}, {and}
  \bibinfo{person}{M. Kitazaki}.} \bibinfo{year}{2020}\natexlab{}.
\newblock \showarticletitle{Perception of Walking Self-body Avatar Enhances
  Virtual-walking Sensation}. In \bibinfo{booktitle}{\emph{IEEE Conf. on
  Virtual Reality and 3D User Interfaces Abstracts and Workshops}}.
\newblock


\bibitem[\protect\citeauthoryear{Matsumoto, Langbehn, Narumi, and
  Steinicke}{Matsumoto et~al\mbox{.}}{2020}]%
        {matsumoto2020detection}
\bibfield{author}{\bibinfo{person}{K. Matsumoto}, \bibinfo{person}{E.
  Langbehn}, \bibinfo{person}{T. Narumi}, {and} \bibinfo{person}{F.
  Steinicke}.} \bibinfo{year}{2020}\natexlab{}.
\newblock \showarticletitle{Detection Thresholds for Vertical Gains in VR and
  Drone-based Telepresence Systems}. In \bibinfo{booktitle}{\emph{IEEE
  Conference on Virtual Reality and 3D User Interfaces (VR)}}.
  \bibinfo{pages}{101--107}.
\newblock


\bibitem[\protect\citeauthoryear{Matsumoto, Narumi, Ban, Yanase, Tanikawa, and
  Hirose}{Matsumoto et~al\mbox{.}}{2019}]%
        {matsumoto2019unlimited}
\bibfield{author}{\bibinfo{person}{K. Matsumoto}, \bibinfo{person}{T. Narumi},
  \bibinfo{person}{Y. Ban}, \bibinfo{person}{Y. Yanase}, \bibinfo{person}{T.
  Tanikawa}, {and} \bibinfo{person}{M. Hirose}.}
  \bibinfo{year}{2019}\natexlab{}.
\newblock \showarticletitle{Unlimited Corridor: A Visuo-haptic Redirection
  System}. In \bibinfo{booktitle}{\emph{International Conference on
  Virtual-Reality Continuum and its Applications in Industry}}.
  \bibinfo{pages}{1--9}.
\newblock


\bibitem[\protect\citeauthoryear{McGill, Ng, and Brewster}{McGill
  et~al\mbox{.}}{2017}]%
        {mcgill2017passenger}
\bibfield{author}{\bibinfo{person}{Mark McGill}, \bibinfo{person}{Alexander
  Ng}, {and} \bibinfo{person}{Stephen Brewster}.}
  \bibinfo{year}{2017}\natexlab{}.
\newblock \showarticletitle{I am the passenger: how visual motion cues can
  influence sickness for in-car VR}. In \bibinfo{booktitle}{\emph{Proceedings
  of the Conference on Human Factors in Computing Systems}}.
  \bibinfo{pages}{5655--5668}.
\newblock


\bibitem[\protect\citeauthoryear{Melo, Gon{\c{c}}alves, Monteiro, Coelho,
  Vasconcelos-Raposo, and Bessa}{Melo et~al\mbox{.}}{2020}]%
        {melo2020multisensory}
\bibfield{author}{\bibinfo{person}{M. Melo}, \bibinfo{person}{G.
  Gon{\c{c}}alves}, \bibinfo{person}{P. Monteiro}, \bibinfo{person}{H. Coelho},
  \bibinfo{person}{J. Vasconcelos-Raposo}, {and} \bibinfo{person}{M. Bessa}.}
  \bibinfo{year}{2020}\natexlab{}.
\newblock \showarticletitle{Do Multisensory stimuli benefit the virtual reality
  experience? A systematic review}.
\newblock \bibinfo{journal}{\emph{IEEE Trans. on Vis. and Computer Graphics}}
  (\bibinfo{year}{2020}).
\newblock


\bibitem[\protect\citeauthoryear{Meyer, Shao, White, Hopkins, and
  Robotham}{Meyer et~al\mbox{.}}{2013}]%
        {meyer2013modulation}
\bibfield{author}{\bibinfo{person}{Georg~F Meyer}, \bibinfo{person}{Fei Shao},
  \bibinfo{person}{Mark~D White}, \bibinfo{person}{Carl Hopkins}, {and}
  \bibinfo{person}{Antony~J Robotham}.} \bibinfo{year}{2013}\natexlab{}.
\newblock \showarticletitle{Modulation of visually evoked postural responses by
  contextual visual, haptic and auditory information: a ‘virtual reality
  check’}.
\newblock \bibinfo{journal}{\emph{PloS one}} \bibinfo{volume}{8},
  \bibinfo{number}{6} (\bibinfo{year}{2013}).
\newblock


\bibitem[\protect\citeauthoryear{Min, Zhai, Zhou, Zhang, Yang, and Guan}{Min
  et~al\mbox{.}}{2020}]%
        {min2020multimodal}
\bibfield{author}{\bibinfo{person}{X. Min}, \bibinfo{person}{G. Zhai},
  \bibinfo{person}{J. Zhou}, \bibinfo{person}{X. Zhang}, \bibinfo{person}{X.
  Yang}, {and} \bibinfo{person}{X. Guan}.} \bibinfo{year}{2020}\natexlab{}.
\newblock \showarticletitle{A Multimodal Saliency Model for Videos With High
  Audio-Visual Correspondence}.
\newblock \bibinfo{journal}{\emph{IEEE Trans. on Image Processing}}
  \bibinfo{volume}{29} (\bibinfo{year}{2020}), \bibinfo{pages}{3805--3819}.
\newblock


\bibitem[\protect\citeauthoryear{Miranda}{Miranda}{2012}]%
        {miranda2012taste}
\bibfield{author}{\bibinfo{person}{Mar{\'\i}a~Isabel Miranda}.}
  \bibinfo{year}{2012}\natexlab{}.
\newblock \showarticletitle{Taste and odor recognition memory: the emotional
  flavor of life}.
\newblock \bibinfo{journal}{\emph{Reviews in the Neurosciences}}
  \bibinfo{volume}{23}, \bibinfo{number}{5-6} (\bibinfo{year}{2012}),
  \bibinfo{pages}{481--499}.
\newblock


\bibitem[\protect\citeauthoryear{Morgado, Nvasconcelos, Langlois, and
  Wang}{Morgado et~al\mbox{.}}{2018}]%
        {morgado2018self}
\bibfield{author}{\bibinfo{person}{Pedro Morgado}, \bibinfo{person}{Nuno
  Nvasconcelos}, \bibinfo{person}{Timothy Langlois}, {and}
  \bibinfo{person}{Oliver Wang}.} \bibinfo{year}{2018}\natexlab{}.
\newblock \showarticletitle{Self-supervised generation of spatial audio for 360
  video}. In \bibinfo{booktitle}{\emph{Advances in Neural Information
  Processing Systems}}. \bibinfo{pages}{362--372}.
\newblock


\bibitem[\protect\citeauthoryear{Mueller, Kari, Khot, Li, Wang, Mehta, and
  Arnold}{Mueller et~al\mbox{.}}{2018}]%
        {mueller2018towards}
\bibfield{author}{\bibinfo{person}{F. Mueller}, \bibinfo{person}{T. Kari},
  \bibinfo{person}{R. Khot}, \bibinfo{person}{Z. Li}, \bibinfo{person}{Y.
  Wang}, \bibinfo{person}{Y. Mehta}, {and} \bibinfo{person}{P. Arnold}.}
  \bibinfo{year}{2018}\natexlab{}.
\newblock \showarticletitle{Towards experiencing eating as a form of play}. In
  \bibinfo{booktitle}{\emph{Proc. of the Symposium on Computer-Human
  Interaction in Play Companion Extended Abstracts}}.
\newblock


\bibitem[\protect\citeauthoryear{M{\"u}hlberger, Weik, Pauli, and
  Wiedemann}{M{\"u}hlberger et~al\mbox{.}}{2006}]%
        {muhlberger2006one}
\bibfield{author}{\bibinfo{person}{A. M{\"u}hlberger}, \bibinfo{person}{A.
  Weik}, \bibinfo{person}{P. Pauli}, {and} \bibinfo{person}{G. Wiedemann}.}
  \bibinfo{year}{2006}\natexlab{}.
\newblock \showarticletitle{One-session virtual reality exposure treatment for
  fear of flying: 1-year follow-up and graduation flight accompaniment
  effects}.
\newblock \bibinfo{journal}{\emph{Psychotherapy Research}}
  \bibinfo{volume}{16}, \bibinfo{number}{1} (\bibinfo{year}{2006}).
\newblock


\bibitem[\protect\citeauthoryear{Muhlberger, Wiedemann, and Pauli}{Muhlberger
  et~al\mbox{.}}{2003}]%
        {muhlberger2003efficacy}
\bibfield{author}{\bibinfo{person}{A Muhlberger}, \bibinfo{person}{Georg
  Wiedemann}, {and} \bibinfo{person}{Paul Pauli}.}
  \bibinfo{year}{2003}\natexlab{}.
\newblock \showarticletitle{Efficacy of a one-session virtual reality exposure
  treatment for fear of flying}.
\newblock \bibinfo{journal}{\emph{Psychotherapy Research}}
  \bibinfo{volume}{13}, \bibinfo{number}{3} (\bibinfo{year}{2003}),
  \bibinfo{pages}{323--336}.
\newblock


\bibitem[\protect\citeauthoryear{Nakajima, Sugimoto, and Kawamoto}{Nakajima
  et~al\mbox{.}}{2013}]%
        {nakajima2013incorporating}
\bibfield{author}{\bibinfo{person}{Jiro Nakajima}, \bibinfo{person}{Akihiro
  Sugimoto}, {and} \bibinfo{person}{Kazuhiko Kawamoto}.}
  \bibinfo{year}{2013}\natexlab{}.
\newblock \showarticletitle{Incorporating audio signals into constructing a
  visual saliency map}. In \bibinfo{booktitle}{\emph{Pacific-Rim Symposium on
  Image and Video Technology}}. Springer, \bibinfo{pages}{468--480}.
\newblock


\bibitem[\protect\citeauthoryear{Nesbitt and Hoskens}{Nesbitt and
  Hoskens}{2008}]%
        {nesbitt2008multi}
\bibfield{author}{\bibinfo{person}{Keith~V Nesbitt} {and} \bibinfo{person}{Ian
  Hoskens}.} \bibinfo{year}{2008}\natexlab{}.
\newblock \showarticletitle{Multi-sensory game interface improves player
  satisfaction but not performance}. In \bibinfo{booktitle}{\emph{Proceedings
  of the ninth conference on Australasian user interface-Volume 76}}.
  \bibinfo{pages}{13--18}.
\newblock


\bibitem[\protect\citeauthoryear{Neyret, Navarro, Beacco, Oliva, Bourdin,
  Valenzuela, Barberia, and Slater}{Neyret et~al\mbox{.}}{2020}]%
        {neyret2020embodied}
\bibfield{author}{\bibinfo{person}{Sol{\`e}ne Neyret}, \bibinfo{person}{Xavi
  Navarro}, \bibinfo{person}{Alejandro Beacco}, \bibinfo{person}{Ramon Oliva},
  \bibinfo{person}{Pierre Bourdin}, \bibinfo{person}{Jose Valenzuela},
  \bibinfo{person}{Itxaso Barberia}, {and} \bibinfo{person}{Mel Slater}.}
  \bibinfo{year}{2020}\natexlab{}.
\newblock \showarticletitle{An embodied perspective as a victim of sexual
  harassment in virtual reality reduces action conformity in a later milgram
  obedience scenario}.
\newblock \bibinfo{journal}{\emph{Scientific Reports}} \bibinfo{volume}{10},
  \bibinfo{number}{1} (\bibinfo{year}{2020}), \bibinfo{pages}{1--18}.
\newblock


\bibitem[\protect\citeauthoryear{Nidiffer, Diederich, Ramachandran, and
  Wallace}{Nidiffer et~al\mbox{.}}{2018}]%
        {nidiffer2018}
\bibfield{author}{\bibinfo{person}{Aaron~R Nidiffer}, \bibinfo{person}{Adele
  Diederich}, \bibinfo{person}{Ramnarayan Ramachandran}, {and}
  \bibinfo{person}{Mark~T Wallace}.} \bibinfo{year}{2018}\natexlab{}.
\newblock \showarticletitle{Multisensory perception reflects individual
  differences in processing temporal correlations}.
\newblock \bibinfo{journal}{\emph{Scientific Reports}} \bibinfo{volume}{8},
  \bibinfo{number}{1} (\bibinfo{year}{2018}), \bibinfo{pages}{1--15}.
\newblock


\bibitem[\protect\citeauthoryear{Nielsen, M{\o}ller, Hartmeyer, Ljung, Nilsson,
  Nordahl, and Serafin}{Nielsen et~al\mbox{.}}{2016}]%
        {nielsen2016missing}
\bibfield{author}{\bibinfo{person}{Lasse~T Nielsen}, \bibinfo{person}{Matias~B
  M{\o}ller}, \bibinfo{person}{Sune~D Hartmeyer}, \bibinfo{person}{Troels~CM
  Ljung}, \bibinfo{person}{Niels~C Nilsson}, \bibinfo{person}{Rolf Nordahl},
  {and} \bibinfo{person}{Stefania Serafin}.} \bibinfo{year}{2016}\natexlab{}.
\newblock \showarticletitle{Missing the point: an exploration of how to guide
  users' attention during cinematic virtual reality}. In
  \bibinfo{booktitle}{\emph{Proceedings of the 22nd ACM Conference on Virtual
  Reality Software and Technology}}. \bibinfo{pages}{229--232}.
\newblock


\bibitem[\protect\citeauthoryear{Nilsson, Nordahl, Sikstr{\"o}m, Turchet, and
  Serafin}{Nilsson et~al\mbox{.}}{2012}]%
        {nilsson2012haptically}
\bibfield{author}{\bibinfo{person}{Niels~C Nilsson}, \bibinfo{person}{Rolf
  Nordahl}, \bibinfo{person}{Erik Sikstr{\"o}m}, \bibinfo{person}{Luca
  Turchet}, {and} \bibinfo{person}{Stefania Serafin}.}
  \bibinfo{year}{2012}\natexlab{}.
\newblock \showarticletitle{Haptically induced illusory self-motion and the
  influence of context of motion}. In \bibinfo{booktitle}{\emph{International
  Conference on Human Haptic Sensing and Touch Enabled Computer Applications}}.
  Springer, \bibinfo{pages}{349--360}.
\newblock


\bibitem[\protect\citeauthoryear{Nilsson, Peck, Bruder, Hodgson, Serafin,
  Whitton, Steinicke, and Rosenberg}{Nilsson et~al\mbox{.}}{2018}]%
        {nilsson201815}
\bibfield{author}{\bibinfo{person}{Niels~Christian Nilsson},
  \bibinfo{person}{Tabitha Peck}, \bibinfo{person}{Gerd Bruder},
  \bibinfo{person}{Eri Hodgson}, \bibinfo{person}{Stefania Serafin},
  \bibinfo{person}{Mary Whitton}, \bibinfo{person}{Frank Steinicke}, {and}
  \bibinfo{person}{Evan~Suma Rosenberg}.} \bibinfo{year}{2018}\natexlab{}.
\newblock \showarticletitle{15 years of research on redirected walking in
  immersive virtual environments}.
\newblock \bibinfo{journal}{\emph{IEEE Computer Graphics and Applications}}
  \bibinfo{volume}{38}, \bibinfo{number}{2} (\bibinfo{year}{2018}),
  \bibinfo{pages}{44--56}.
\newblock


\bibitem[\protect\citeauthoryear{Nilsson, Suma, Nordahl, Bolas, and
  Serafin}{Nilsson et~al\mbox{.}}{2016}]%
        {nilsson2016estimation}
\bibfield{author}{\bibinfo{person}{Niels~Christian Nilsson},
  \bibinfo{person}{Evan Suma}, \bibinfo{person}{Rolf Nordahl},
  \bibinfo{person}{Mark Bolas}, {and} \bibinfo{person}{Stefania Serafin}.}
  \bibinfo{year}{2016}\natexlab{}.
\newblock \showarticletitle{Estimation of detection thresholds for audiovisual
  rotation gains}. In \bibinfo{booktitle}{\emph{2016 IEEE Virtual Reality
  (VR)}}. IEEE, \bibinfo{pages}{241--242}.
\newblock


\bibitem[\protect\citeauthoryear{Noesselt, Bergmann, Hake, Heinze, and
  Fendrich}{Noesselt et~al\mbox{.}}{2008}]%
        {noesselt2008sound}
\bibfield{author}{\bibinfo{person}{Toemme Noesselt}, \bibinfo{person}{Daniel
  Bergmann}, \bibinfo{person}{Maria Hake}, \bibinfo{person}{Hans-Jochen
  Heinze}, {and} \bibinfo{person}{Robert Fendrich}.}
  \bibinfo{year}{2008}\natexlab{}.
\newblock \showarticletitle{Sound increases the saliency of visual events}.
\newblock \bibinfo{journal}{\emph{Brain Research}}  \bibinfo{volume}{1220}
  (\bibinfo{year}{2008}), \bibinfo{pages}{157--163}.
\newblock


\bibitem[\protect\citeauthoryear{Nogalski and Fohl}{Nogalski and Fohl}{2016}]%
        {nogalski2016acoustic}
\bibfield{author}{\bibinfo{person}{M. Nogalski} {and} \bibinfo{person}{W.
  Fohl}.} \bibinfo{year}{2016}\natexlab{}.
\newblock \showarticletitle{Acoustic redirected walking with auditory cues by
  means of wave field synthesis}. In \bibinfo{booktitle}{\emph{IEEE Virtual
  Reality}}.
\newblock


\bibitem[\protect\citeauthoryear{Nogalski and Fohl}{Nogalski and Fohl}{2017}]%
        {nogalski2017curvature}
\bibfield{author}{\bibinfo{person}{Malte Nogalski} {and}
  \bibinfo{person}{Wolfgang Fohl}.} \bibinfo{year}{2017}\natexlab{}.
\newblock \showarticletitle{Curvature gains in redirected walking: A closer
  look}. In \bibinfo{booktitle}{\emph{2017 IEEE Virtual Reality (VR)}}. IEEE,
  \bibinfo{pages}{267--268}.
\newblock


\bibitem[\protect\citeauthoryear{Normand, Giannopoulos, Spanlang, and
  Slater}{Normand et~al\mbox{.}}{2011}]%
        {normand2011multisensory}
\bibfield{author}{\bibinfo{person}{Jean-Marie Normand}, \bibinfo{person}{Elias
  Giannopoulos}, \bibinfo{person}{Bernhard Spanlang}, {and}
  \bibinfo{person}{Mel Slater}.} \bibinfo{year}{2011}\natexlab{}.
\newblock \showarticletitle{Multisensory stimulation can induce an illusion of
  larger belly size in immersive virtual reality}.
\newblock \bibinfo{journal}{\emph{PloS one}} \bibinfo{volume}{6},
  \bibinfo{number}{1} (\bibinfo{year}{2011}).
\newblock


\bibitem[\protect\citeauthoryear{Paek}{Paek}{2012}]%
        {paek2012impact}
\bibfield{author}{\bibinfo{person}{Seungoh Paek}.}
  \bibinfo{year}{2012}\natexlab{}.
\newblock \emph{\bibinfo{title}{The impact of multimodal virtual manipulatives
  on young children's mathematics learning}}.
\newblock \bibinfo{thesistype}{Ph.D. Dissertation}. \bibinfo{school}{Teachers
  College, Columbia University}.
\newblock


\bibitem[\protect\citeauthoryear{Pavel, Hartmann, and Agrawala}{Pavel
  et~al\mbox{.}}{2017}]%
        {pavel2017shot}
\bibfield{author}{\bibinfo{person}{Amy Pavel}, \bibinfo{person}{Bj{\"o}rn
  Hartmann}, {and} \bibinfo{person}{Maneesh Agrawala}.}
  \bibinfo{year}{2017}\natexlab{}.
\newblock \showarticletitle{Shot orientation controls for interactive
  cinematography with 360 video}. In \bibinfo{booktitle}{\emph{Proceedings of
  the 30th Annual ACM Symposium on User Interface Software and Technology}}.
\newblock


\bibitem[\protect\citeauthoryear{Peatfield, Mueller, Ruhnau, and
  Weisz}{Peatfield et~al\mbox{.}}{2015}]%
        {peatfield2015}
\bibfield{author}{\bibinfo{person}{Nicholas Peatfield}, \bibinfo{person}{Nadia
  Mueller}, \bibinfo{person}{Phillipp Ruhnau}, {and} \bibinfo{person}{Nathan
  Weisz}.} \bibinfo{year}{2015}\natexlab{}.
\newblock \showarticletitle{Rubin-vase illusion perception is predicted by
  prestimulus activity and connectivity}.
\newblock \bibinfo{journal}{\emph{Journal of vision}} \bibinfo{volume}{15},
  \bibinfo{number}{12} (\bibinfo{year}{2015}), \bibinfo{pages}{429--429}.
\newblock


\bibitem[\protect\citeauthoryear{Peck, Fuchs, and Whitton}{Peck
  et~al\mbox{.}}{2009}]%
        {peck2009evaluation}
\bibfield{author}{\bibinfo{person}{Tabitha~C Peck}, \bibinfo{person}{Henry
  Fuchs}, {and} \bibinfo{person}{Mary~C Whitton}.}
  \bibinfo{year}{2009}\natexlab{}.
\newblock \showarticletitle{Evaluation of reorientation techniques and
  distractors for walking in large virtual environments}.
\newblock \bibinfo{journal}{\emph{IEEE Trans. on Visualization and Computer
  Graphics}} \bibinfo{volume}{15}, \bibinfo{number}{3} (\bibinfo{year}{2009}).
\newblock


\bibitem[\protect\citeauthoryear{Petkova and Ehrsson}{Petkova and
  Ehrsson}{2008}]%
        {petkova2008}
\bibfield{author}{\bibinfo{person}{V. Petkova} {and} \bibinfo{person}{H.
  Ehrsson}.} \bibinfo{year}{2008}\natexlab{}.
\newblock \showarticletitle{If I were you: perceptual illusion of body
  swapping}.
\newblock \bibinfo{journal}{\emph{PloS one}} \bibinfo{volume}{3},
  \bibinfo{number}{12} (\bibinfo{year}{2008}).
\newblock


\bibitem[\protect\citeauthoryear{Pezent, O'Malley, Israr, Samad, Robinson,
  Agarwal, Benko, and Colonnese}{Pezent et~al\mbox{.}}{2020}]%
        {pezent2020explorations}
\bibfield{author}{\bibinfo{person}{Evan Pezent}, \bibinfo{person}{Marcia~K
  O'Malley}, \bibinfo{person}{Ali Israr}, \bibinfo{person}{Majed Samad},
  \bibinfo{person}{Shea Robinson}, \bibinfo{person}{Priyanshu Agarwal},
  \bibinfo{person}{Hrvoje Benko}, {and} \bibinfo{person}{Nicholas Colonnese}.}
  \bibinfo{year}{2020}\natexlab{}.
\newblock \showarticletitle{Explorations of Wrist Haptic Feedback for AR/VR
  Interactions with Tasbi}. In \bibinfo{booktitle}{\emph{Extended Abstracts of
  the Conference on Human Factors in Computing Systems}}.
  \bibinfo{pages}{1--4}.
\newblock


\bibitem[\protect\citeauthoryear{Poupyrev, Ichikawa, Weghorst, and
  Billinghurst}{Poupyrev et~al\mbox{.}}{1998}]%
        {poupyrev1998egocentric}
\bibfield{author}{\bibinfo{person}{Ivan Poupyrev}, \bibinfo{person}{Tadao
  Ichikawa}, \bibinfo{person}{Suzanne Weghorst}, {and} \bibinfo{person}{Mark
  Billinghurst}.} \bibinfo{year}{1998}\natexlab{}.
\newblock \showarticletitle{Egocentric object manipulation in virtual
  environments: empirical evaluation of interaction techniques}. In
  \bibinfo{booktitle}{\emph{Computer Graphics Forum}},
  Vol.~\bibinfo{volume}{17}.
\newblock


\bibitem[\protect\citeauthoryear{Powers~Iii, Hillock-Dunn, and
  Wallace}{Powers~Iii et~al\mbox{.}}{2016}]%
        {powers2016}
\bibfield{author}{\bibinfo{person}{Albert~R Powers~Iii},
  \bibinfo{person}{Andrea Hillock-Dunn}, {and} \bibinfo{person}{Mark~T
  Wallace}.} \bibinfo{year}{2016}\natexlab{}.
\newblock \showarticletitle{Generalization of multisensory perceptual
  learning}.
\newblock \bibinfo{journal}{\emph{Scientific Reports}}  \bibinfo{volume}{6}
  (\bibinfo{year}{2016}), \bibinfo{pages}{23374}.
\newblock


\bibitem[\protect\citeauthoryear{Poyade}{Poyade}{2013}]%
        {poyade2013motor}
\bibfield{author}{\bibinfo{person}{Matthieu Poyade}.}
  \bibinfo{year}{2013}\natexlab{}.
\newblock \showarticletitle{Motor skill training using virtual reality and
  haptic interaction--A case study in industrial maintenance}.
\newblock \bibinfo{journal}{\emph{M{\'A}LAGA}} (\bibinfo{year}{2013}).
\newblock


\bibitem[\protect\citeauthoryear{Pozeg, Galli, and Blanke}{Pozeg
  et~al\mbox{.}}{2015}]%
        {pozeg2015those}
\bibfield{author}{\bibinfo{person}{Polona Pozeg}, \bibinfo{person}{Giulia
  Galli}, {and} \bibinfo{person}{Olaf Blanke}.}
  \bibinfo{year}{2015}\natexlab{}.
\newblock \showarticletitle{Those are your legs: the effect of visuo-spatial
  viewpoint on visuo-tactile integration and body ownership}.
\newblock \bibinfo{journal}{\emph{Frontiers in psychology}}
  \bibinfo{volume}{6} (\bibinfo{year}{2015}), \bibinfo{pages}{1749}.
\newblock


\bibitem[\protect\citeauthoryear{Prange, Barz, and Sonntag}{Prange
  et~al\mbox{.}}{2018}]%
        {prange2018medical}
\bibfield{author}{\bibinfo{person}{Alexander Prange}, \bibinfo{person}{Michael
  Barz}, {and} \bibinfo{person}{Daniel Sonntag}.}
  \bibinfo{year}{2018}\natexlab{}.
\newblock \showarticletitle{Medical 3d images in multimodal virtual reality}.
  In \bibinfo{booktitle}{\emph{Proceedings of the 23rd International Conference
  on Intelligent User Interfaces Companion}}. \bibinfo{pages}{1--2}.
\newblock


\bibitem[\protect\citeauthoryear{Prinz}{Prinz}{2006}]%
        {Prinz2006-PRIITM}
\bibfield{author}{\bibinfo{person}{Jesse~J. Prinz}.}
  \bibinfo{year}{2006}\natexlab{}.
\newblock \showarticletitle{Is the Mind Really Modular?}
\newblock In \bibinfo{booktitle}{\emph{Contemporary Debates in Cognitive
  Science}}. \bibinfo{pages}{22--36}.
\newblock


\bibitem[\protect\citeauthoryear{Ramachandran and
  Rogers-Ramachandran}{Ramachandran and Rogers-Ramachandran}{1996}]%
        {ramachandran1996}
\bibfield{author}{\bibinfo{person}{Vilayanur~S Ramachandran} {and}
  \bibinfo{person}{Diane Rogers-Ramachandran}.}
  \bibinfo{year}{1996}\natexlab{}.
\newblock \showarticletitle{Synaesthesia in phantom limbs induced with
  mirrors}.
\newblock \bibinfo{journal}{\emph{Proceedings of the Royal Society of London.
  Series B: Biological Sciences}} \bibinfo{volume}{263}, \bibinfo{number}{1369}
  (\bibinfo{year}{1996}), \bibinfo{pages}{377--386}.
\newblock


\bibitem[\protect\citeauthoryear{Ranasinghe, Cheok, Nakatsu, and Do}{Ranasinghe
  et~al\mbox{.}}{2013}]%
        {ranasinghe2013simulating}
\bibfield{author}{\bibinfo{person}{Nimesha Ranasinghe}, \bibinfo{person}{Adrian
  Cheok}, \bibinfo{person}{Ryohei Nakatsu}, {and} \bibinfo{person}{Ellen
  Yi-Luen Do}.} \bibinfo{year}{2013}\natexlab{}.
\newblock \showarticletitle{Simulating the sensation of taste for immersive
  experiences}. In \bibinfo{booktitle}{\emph{Proceedings of the ACM
  International Workshop on Immersive Media Experiences}}.
\newblock


\bibitem[\protect\citeauthoryear{Ranasinghe, Jain, Thi Ngoc~Tram, Koh, Tolley,
  Karwita, Lien-Ya, Liangkun, Shamaiah, Eason Wai~Tung,
  et~al\mbox{.}}{Ranasinghe et~al\mbox{.}}{2018}]%
        {ranasinghe2018season}
\bibfield{author}{\bibinfo{person}{N. Ranasinghe}, \bibinfo{person}{P. Jain},
  \bibinfo{person}{N. Thi Ngoc~Tram}, \bibinfo{person}{K. Koh},
  \bibinfo{person}{D. Tolley}, \bibinfo{person}{S. Karwita},
  \bibinfo{person}{L. Lien-Ya}, \bibinfo{person}{Y. Liangkun},
  \bibinfo{person}{K. Shamaiah}, \bibinfo{person}{C. Eason Wai~Tung},
  {et~al\mbox{.}}} \bibinfo{year}{2018}\natexlab{}.
\newblock \showarticletitle{Season traveller: Multisensory narration for
  enhancing the virtual reality experience}. In
  \bibinfo{booktitle}{\emph{Proceedings of the Conference on Human Factors in
  Computing Systems}}. \bibinfo{pages}{1--13}.
\newblock


\bibitem[\protect\citeauthoryear{Rewkowski, Rungta, Whitton, and Lin}{Rewkowski
  et~al\mbox{.}}{2019}]%
        {rewkowski2019evaluating}
\bibfield{author}{\bibinfo{person}{N. Rewkowski}, \bibinfo{person}{A. Rungta},
  \bibinfo{person}{M. Whitton}, {and} \bibinfo{person}{M. Lin}.}
  \bibinfo{year}{2019}\natexlab{}.
\newblock \showarticletitle{Evaluating the Effectiveness of Redirected Walking
  with Auditory Distractors for Navigation in Virtual Environments}. In
  \bibinfo{booktitle}{\emph{IEEE Conf. on Virtual Reality and 3D User
  Interfaces}}.
\newblock


\bibitem[\protect\citeauthoryear{Richard, Tijou, and Richard}{Richard
  et~al\mbox{.}}{2006}]%
        {richard2006bmulti}
\bibfield{author}{\bibinfo{person}{Emmanuelle Richard},
  \bibinfo{person}{Ang{\`e}le Tijou}, {and} \bibinfo{person}{Paul Richard}.}
  \bibinfo{year}{2006}\natexlab{}.
\newblock \showarticletitle{Multi-modal virtual environments for education:
  From illusion to immersion}. In \bibinfo{booktitle}{\emph{International
  Conference on Technologies for E-Learning and Digital Entertainment}}.
  Springer.
\newblock


\bibitem[\protect\citeauthoryear{Richardt, Tompkin, and Wetzstein}{Richardt
  et~al\mbox{.}}{2020}]%
        {richardt2020capture}
\bibfield{author}{\bibinfo{person}{Christian Richardt}, \bibinfo{person}{James
  Tompkin}, {and} \bibinfo{person}{Gordon Wetzstein}.}
  \bibinfo{year}{2020}\natexlab{}.
\newblock \showarticletitle{Capture, Reconstruction, and Representation of the
  Visual Real World for Virtual Reality}.
\newblock In \bibinfo{booktitle}{\emph{Real VR--Immersive Digital Reality}}.
  \bibinfo{publisher}{Springer}, \bibinfo{pages}{3--32}.
\newblock


\bibitem[\protect\citeauthoryear{Riecke, V{\"a}ljam{\"a}e, and
  Schulte-Pelkum}{Riecke et~al\mbox{.}}{2009}]%
        {riecke2009moving}
\bibfield{author}{\bibinfo{person}{Bernhard~E Riecke},
  \bibinfo{person}{Aleksander V{\"a}ljam{\"a}e}, {and}
  \bibinfo{person}{J{\"o}rg Schulte-Pelkum}.} \bibinfo{year}{2009}\natexlab{}.
\newblock \showarticletitle{Moving sounds enhance the visually-induced
  self-motion illusion (circular vection) in virtual reality}.
\newblock \bibinfo{journal}{\emph{ACM Trans. on Applied Perception}}
  \bibinfo{volume}{6}, \bibinfo{number}{2} (\bibinfo{year}{2009}).
\newblock


\bibitem[\protect\citeauthoryear{Rock and Victor}{Rock and Victor}{1964}]%
        {rock64}
\bibfield{author}{\bibinfo{person}{Irvin Rock} {and} \bibinfo{person}{Jack
  Victor}.} \bibinfo{year}{1964}\natexlab{}.
\newblock \showarticletitle{Vision and Touch: An Experimentally Created
  Conflict between the Two Senses}.
\newblock \bibinfo{journal}{\emph{Science}} \bibinfo{volume}{143},
  \bibinfo{number}{3606} (\bibinfo{year}{1964}), \bibinfo{pages}{594--596}.
\newblock
\showISSN{0036-8075}


\bibitem[\protect\citeauthoryear{Rosen, Soltanian, Redett, and Laub}{Rosen
  et~al\mbox{.}}{1996}]%
        {rosen1996evolution}
\bibfield{author}{\bibinfo{person}{Joseph~M Rosen}, \bibinfo{person}{Hooman
  Soltanian}, \bibinfo{person}{Richard~J Redett}, {and}
  \bibinfo{person}{Donald~R Laub}.} \bibinfo{year}{1996}\natexlab{}.
\newblock \showarticletitle{Evolution of virtual reality [Medicine]}.
\newblock \bibinfo{journal}{\emph{IEEE Engineering in Medicine and Biology
  Magazine}} \bibinfo{volume}{15}, \bibinfo{number}{2} (\bibinfo{year}{1996}),
  \bibinfo{pages}{16--22}.
\newblock


\bibitem[\protect\citeauthoryear{Rosenkvist, Eriksen, Koehlert, Valimaa,
  Vittrup, Andreasen, and Palamas}{Rosenkvist et~al\mbox{.}}{2019}]%
        {rosenkvist2019hearing}
\bibfield{author}{\bibinfo{person}{Amalie Rosenkvist},
  \bibinfo{person}{David~Sebastian Eriksen}, \bibinfo{person}{Jeppe Koehlert},
  \bibinfo{person}{Miicha Valimaa}, \bibinfo{person}{Mikkel~Brogaard Vittrup},
  \bibinfo{person}{Anastasia Andreasen}, {and} \bibinfo{person}{George
  Palamas}.} \bibinfo{year}{2019}\natexlab{}.
\newblock \showarticletitle{Hearing with Eyes in Virtual Reality}. In
  \bibinfo{booktitle}{\emph{2019 IEEE Conference on Virtual Reality and 3D User
  Interfaces (VR)}}. IEEE, \bibinfo{pages}{1349--1350}.
\newblock


\bibitem[\protect\citeauthoryear{Rothe, Buschek, and Hu{\ss}mann}{Rothe
  et~al\mbox{.}}{2019}]%
        {rothe2019guidance}
\bibfield{author}{\bibinfo{person}{Sylvia Rothe}, \bibinfo{person}{Daniel
  Buschek}, {and} \bibinfo{person}{Heinrich Hu{\ss}mann}.}
  \bibinfo{year}{2019}\natexlab{}.
\newblock \showarticletitle{Guidance in cinematic virtual reality-taxonomy,
  research status and challenges}.
\newblock \bibinfo{journal}{\emph{Multimodal Technologies and Interaction}}
  \bibinfo{volume}{3}, \bibinfo{number}{1} (\bibinfo{year}{2019}),
  \bibinfo{pages}{19}.
\newblock


\bibitem[\protect\citeauthoryear{Rothe and Hu{\ss}mann}{Rothe and
  Hu{\ss}mann}{2018}]%
        {rothe2018guiding}
\bibfield{author}{\bibinfo{person}{Sylvia Rothe} {and}
  \bibinfo{person}{Heinrich Hu{\ss}mann}.} \bibinfo{year}{2018}\natexlab{}.
\newblock \showarticletitle{Guiding the viewer in cinematic virtual reality by
  diegetic cues}. In \bibinfo{booktitle}{\emph{International Conference on
  Augmented Reality, Virtual Reality and Computer Graphics}}. Springer,
  \bibinfo{pages}{101--117}.
\newblock


\bibitem[\protect\citeauthoryear{Rothe, Hu{\ss}mann, and Allary}{Rothe
  et~al\mbox{.}}{2017}]%
        {rothe2017diegetic}
\bibfield{author}{\bibinfo{person}{Sylvia Rothe}, \bibinfo{person}{Heinrich
  Hu{\ss}mann}, {and} \bibinfo{person}{Mathias Allary}.}
  \bibinfo{year}{2017}\natexlab{}.
\newblock \showarticletitle{Diegetic cues for guiding the viewer in cinematic
  virtual reality}. In \bibinfo{booktitle}{\emph{Proceedings of the 23rd ACM
  Symposium on Virtual Reality Software and Technology}}.
  \bibinfo{pages}{1--2}.
\newblock


\bibitem[\protect\citeauthoryear{Rubio-Tamayo, Gertrudix~Barrio, and
  Garc{\'\i}a~Garc{\'\i}a}{Rubio-Tamayo et~al\mbox{.}}{2017}]%
        {rubio2017immersive}
\bibfield{author}{\bibinfo{person}{J. Rubio-Tamayo}, \bibinfo{person}{M.
  Gertrudix~Barrio}, {and} \bibinfo{person}{F. Garc{\'\i}a~Garc{\'\i}a}.}
  \bibinfo{year}{2017}\natexlab{}.
\newblock \showarticletitle{Immersive environments and virtual reality:
  Systematic review and advances in communication, interaction and simulation}.
\newblock \bibinfo{journal}{\emph{Multimodal Tech. and Interact.}}
  (\bibinfo{year}{2017}).
\newblock


\bibitem[\protect\citeauthoryear{Ruotolo, Maffei, Di~Gabriele, Iachini,
  Masullo, Ruggiero, and Senese}{Ruotolo et~al\mbox{.}}{2013}]%
        {ruotolo2013immersive}
\bibfield{author}{\bibinfo{person}{Francesco Ruotolo}, \bibinfo{person}{Luigi
  Maffei}, \bibinfo{person}{Maria Di~Gabriele}, \bibinfo{person}{Tina Iachini},
  \bibinfo{person}{Massimiliano Masullo}, \bibinfo{person}{Gennaro Ruggiero},
  {and} \bibinfo{person}{Vincenzo~Paolo Senese}.}
  \bibinfo{year}{2013}\natexlab{}.
\newblock \showarticletitle{Immersive virtual reality and environmental noise
  assessment: An innovative audio--visual approach}.
\newblock \bibinfo{journal}{\emph{Environmental Impact Assessment Review}}
  \bibinfo{volume}{41} (\bibinfo{year}{2013}), \bibinfo{pages}{10--20}.
\newblock


\bibitem[\protect\citeauthoryear{Sadowski and Stanney}{Sadowski and
  Stanney}{2002}]%
        {sadowski2002presence}
\bibfield{author}{\bibinfo{person}{Wallace Sadowski} {and} \bibinfo{person}{Kay
  Stanney}.} \bibinfo{year}{2002}\natexlab{}.
\newblock \showarticletitle{Presence in virtual environments.}
\newblock \bibinfo{journal}{\emph{Human factors and ergonomics. Handbook of
  virtual environments: Design, implementation, and applications (p.
  791–806).}} (\bibinfo{year}{2002}).
\newblock


\bibitem[\protect\citeauthoryear{Sakhardande, Murugan, and Pillai}{Sakhardande
  et~al\mbox{.}}{2020}]%
        {sakhardande2020exploring}
\bibfield{author}{\bibinfo{person}{P. Sakhardande}, \bibinfo{person}{A.
  Murugan}, {and} \bibinfo{person}{J. Pillai}.}
  \bibinfo{year}{2020}\natexlab{}.
\newblock \showarticletitle{Exploring Effect Of Different External Stimuli On
  Body Association In VR}. In \bibinfo{booktitle}{\emph{IEEE Conference on
  Virtual Reality and 3D User Interfaces Abstracts and Workshops (VRW)}}.
  \bibinfo{pages}{689--690}.
\newblock


\bibitem[\protect\citeauthoryear{Salazar, Pacchierotti, de~Tinguy, Maciel, and
  Marchal}{Salazar et~al\mbox{.}}{2020}]%
        {salazar2020altering}
\bibfield{author}{\bibinfo{person}{S. Salazar}, \bibinfo{person}{C.
  Pacchierotti}, \bibinfo{person}{X. de Tinguy}, \bibinfo{person}{A. Maciel},
  {and} \bibinfo{person}{M. Marchal}.} \bibinfo{year}{2020}\natexlab{}.
\newblock \showarticletitle{Altering the stiffness, friction, and shape
  perception of tangible objects in virtual reality using wearable haptics}.
\newblock \bibinfo{journal}{\emph{IEEE Trans. on Haptics}}
  \bibinfo{volume}{13}, \bibinfo{number}{1} (\bibinfo{year}{2020}),
  \bibinfo{pages}{167--174}.
\newblock


\bibitem[\protect\citeauthoryear{Salselas, Penha, and Bernardes}{Salselas
  et~al\mbox{.}}{2020}]%
        {salselas2020}
\bibfield{author}{\bibinfo{person}{In{\^e}s Salselas}, \bibinfo{person}{Rui
  Penha}, {and} \bibinfo{person}{Gilberto Bernardes}.}
  \bibinfo{year}{2020}\natexlab{}.
\newblock \showarticletitle{Sound design inducing attention in the context of
  audiovisual immersive environments}.
\newblock \bibinfo{journal}{\emph{Personal and Ubiquitous Computing}}
  (\bibinfo{year}{2020}), \bibinfo{pages}{1--12}.
\newblock


\bibitem[\protect\citeauthoryear{Samad, Gatti, Hermes, Benko, and Parise}{Samad
  et~al\mbox{.}}{2019}]%
        {samad2019pseudo}
\bibfield{author}{\bibinfo{person}{M. Samad}, \bibinfo{person}{E. Gatti},
  \bibinfo{person}{A. Hermes}, \bibinfo{person}{H. Benko}, {and}
  \bibinfo{person}{C. Parise}.} \bibinfo{year}{2019}\natexlab{}.
\newblock \showarticletitle{Pseudo-haptic weight: Changing the perceived weight
  of virtual objects by manipulating control-display ratio}. In
  \bibinfo{booktitle}{\emph{Proc. of the Conf. on Human Factors in Computing
  Systems}}.
\newblock


\bibitem[\protect\citeauthoryear{Sanchez-Vives and Slater}{Sanchez-Vives and
  Slater}{2005}]%
        {sanchez2005presence}
\bibfield{author}{\bibinfo{person}{Maria~V Sanchez-Vives} {and}
  \bibinfo{person}{Mel Slater}.} \bibinfo{year}{2005}\natexlab{}.
\newblock \showarticletitle{From presence to consciousness through virtual
  reality}.
\newblock \bibinfo{journal}{\emph{Nature Reviews Neuroscience}}
  \bibinfo{volume}{6}, \bibinfo{number}{4} (\bibinfo{year}{2005}),
  \bibinfo{pages}{332--339}.
\newblock


\bibitem[\protect\citeauthoryear{Sano, Ichinose, Wake, Osumi, Sumitani,
  Kumagaya, and Kuniyoshi}{Sano et~al\mbox{.}}{2015}]%
        {sano2015reliability}
\bibfield{author}{\bibinfo{person}{Yuko Sano}, \bibinfo{person}{Akimichi
  Ichinose}, \bibinfo{person}{Naoki Wake}, \bibinfo{person}{Michihiro Osumi},
  \bibinfo{person}{Masahiko Sumitani}, \bibinfo{person}{Shin-ichiro Kumagaya},
  {and} \bibinfo{person}{Yasuo Kuniyoshi}.} \bibinfo{year}{2015}\natexlab{}.
\newblock \showarticletitle{Reliability of phantom pain relief in
  neurorehabilitation using a multimodal virtual reality system}. In
  \bibinfo{booktitle}{\emph{Conference of the IEEE Engineering in Medicine and
  Biology Society (EMBC)}}.
\newblock


\bibitem[\protect\citeauthoryear{Sarlat, Warusfel, and Viaud-Delmon}{Sarlat
  et~al\mbox{.}}{2006}]%
        {sarlat2006ventriloquism}
\bibfield{author}{\bibinfo{person}{Ludivine Sarlat}, \bibinfo{person}{Olivier
  Warusfel}, {and} \bibinfo{person}{Isabelle Viaud-Delmon}.}
  \bibinfo{year}{2006}\natexlab{}.
\newblock \showarticletitle{Ventriloquism aftereffects occur in the rear
  hemisphere}.
\newblock \bibinfo{journal}{\emph{Neuroscience letters}} \bibinfo{volume}{404},
  \bibinfo{number}{3} (\bibinfo{year}{2006}), \bibinfo{pages}{324--329}.
\newblock


\bibitem[\protect\citeauthoryear{Satava and Jones}{Satava and Jones}{1998}]%
        {satava1998current}
\bibfield{author}{\bibinfo{person}{R. Satava} {and} \bibinfo{person}{S.
  Jones}.} \bibinfo{year}{1998}\natexlab{}.
\newblock \showarticletitle{Current and future applications of virtual reality
  for medicine}.
\newblock \bibinfo{journal}{\emph{Proc. IEEE}} \bibinfo{volume}{86},
  \bibinfo{number}{3} (\bibinfo{year}{1998}).
\newblock


\bibitem[\protect\citeauthoryear{Schmitz, MacQuarrie, Julier,
  et~al\mbox{.}}{Schmitz et~al\mbox{.}}{2020}]%
        {schmitz2020directing}
\bibfield{author}{\bibinfo{person}{A. Schmitz}, \bibinfo{person}{A.
  MacQuarrie}, \bibinfo{person}{S. Julier}, {et~al\mbox{.}}}
  \bibinfo{year}{2020}\natexlab{}.
\newblock \showarticletitle{Directing versus Attracting Attention: Exploring
  the Effectiveness of Central and Peripheral Cues in Panoramic Videos}. In
  \bibinfo{booktitle}{\emph{IEEE Conf. on Virtual Reality and 3D User
  Interfaces}}.
\newblock


\bibitem[\protect\citeauthoryear{Schuemie, Van Der~Straaten, Krijn, and Van
  Der~Mast}{Schuemie et~al\mbox{.}}{2001}]%
        {schuemie2001research}
\bibfield{author}{\bibinfo{person}{Martijn~J Schuemie}, \bibinfo{person}{Peter
  Van Der~Straaten}, \bibinfo{person}{Merel Krijn}, {and}
  \bibinfo{person}{Charles~APG Van Der~Mast}.} \bibinfo{year}{2001}\natexlab{}.
\newblock \showarticletitle{Research on presence in virtual reality: A survey}.
\newblock \bibinfo{journal}{\emph{CyberPsychology \& Behavior}}
  \bibinfo{volume}{4}, \bibinfo{number}{2} (\bibinfo{year}{2001}),
  \bibinfo{pages}{183--201}.
\newblock


\bibitem[\protect\citeauthoryear{Seinfeld, Feuchtner, Pinzek, and
  M{\"u}ller}{Seinfeld et~al\mbox{.}}{2020}]%
        {seinfeld2020impact}
\bibfield{author}{\bibinfo{person}{Sofia Seinfeld}, \bibinfo{person}{Tiare
  Feuchtner}, \bibinfo{person}{Johannes Pinzek}, {and}
  \bibinfo{person}{J{\"o}rg M{\"u}ller}.} \bibinfo{year}{2020}\natexlab{}.
\newblock \showarticletitle{Impact of Information Placement and User
  Representations in VR on Performance and Embodiment}.
\newblock \bibinfo{journal}{\emph{arXiv preprint arXiv:2002.12007}}
  (\bibinfo{year}{2020}).
\newblock


\bibitem[\protect\citeauthoryear{Seitz, Kim, and Shams}{Seitz
  et~al\mbox{.}}{2006}]%
        {seitz2006sound}
\bibfield{author}{\bibinfo{person}{Aaron~R Seitz}, \bibinfo{person}{Robyn Kim},
  {and} \bibinfo{person}{Ladan Shams}.} \bibinfo{year}{2006}\natexlab{}.
\newblock \showarticletitle{Sound facilitates visual learning}.
\newblock \bibinfo{journal}{\emph{Current Biology}} \bibinfo{volume}{16},
  \bibinfo{number}{14} (\bibinfo{year}{2006}).
\newblock


\bibitem[\protect\citeauthoryear{Serafin and Serafin}{Serafin and
  Serafin}{2004}]%
        {serafin2004sound}
\bibfield{author}{\bibinfo{person}{G Serafin} {and} \bibinfo{person}{S
  Serafin}.} \bibinfo{year}{2004}\natexlab{}.
\newblock \showarticletitle{Sound design to enhance presence in photorealistic
  virtual reality}. Georgia Inst. of Tech.
\newblock


\bibitem[\protect\citeauthoryear{Serafin, Geronazzo, Erkut, Nilsson, and
  Nordahl}{Serafin et~al\mbox{.}}{2018}]%
        {serafin2018sonic}
\bibfield{author}{\bibinfo{person}{Stefania Serafin}, \bibinfo{person}{Michele
  Geronazzo}, \bibinfo{person}{Cumhur Erkut}, \bibinfo{person}{Niels~C
  Nilsson}, {and} \bibinfo{person}{Rolf Nordahl}.}
  \bibinfo{year}{2018}\natexlab{}.
\newblock \showarticletitle{Sonic interactions in virtual reality: state of the
  art, current challenges, and future directions}.
\newblock \bibinfo{journal}{\emph{IEEE Comp. Graph. and App.}}
  \bibinfo{volume}{38}, \bibinfo{number}{2} (\bibinfo{year}{2018}).
\newblock


\bibitem[\protect\citeauthoryear{Serafin, Nilsson, Sikstrom, De~Goetzen, and
  Nordahl}{Serafin et~al\mbox{.}}{2013}]%
        {serafin2013estimation}
\bibfield{author}{\bibinfo{person}{Stefania Serafin}, \bibinfo{person}{Niels~C
  Nilsson}, \bibinfo{person}{Erik Sikstrom}, \bibinfo{person}{Amalia
  De~Goetzen}, {and} \bibinfo{person}{Rolf Nordahl}.}
  \bibinfo{year}{2013}\natexlab{}.
\newblock \showarticletitle{Estimation of detection thresholds for acoustic
  based redirected walking techniques}. In \bibinfo{booktitle}{\emph{2013 IEEE
  Virtual Reality (VR)}}. IEEE, \bibinfo{pages}{161--162}.
\newblock


\bibitem[\protect\citeauthoryear{Serrano, Martin, Gutierrez, Myszkowski, and
  Masia}{Serrano et~al\mbox{.}}{2020}]%
        {Serrano2020_VR-LateralMotion}
\bibfield{author}{\bibinfo{person}{Ana Serrano}, \bibinfo{person}{Daniel
  Martin}, \bibinfo{person}{Diego Gutierrez}, \bibinfo{person}{Karol
  Myszkowski}, {and} \bibinfo{person}{Belen Masia}.}
  \bibinfo{year}{2020}\natexlab{}.
\newblock \showarticletitle{Imperceptible manipulation of lateral camera motion
  for improved virtual reality applications}.
\newblock \bibinfo{journal}{\emph{ACM Trans. on Graphics}}
  \bibinfo{volume}{39}, \bibinfo{number}{6} (\bibinfo{year}{2020}).
\newblock


\bibitem[\protect\citeauthoryear{Serrano, Sitzmann, Ruiz-Borau, Wetzstein,
  Gutierrez, and Masia}{Serrano et~al\mbox{.}}{2017}]%
        {Serrano_VR-cine_SIGGRAPH2017}
\bibfield{author}{\bibinfo{person}{Ana Serrano}, \bibinfo{person}{Vincent
  Sitzmann}, \bibinfo{person}{Jaime Ruiz-Borau}, \bibinfo{person}{Gordon
  Wetzstein}, \bibinfo{person}{Diego Gutierrez}, {and} \bibinfo{person}{Belen
  Masia}.} \bibinfo{year}{2017}\natexlab{}.
\newblock \showarticletitle{Movie Editing and Cognitive Event Segmentation in
  Virtual Reality Video}.
\newblock \bibinfo{journal}{\emph{ACM Trans. on Graph. (SIGGRAPH)}}
  \bibinfo{volume}{36}, \bibinfo{number}{4} (\bibinfo{year}{2017}).
\newblock


\bibitem[\protect\citeauthoryear{Seth, Vance, and Oliver}{Seth
  et~al\mbox{.}}{2011}]%
        {seth2011virtual}
\bibfield{author}{\bibinfo{person}{Abhishek Seth}, \bibinfo{person}{Judy~M
  Vance}, {and} \bibinfo{person}{James~H Oliver}.}
  \bibinfo{year}{2011}\natexlab{}.
\newblock \showarticletitle{Virtual reality for assembly methods prototyping: a
  review}.
\newblock \bibinfo{journal}{\emph{Virtual reality}} \bibinfo{volume}{15},
  \bibinfo{number}{1} (\bibinfo{year}{2011}), \bibinfo{pages}{5--20}.
\newblock


\bibitem[\protect\citeauthoryear{Shams and Kim}{Shams and Kim}{2010}]%
        {shams2010crossmodal}
\bibfield{author}{\bibinfo{person}{Ladan Shams} {and} \bibinfo{person}{Robyn
  Kim}.} \bibinfo{year}{2010}\natexlab{}.
\newblock \showarticletitle{Crossmodal influences on visual perception}.
\newblock \bibinfo{journal}{\emph{Physics of life reviews}}
  \bibinfo{volume}{7}, \bibinfo{number}{3} (\bibinfo{year}{2010}).
\newblock


\bibitem[\protect\citeauthoryear{Shams, Ma, and Beierholm}{Shams
  et~al\mbox{.}}{2005}]%
        {shams2005sound}
\bibfield{author}{\bibinfo{person}{L. Shams}, \bibinfo{person}{W. Ma}, {and}
  \bibinfo{person}{U. Beierholm}.} \bibinfo{year}{2005}\natexlab{}.
\newblock \showarticletitle{Sound-induced flash illusion as an optimal
  percept}.
\newblock \bibinfo{journal}{\emph{Neuroreport}} (\bibinfo{year}{2005}).
\newblock


\bibitem[\protect\citeauthoryear{Shams and Seitz}{Shams and Seitz}{2008}]%
        {shams2008benefits}
\bibfield{author}{\bibinfo{person}{Ladan Shams} {and} \bibinfo{person}{Aaron~R
  Seitz}.} \bibinfo{year}{2008}\natexlab{}.
\newblock \showarticletitle{Benefits of multisensory learning}.
\newblock \bibinfo{journal}{\emph{Trends in cognitive sciences}}
  \bibinfo{volume}{12}, \bibinfo{number}{11} (\bibinfo{year}{2008}).
\newblock


\bibitem[\protect\citeauthoryear{Shapiro and Todorovic}{Shapiro and
  Todorovic}{2016}]%
        {shapiro2016}
\bibfield{author}{\bibinfo{person}{Arthur~G Shapiro} {and}
  \bibinfo{person}{Dejan Todorovic}.} \bibinfo{year}{2016}\natexlab{}.
\newblock \bibinfo{booktitle}{\emph{The Oxford compendium of visual
  illusions}}.
\newblock \bibinfo{publisher}{Oxford University Press}.
\newblock


\bibitem[\protect\citeauthoryear{Shiban, Pauli, and M{\"u}hlberger}{Shiban
  et~al\mbox{.}}{2013}]%
        {shiban2013effect}
\bibfield{author}{\bibinfo{person}{Youssef Shiban}, \bibinfo{person}{Paul
  Pauli}, {and} \bibinfo{person}{Andreas M{\"u}hlberger}.}
  \bibinfo{year}{2013}\natexlab{}.
\newblock \showarticletitle{Effect of multiple context exposure on renewal in
  spider phobia}.
\newblock \bibinfo{journal}{\emph{Behaviour Research and Therapy}}
  \bibinfo{volume}{51}, \bibinfo{number}{2} (\bibinfo{year}{2013}),
  \bibinfo{pages}{68--74}.
\newblock


\bibitem[\protect\citeauthoryear{Siddig, Ragano, Jahromi, and Hines}{Siddig
  et~al\mbox{.}}{2019a}]%
        {abubakr2019}
\bibfield{author}{\bibinfo{person}{A. Siddig}, \bibinfo{person}{A. Ragano},
  \bibinfo{person}{H. Jahromi}, {and} \bibinfo{person}{A. Hines}.}
  \bibinfo{year}{2019}\natexlab{a}.
\newblock \showarticletitle{Fusion Confusion: Exploring Ambisonic Spatial
  Localisation for Audio-Visual Immersion Using the McGurk Effect}. In
  \bibinfo{booktitle}{\emph{ACM Workshop on Immersive Mixed and Virtual Env.
  Systems}}.
\newblock


\bibitem[\protect\citeauthoryear{Siddig, Sun, Parker, and Hines}{Siddig
  et~al\mbox{.}}{2019b}]%
        {siddig2019}
\bibfield{author}{\bibinfo{person}{A. Siddig}, \bibinfo{person}{P. Sun},
  \bibinfo{person}{M. Parker}, {and} \bibinfo{person}{A. Hines}.}
  \bibinfo{year}{2019}\natexlab{b}.
\newblock \showarticletitle{Perception Deception: Audio-Visual Mismatch in
  Virtual Reality Using the McGurk Effect}.
\newblock  (\bibinfo{year}{2019}).
\newblock


\bibitem[\protect\citeauthoryear{Sitzmann, Serrano, Pavel, Agrawala, Gutierrez,
  Masia, and Wetzstein}{Sitzmann et~al\mbox{.}}{2017}]%
        {Sitzmann_TVCG_VR-saliency}
\bibfield{author}{\bibinfo{person}{Vincent Sitzmann}, \bibinfo{person}{Ana
  Serrano}, \bibinfo{person}{Amy Pavel}, \bibinfo{person}{Maneesh Agrawala},
  \bibinfo{person}{Diego Gutierrez}, \bibinfo{person}{Belen Masia}, {and}
  \bibinfo{person}{Gordon Wetzstein}.} \bibinfo{year}{2017}\natexlab{}.
\newblock \showarticletitle{How do people explore virtual environments?}
\newblock \bibinfo{journal}{\emph{IEEE Trans. on Visualization and Computer
  Graphics}} (\bibinfo{year}{2017}).
\newblock


\bibitem[\protect\citeauthoryear{Slater}{Slater}{2009}]%
        {slater2009place}
\bibfield{author}{\bibinfo{person}{Mel Slater}.}
  \bibinfo{year}{2009}\natexlab{}.
\newblock \showarticletitle{Place illusion and plausibility can lead to
  realistic behaviour in immersive virtual environments}.
\newblock \bibinfo{journal}{\emph{Philosophical Trans. of the Royal Society B:
  Biological Sciences}} \bibinfo{volume}{364}, \bibinfo{number}{1535}
  (\bibinfo{year}{2009}), \bibinfo{pages}{3549--3557}.
\newblock


\bibitem[\protect\citeauthoryear{Slater, Khanna, Mortensen, and Yu}{Slater
  et~al\mbox{.}}{2009}]%
        {slater2009visual}
\bibfield{author}{\bibinfo{person}{Mel Slater}, \bibinfo{person}{Pankaj
  Khanna}, \bibinfo{person}{Jesper Mortensen}, {and} \bibinfo{person}{Insu
  Yu}.} \bibinfo{year}{2009}\natexlab{}.
\newblock \showarticletitle{Visual realism enhances realistic response in an
  immersive virtual environment}.
\newblock \bibinfo{journal}{\emph{IEEE Computer Graphics and Applications}}
  \bibinfo{volume}{29}, \bibinfo{number}{3} (\bibinfo{year}{2009}),
  \bibinfo{pages}{76--84}.
\newblock


\bibitem[\protect\citeauthoryear{Slater and Usoh}{Slater and Usoh}{1993}]%
        {slater1993presence}
\bibfield{author}{\bibinfo{person}{Mel Slater} {and} \bibinfo{person}{Martin
  Usoh}.} \bibinfo{year}{1993}\natexlab{}.
\newblock \showarticletitle{Presence in immersive virtual environments}. In
  \bibinfo{booktitle}{\emph{Proceedings of IEEE Virtual Reality Annual
  International Symposium}}. IEEE, \bibinfo{pages}{90--96}.
\newblock


\bibitem[\protect\citeauthoryear{Spanlang}{Spanlang}{2014}]%
        {spanlang2014build}
\bibfield{author}{\bibinfo{person}{Bernhard et~al. Spanlang}.}
  \bibinfo{year}{2014}\natexlab{}.
\newblock \showarticletitle{How to build an embodiment lab: achieving body
  representation illusions in virtual reality}.
\newblock \bibinfo{journal}{\emph{Frontiers in Robotics and AI}}
  \bibinfo{volume}{1} (\bibinfo{year}{2014}), \bibinfo{pages}{9}.
\newblock


\bibitem[\protect\citeauthoryear{Spence, Lee, and Van~der Stoep}{Spence
  et~al\mbox{.}}{2017}]%
        {spence2017}
\bibfield{author}{\bibinfo{person}{Charles Spence}, \bibinfo{person}{Jae Lee},
  {and} \bibinfo{person}{Nathan Van~der Stoep}.}
  \bibinfo{year}{2017}\natexlab{}.
\newblock \showarticletitle{Responding to sounds from unseen locations:
  Crossmodal attentional orienting in response to sounds presented from the
  rear}.
\newblock \bibinfo{journal}{\emph{European Journal of Neuroscience}}
  \bibinfo{volume}{51}, \bibinfo{number}{5} (\bibinfo{year}{2017}).
\newblock


\bibitem[\protect\citeauthoryear{Spence and Parise}{Spence and Parise}{2010}]%
        {spence2010prior}
\bibfield{author}{\bibinfo{person}{Charles Spence} {and}
  \bibinfo{person}{Cesare Parise}.} \bibinfo{year}{2010}\natexlab{}.
\newblock \showarticletitle{Prior-entry: A review}.
\newblock \bibinfo{journal}{\emph{Consciousness and cognition}}
  \bibinfo{volume}{19}, \bibinfo{number}{1} (\bibinfo{year}{2010}),
  \bibinfo{pages}{364--379}.
\newblock


\bibitem[\protect\citeauthoryear{Spence, Senkowski, and R{\"o}der}{Spence
  et~al\mbox{.}}{2009}]%
        {spence2009crossmodal}
\bibfield{author}{\bibinfo{person}{Charles Spence}, \bibinfo{person}{Daniel
  Senkowski}, {and} \bibinfo{person}{Brigitte R{\"o}der}.}
  \bibinfo{year}{2009}\natexlab{}.
\newblock \bibinfo{title}{Crossmodal processing}.
\newblock
\newblock


\bibitem[\protect\citeauthoryear{Steuer}{Steuer}{1992}]%
        {steuer1992defining}
\bibfield{author}{\bibinfo{person}{Jonathan Steuer}.}
  \bibinfo{year}{1992}\natexlab{}.
\newblock \showarticletitle{Defining virtual reality: Dimensions determining
  telepresence}.
\newblock \bibinfo{journal}{\emph{Journal of Comm.}} \bibinfo{volume}{42},
  \bibinfo{number}{4} (\bibinfo{year}{1992}).
\newblock


\bibitem[\protect\citeauthoryear{Stoj{\v{s}}i{\'c}, Ivkov-D{\v{z}}igurski, and
  Mari{\v{c}}i{\'c}}{Stoj{\v{s}}i{\'c} et~al\mbox{.}}{2019}]%
        {stojvsic2019virtual}
\bibfield{author}{\bibinfo{person}{Ivan Stoj{\v{s}}i{\'c}},
  \bibinfo{person}{An{\dj}elija Ivkov-D{\v{z}}igurski}, {and}
  \bibinfo{person}{Olja Mari{\v{c}}i{\'c}}.} \bibinfo{year}{2019}\natexlab{}.
\newblock \showarticletitle{Virtual reality as a learning tool: How and where
  to start with immersive teaching}.
\newblock In \bibinfo{booktitle}{\emph{Didactics of Smart Pedagogy}}.
  \bibinfo{publisher}{Springer}, \bibinfo{pages}{353--369}.
\newblock


\bibitem[\protect\citeauthoryear{Strandholt, Dogaru, Nilsson,
  et~al\mbox{.}}{Strandholt et~al\mbox{.}}{2020}]%
        {strandholt2020knock}
\bibfield{author}{\bibinfo{person}{P. Strandholt}, \bibinfo{person}{O. Dogaru},
  \bibinfo{person}{N. Nilsson}, {et~al\mbox{.}}}
  \bibinfo{year}{2020}\natexlab{}.
\newblock \showarticletitle{Knock on Wood: Combining Redirected Touching and
  Physical Props for Tool-Based Interaction in Virtual Reality}. In
  \bibinfo{booktitle}{\emph{Proc. of the Conf. on Human Factors in Computing
  Systems}}.
\newblock


\bibitem[\protect\citeauthoryear{Strasnick, Holz, Ofek, Sinclair, and
  Benko}{Strasnick et~al\mbox{.}}{2018}]%
        {strasnick2018haptic}
\bibfield{author}{\bibinfo{person}{E. Strasnick}, \bibinfo{person}{C. Holz},
  \bibinfo{person}{E. Ofek}, \bibinfo{person}{M. Sinclair}, {and}
  \bibinfo{person}{H. Benko}.} \bibinfo{year}{2018}\natexlab{}.
\newblock \showarticletitle{Haptic links: Bimanual haptics for virtual reality
  using variable stiffness actuation}. In \bibinfo{booktitle}{\emph{Proceedings
  of the Conference on Human Factors in Computing Systems}}.
\newblock


\bibitem[\protect\citeauthoryear{Sun, Patney, Wei, Shapira, Lu, Asente, Zhu,
  McGuire, Luebke, and Kaufman}{Sun et~al\mbox{.}}{2018}]%
        {sun2018}
\bibfield{author}{\bibinfo{person}{Q. Sun}, \bibinfo{person}{A. Patney},
  \bibinfo{person}{L. Wei}, \bibinfo{person}{O. Shapira}, \bibinfo{person}{J.
  Lu}, \bibinfo{person}{P. Asente}, \bibinfo{person}{S. Zhu},
  \bibinfo{person}{M. McGuire}, \bibinfo{person}{D. Luebke}, {and}
  \bibinfo{person}{A. Kaufman}.} \bibinfo{year}{2018}\natexlab{}.
\newblock \showarticletitle{Towards virtual reality infinite walking: Dynamic
  saccadic redirection}.
\newblock \bibinfo{journal}{\emph{ACM Trans. on Graph. (TOG)}}
  \bibinfo{volume}{37}, \bibinfo{number}{4} (\bibinfo{year}{2018}),
  \bibinfo{pages}{67}.
\newblock


\bibitem[\protect\citeauthoryear{Swapp, Pawar, and Loscos}{Swapp
  et~al\mbox{.}}{2006}]%
        {swapp2006}
\bibfield{author}{\bibinfo{person}{David Swapp}, \bibinfo{person}{Vijay Pawar},
  {and} \bibinfo{person}{C{\'e}line Loscos}.} \bibinfo{year}{2006}\natexlab{}.
\newblock \showarticletitle{Interaction with co-located haptic feedback in
  virtual reality}.
\newblock \bibinfo{journal}{\emph{Virtual Reality}} \bibinfo{volume}{10},
  \bibinfo{number}{1} (\bibinfo{year}{2006}), \bibinfo{pages}{24--30}.
\newblock


\bibitem[\protect\citeauthoryear{Sz{\`a}kely and Satava}{Sz{\`a}kely and
  Satava}{1999}]%
        {satava1999virtual}
\bibfield{author}{\bibinfo{person}{G. Sz{\`a}kely} {and} \bibinfo{person}{R.
  Satava}.} \bibinfo{year}{1999}\natexlab{}.
\newblock \showarticletitle{Virtual reality in medicine}.
\newblock \bibinfo{journal}{\emph{BMJ}} \bibinfo{volume}{319},
  \bibinfo{number}{7220} (\bibinfo{year}{1999}), \bibinfo{pages}{1305}.
\newblock
\showISSN{0959-8138}


\bibitem[\protect\citeauthoryear{Taljaard}{Taljaard}{2016}]%
        {taljaard2016review}
\bibfield{author}{\bibinfo{person}{Johann Taljaard}.}
  \bibinfo{year}{2016}\natexlab{}.
\newblock \showarticletitle{A Review of Multi-Sensory Technologies in a
  Science, Technology, Engineering, Arts and Mathematics (STEAM) Classroom.}
\newblock \bibinfo{journal}{\emph{Journal of Learning Design}}
  \bibinfo{volume}{9}, \bibinfo{number}{2} (\bibinfo{year}{2016}),
  \bibinfo{pages}{46--55}.
\newblock


\bibitem[\protect\citeauthoryear{Tang, Patel, Guo, and Prabhakaran}{Tang
  et~al\mbox{.}}{2010}]%
        {tang2010multimodal}
\bibfield{author}{\bibinfo{person}{Ziying Tang}, \bibinfo{person}{Anant Patel},
  \bibinfo{person}{Xiaohu Guo}, {and} \bibinfo{person}{Balakrishnan
  Prabhakaran}.} \bibinfo{year}{2010}\natexlab{}.
\newblock \showarticletitle{A multimodal virtual environment for interacting
  with 3d deformable models}. In \bibinfo{booktitle}{\emph{Proceedings of the
  ACM International Conference on Multimedia}}.
\newblock


\bibitem[\protect\citeauthoryear{Tinsley, Molodtsov, Prevedel, Wartmann,
  Espigul{\'e}-Pons, Lauwers, and Vaziri}{Tinsley et~al\mbox{.}}{2016}]%
        {tinsley2016direct}
\bibfield{author}{\bibinfo{person}{Jonathan~N Tinsley},
  \bibinfo{person}{Maxim~I Molodtsov}, \bibinfo{person}{Robert Prevedel},
  \bibinfo{person}{David Wartmann}, \bibinfo{person}{Jofre Espigul{\'e}-Pons},
  \bibinfo{person}{Mattias Lauwers}, {and} \bibinfo{person}{Alipasha Vaziri}.}
  \bibinfo{year}{2016}\natexlab{}.
\newblock \showarticletitle{Direct detection of a single photon by humans}.
\newblock \bibinfo{journal}{\emph{Nature Communications}} \bibinfo{volume}{7},
  \bibinfo{number}{1} (\bibinfo{year}{2016}), \bibinfo{pages}{1--9}.
\newblock


\bibitem[\protect\citeauthoryear{Treue}{Treue}{2003}]%
        {treue2003visual}
\bibfield{author}{\bibinfo{person}{S. Treue}.} \bibinfo{year}{2003}\natexlab{}.
\newblock \showarticletitle{Visual attention: the where, what, how and why of
  saliency}.
\newblock \bibinfo{journal}{\emph{Current Opinion in Neurobiology}}
  (\bibinfo{year}{2003}).
\newblock


\bibitem[\protect\citeauthoryear{Tsakiris}{Tsakiris}{2017}]%
        {tsakiris2017multisensory}
\bibfield{author}{\bibinfo{person}{Manos Tsakiris}.}
  \bibinfo{year}{2017}\natexlab{}.
\newblock \showarticletitle{The multisensory basis of the self: from body to
  identity to others}.
\newblock \bibinfo{journal}{\emph{The Quarterly Journal of Experimental
  Psychology}} \bibinfo{volume}{70}, \bibinfo{number}{4}
  (\bibinfo{year}{2017}), \bibinfo{pages}{597--609}.
\newblock


\bibitem[\protect\citeauthoryear{Tuthill and Azim}{Tuthill and Azim}{2018}]%
        {tuthill2018proprioception}
\bibfield{author}{\bibinfo{person}{John~C Tuthill} {and} \bibinfo{person}{Eiman
  Azim}.} \bibinfo{year}{2018}\natexlab{}.
\newblock \showarticletitle{Proprioception}.
\newblock \bibinfo{journal}{\emph{Current Biology}} \bibinfo{volume}{28},
  \bibinfo{number}{5} (\bibinfo{year}{2018}), \bibinfo{pages}{R194--R203}.
\newblock


\bibitem[\protect\citeauthoryear{Valori, McKenna-Plumley, Bayramova,
  Zandonella~C., Alto{\`e}, and Farroni}{Valori et~al\mbox{.}}{2020}]%
        {valori2020proprioceptive}
\bibfield{author}{\bibinfo{person}{I. Valori}, \bibinfo{person}{P.
  McKenna-Plumley}, \bibinfo{person}{R. Bayramova}, \bibinfo{person}{Claudio
  Zandonella~C.}, \bibinfo{person}{G. Alto{\`e}}, {and} \bibinfo{person}{T.
  Farroni}.} \bibinfo{year}{2020}\natexlab{}.
\newblock \showarticletitle{Proprioceptive accuracy in Immersive Virtual
  Reality: A developmental perspective}.
\newblock \bibinfo{journal}{\emph{PloS one}} \bibinfo{volume}{15},
  \bibinfo{number}{1} (\bibinfo{year}{2020}).
\newblock


\bibitem[\protect\citeauthoryear{Van~der Burg, Olivers, Bronkhorst, and
  Theeuwes}{Van~der Burg et~al\mbox{.}}{2008}]%
        {van2008pip}
\bibfield{author}{\bibinfo{person}{E. Van~der Burg}, \bibinfo{person}{C.~NL
  Olivers}, \bibinfo{person}{A. Bronkhorst}, {and} \bibinfo{person}{J.
  Theeuwes}.} \bibinfo{year}{2008}\natexlab{}.
\newblock \showarticletitle{Pip and pop: nonspatial auditory signals improve
  spatial visual search.}
\newblock \bibinfo{journal}{\emph{Journal of Experimental Psychology: Human
  Perception and Performance}} \bibinfo{volume}{34}, \bibinfo{number}{5}
  (\bibinfo{year}{2008}).
\newblock


\bibitem[\protect\citeauthoryear{Van~der Burg, Olivers, Bronkhorst, and
  Theeuwes}{Van~der Burg et~al\mbox{.}}{2009}]%
        {van2009poke}
\bibfield{author}{\bibinfo{person}{Erik Van~der Burg},
  \bibinfo{person}{Christian~NL Olivers}, \bibinfo{person}{Adelbert~W
  Bronkhorst}, {and} \bibinfo{person}{Jan Theeuwes}.}
  \bibinfo{year}{2009}\natexlab{}.
\newblock \showarticletitle{Poke and pop: Tactile--visual synchrony increases
  visual saliency}.
\newblock \bibinfo{journal}{\emph{Neuroscience letters}} \bibinfo{volume}{450},
  \bibinfo{number}{1} (\bibinfo{year}{2009}), \bibinfo{pages}{60--64}.
\newblock


\bibitem[\protect\citeauthoryear{Van~der Meijden and Schijven}{Van~der Meijden
  and Schijven}{2009}]%
        {van2009value}
\bibfield{author}{\bibinfo{person}{Olivier~AJ Van~der Meijden} {and}
  \bibinfo{person}{Marlies~P Schijven}.} \bibinfo{year}{2009}\natexlab{}.
\newblock \showarticletitle{The value of haptic feedback in conventional and
  robot-assisted minimal invasive surgery and virtual reality training: a
  current review}.
\newblock \bibinfo{journal}{\emph{Surgical endoscopy}} \bibinfo{volume}{23},
  \bibinfo{number}{6} (\bibinfo{year}{2009}).
\newblock


\bibitem[\protect\citeauthoryear{Viaud-Delmon, Warusfel, Seguelas, Rio, and
  Jouvent}{Viaud-Delmon et~al\mbox{.}}{2006}]%
        {viaud2006high}
\bibfield{author}{\bibinfo{person}{Isabelle Viaud-Delmon},
  \bibinfo{person}{Olivier Warusfel}, \bibinfo{person}{Angeline Seguelas},
  \bibinfo{person}{Emmanuel Rio}, {and} \bibinfo{person}{Roland Jouvent}.}
  \bibinfo{year}{2006}\natexlab{}.
\newblock \showarticletitle{High sensitivity to multisensory conflicts in
  agoraphobia exhibited by virtual reality}.
\newblock \bibinfo{journal}{\emph{European Psychiatry}} \bibinfo{volume}{21},
  \bibinfo{number}{7} (\bibinfo{year}{2006}).
\newblock


\bibitem[\protect\citeauthoryear{Vorl{\"a}nder, Schr{\"o}der, Pelzer, and
  Wefers}{Vorl{\"a}nder et~al\mbox{.}}{2015}]%
        {vorlander2015virtual}
\bibfield{author}{\bibinfo{person}{Michael Vorl{\"a}nder},
  \bibinfo{person}{Dirk Schr{\"o}der}, \bibinfo{person}{S{\"o}nke Pelzer},
  {and} \bibinfo{person}{Frank Wefers}.} \bibinfo{year}{2015}\natexlab{}.
\newblock \showarticletitle{Virtual reality for architectural acoustics}.
\newblock \bibinfo{journal}{\emph{Journal of Building Performance Simulation}}
  \bibinfo{volume}{8}, \bibinfo{number}{1} (\bibinfo{year}{2015}),
  \bibinfo{pages}{15--25}.
\newblock


\bibitem[\protect\citeauthoryear{Walker and Lindsay}{Walker and
  Lindsay}{2003}]%
        {walker2003effect}
\bibfield{author}{\bibinfo{person}{Bruce~N Walker} {and} \bibinfo{person}{Jeff
  Lindsay}.} \bibinfo{year}{2003}\natexlab{}.
\newblock \showarticletitle{Effect of beacon sounds on navigation performance
  in a virtual reality environment}. Georgia Institute of Technology.
\newblock


\bibitem[\protect\citeauthoryear{Wallgr{\"u}n, Bagher, Sajjadi, and
  Klippel}{Wallgr{\"u}n et~al\mbox{.}}{2020}]%
        {wallgrun2020comparison}
\bibfield{author}{\bibinfo{person}{J. Wallgr{\"u}n}, \bibinfo{person}{M.
  Bagher}, \bibinfo{person}{P. Sajjadi}, {and} \bibinfo{person}{A. Klippel}.}
  \bibinfo{year}{2020}\natexlab{}.
\newblock \showarticletitle{A Comparison of Visual Attention Guiding Approaches
  for 360° Image-Based VR Tours}. In \bibinfo{booktitle}{\emph{IEEE Conference
  on Virtual Reality and 3D User Interfaces (VR)}}. IEEE,
  \bibinfo{pages}{83--91}.
\newblock


\bibitem[\protect\citeauthoryear{Waltemate, Gall, Roth, Botsch, and
  Latoschik}{Waltemate et~al\mbox{.}}{2018}]%
        {waltemate2018impact}
\bibfield{author}{\bibinfo{person}{T. Waltemate}, \bibinfo{person}{D. Gall},
  \bibinfo{person}{D. Roth}, \bibinfo{person}{M. Botsch}, {and}
  \bibinfo{person}{M. Latoschik}.} \bibinfo{year}{2018}\natexlab{}.
\newblock \showarticletitle{The impact of avatar personalization and immersion
  on virtual body ownership, presence, and emotional response}.
\newblock \bibinfo{journal}{\emph{IEEE Tran. on Vis. and Com. Graph.}}
  (\bibinfo{year}{2018}).
\newblock


\bibitem[\protect\citeauthoryear{Wang, Lubetzky, Gospodarek, TaghaviDilamani,
  and Perlin}{Wang et~al\mbox{.}}{2019}]%
        {wang2019virtual}
\bibfield{author}{\bibinfo{person}{Zhu Wang}, \bibinfo{person}{Anat Lubetzky},
  \bibinfo{person}{Marta Gospodarek}, \bibinfo{person}{Makan TaghaviDilamani},
  {and} \bibinfo{person}{Ken Perlin}.} \bibinfo{year}{2019}\natexlab{}.
\newblock \showarticletitle{Virtual Environments for Rehabilitation of Postural
  Control Dysfunction}.
\newblock \bibinfo{journal}{\emph{arXiv preprint arXiv:1902.10223}}
  (\bibinfo{year}{2019}).
\newblock


\bibitem[\protect\citeauthoryear{Whitmire, Benko, Holz, Ofek, and
  Sinclair}{Whitmire et~al\mbox{.}}{2018}]%
        {whitmire2018haptic}
\bibfield{author}{\bibinfo{person}{E. Whitmire}, \bibinfo{person}{H. Benko},
  \bibinfo{person}{C. Holz}, \bibinfo{person}{E. Ofek}, {and}
  \bibinfo{person}{M. Sinclair}.} \bibinfo{year}{2018}\natexlab{}.
\newblock \showarticletitle{Haptic revolver: Touch, shear, texture, and shape
  rendering on a reconfigurable virtual reality controller}. In
  \bibinfo{booktitle}{\emph{Proc. of the Conf. on Human Factors in Computing
  Systems}}.
\newblock


\bibitem[\protect\citeauthoryear{Wilberz, Leschtschow, Trepkowski, Maiero,
  Kruijff, and Riecke}{Wilberz et~al\mbox{.}}{2020}]%
        {wilberz2020facehaptics}
\bibfield{author}{\bibinfo{person}{A. Wilberz}, \bibinfo{person}{D.
  Leschtschow}, \bibinfo{person}{C. Trepkowski}, \bibinfo{person}{J. Maiero},
  \bibinfo{person}{E. Kruijff}, {and} \bibinfo{person}{B. Riecke}.}
  \bibinfo{year}{2020}\natexlab{}.
\newblock \showarticletitle{Facehaptics: Robot arm based versatile facial
  haptics for immersive environments}. In \bibinfo{booktitle}{\emph{Proc. of
  the Conf. on Human Factors in Computing Systems}}.
\newblock


\bibitem[\protect\citeauthoryear{Wilcox, Allison, Elfassy, and Grelik}{Wilcox
  et~al\mbox{.}}{2006}]%
        {wilcox2006personal}
\bibfield{author}{\bibinfo{person}{Laurie~M Wilcox}, \bibinfo{person}{Robert~S
  Allison}, \bibinfo{person}{Samuel Elfassy}, {and} \bibinfo{person}{Cynthia
  Grelik}.} \bibinfo{year}{2006}\natexlab{}.
\newblock \showarticletitle{Personal space in virtual reality}.
\newblock \bibinfo{journal}{\emph{ACM Trans. on Perception}}
  \bibinfo{volume}{3}, \bibinfo{number}{4} (\bibinfo{year}{2006}),
  \bibinfo{pages}{412--428}.
\newblock


\bibitem[\protect\citeauthoryear{Wilson and Soranzo}{Wilson and
  Soranzo}{2015}]%
        {wilson2015use}
\bibfield{author}{\bibinfo{person}{Christopher~J Wilson} {and}
  \bibinfo{person}{Alessandro Soranzo}.} \bibinfo{year}{2015}\natexlab{}.
\newblock \showarticletitle{The use of virtual reality in psychology: A case
  study in visual perception}.
\newblock \bibinfo{journal}{\emph{Computational and Mathematical Methods in
  Medicine}}  \bibinfo{volume}{2015} (\bibinfo{year}{2015}).
\newblock


\bibitem[\protect\citeauthoryear{Winther, Ravindran, Svendsen, and
  Feuchtner}{Winther et~al\mbox{.}}{2020}]%
        {winther2020design}
\bibfield{author}{\bibinfo{person}{F. Winther}, \bibinfo{person}{L. Ravindran},
  \bibinfo{person}{Kasper~P. Svendsen}, {and} \bibinfo{person}{T. Feuchtner}.}
  \bibinfo{year}{2020}\natexlab{}.
\newblock \showarticletitle{Design and evaluation of a VR training simulation
  for pump maintenance}. In \bibinfo{booktitle}{\emph{Extended Abstracts of the
  Conf. on Human Factors in Computing Systems}}. \bibinfo{pages}{1--8}.
\newblock


\bibitem[\protect\citeauthoryear{Xu, Li, Zhang, and Le~Callet}{Xu
  et~al\mbox{.}}{2020}]%
        {xu2020state}
\bibfield{author}{\bibinfo{person}{Mai Xu}, \bibinfo{person}{Chen Li},
  \bibinfo{person}{Shanyi Zhang}, {and} \bibinfo{person}{Patrick Le~Callet}.}
  \bibinfo{year}{2020}\natexlab{}.
\newblock \showarticletitle{State-of-the-art in 360 video/image processing:
  Perception, assessment and compression}.
\newblock \bibinfo{journal}{\emph{IEEE Journal of Selected Topics in Signal
  Processing}} \bibinfo{volume}{14}, \bibinfo{number}{1}
  (\bibinfo{year}{2020}), \bibinfo{pages}{5--26}.
\newblock


\bibitem[\protect\citeauthoryear{Yu and Brewster}{Yu and Brewster}{2002}]%
        {yu2002multimodal}
\bibfield{author}{\bibinfo{person}{Wai Yu} {and} \bibinfo{person}{Stephen
  Brewster}.} \bibinfo{year}{2002}\natexlab{}.
\newblock \showarticletitle{Multimodal virtual reality versus printed medium in
  visualization for blind people}. In \bibinfo{booktitle}{\emph{Proceedings of
  the International ACM Conference on Assistive Technologies}}.
  \bibinfo{pages}{57--64}.
\newblock


\bibitem[\protect\citeauthoryear{Yuan and Steed}{Yuan and Steed}{2010}]%
        {yuan2010}
\bibfield{author}{\bibinfo{person}{Ye Yuan} {and} \bibinfo{person}{Anthony
  Steed}.} \bibinfo{year}{2010}\natexlab{}.
\newblock \showarticletitle{Is the rubber hand illusion induced by immersive
  virtual reality?}. In \bibinfo{booktitle}{\emph{2010 IEEE Virtual Reality
  Conference (VR)}}. IEEE, \bibinfo{pages}{95--102}.
\newblock


\bibitem[\protect\citeauthoryear{Zhu, Zhai, and Min}{Zhu et~al\mbox{.}}{2018}]%
        {zhu2018prediction}
\bibfield{author}{\bibinfo{person}{Y. Zhu}, \bibinfo{person}{G. Zhai}, {and}
  \bibinfo{person}{X. Min}.} \bibinfo{year}{2018}\natexlab{}.
\newblock \showarticletitle{The prediction of head and eye movement for 360
  degree images}.
\newblock \bibinfo{journal}{\emph{Signal Processing: Image Communication}}
  \bibinfo{volume}{69} (\bibinfo{year}{2018}), \bibinfo{pages}{15--25}.
\newblock


\end{thebibliography}
